\documentclass[11pt]{article}

\usepackage[margin=1.37in]{geometry}
\usepackage[english]{babel}
\usepackage{epsf}
\usepackage[]{natbib}
\usepackage{amsfonts}
\usepackage{amsmath}
\usepackage{amssymb}
\usepackage{amsthm}
\usepackage{hyperref}
\usepackage[noabbrev,capitalize]{cleveref}
\usepackage{natbib}
\setcitestyle{square,numbers}
\usepackage{bbm}
\usepackage[makeroom]{cancel}
\usepackage{tikz}
\usetikzlibrary{positioning}
\definecolor {processblue}{cmyk}{0.96,0.04,0,0}
\usepackage{algorithm}
\usepackage[utf8]{inputenc}
\usepackage[noend]{algpseudocode}
\usepackage{verbatim}
\usepackage{varwidth}
\usepackage{relsize}
\usepackage{textcomp}

\newcommand{\rt}[1]{{\tiny \bf \color{red}{#1}}}

\newcommand{\ra}{\rightarrow}
\newcommand{\cg}{\mathcal{C}}
\newcommand{\cj}{\mathcal{J}}

\newcommand{\xra}[1]{\xrightarrow{#1}}
\newcommand{\myfloor}[1]{\lfloor #1 \rfloor}
\newcommand{\myceil}[1]{\lceil #1 \rceil}
\newcommand{\pp}[1]{\partial #1}
\newcommand{\npc}{$\mathcal{NP}$-\emph{complete}}

\newcommand{\hs}{\hspace{0.1in}}

\newcommand{\ol}{\emph{orbital line }}
\newcommand{\bfz}{{\bf 0}}
\newcommand{\bfo}{{\bf 1}}
\newcommand{\bfx}{{\bf x}}
\newcommand{\siga}{\sigma_{\bf a}}
\newcommand{\ee}{{ }}
\newcommand{\ssp}{\emph{Subset Sum Problem}}
\newcommand{\mc}[1]{\mathcal{#1}}
\newcommand{\phc}{\pp{\hat{c}}}
\newcommand{\pc}{\pp{c}}

\newcommand{\hg}{\mathcal{H}}
\newcommand{\shp}[1]{\alpha_{#1}^{\ast}}
\newcommand{\lop}[1]{\alpha_{#1}^{\square}}
\newcommand{\bshp}[1]{\beta_{#1}^{\ast}}
\newcommand{\blop}[1]{\beta_{#1}^{\square}}
\newcommand{\bg}{\bar{\gamma}}
\newcommand{\zgamma}{\mathring{\gamma}}
\newcommand{\mcu}{\mathcal U}

\newtheoremstyle{theorembb}
  {\topsep}
  {\topsep}
  {\slshape}
  {0pt}
  {\bfseries}
  {. }
  { }
  {\thmname{#1}\thmnumber{ #2}\textnormal{\thmnote{ (#3)}}}

\theoremstyle{theorembb}

\newtheorem{theorem}{\normalfont\scshape Theorem}[section]
\newtheorem{definition}[theorem]{\normalfont\scshape Definition}
\newtheorem{lemma}[theorem]{\normalfont\scshape Lemma}
\newtheorem{corollary}[theorem]{\normalfont\scshape Corollary}
\newtheorem{identity}[theorem]{\normalfont\scshape Identity}
\newtheorem{proposition}[theorem]{\normalfont\scshape Proposition}
\newtheorem{example}[theorem]{\normalfont\scshape Example}
\newtheorem{question}[theorem]{\normalfont\scshape Question}
\newtheorem{observation}[theorem]{\normalfont\scshape Observation}
\newtheorem{remark}[theorem]{\normalfont\scshape Remark}
\newtheorem{notation}[theorem]{\normalfont\scshape Notation}
\newtheorem{claim}[theorem]{\normalfont\scshape Claim}

\numberwithin{equation}{section}

\hypersetup{
    colorlinks=true, 
    linktoc=all,     
    linkcolor=blue,  
}
\hypersetup{linktocpage}

\begin{document}
\title{ Geometry Of The Subset Sum Problem - Part I }
\author{Srinivas Balaji Bollepalli}
\date{\today}
\maketitle

\begin{abstract}
	We announce two breakthrough results concerning important questions in the Theory of Computational Complexity. In this expository paper, 
	a systematic and comprehensive geometric characterization of 
        the Subset Sum Problem is presented. We show the existence of a universal geometric structure, comprised of a family of
        non-decreasing paths in the Cartesian plane, that captures any instance of the problem of size $n$.
        Inspired by the geometric structure, we provide an unconditional, deterministic and
        polynomial time algorithm, albeit with fairly high complexity, thereby showing that ${\mathcal P} = \mathcal{NP}$.
        Furthermore, our algorithm also outputs the number of solutions to the problem in polynomial time, thus leading to $\mathcal{FP} = \mathcal{\#P}$.
	As a bonus, one important consequence of our results, out of many, is that the quantum-polynomial class $\mathcal{BQP} \subseteq \mathcal{P}$.

        Not only this, but we show that when multiple solutions exist, they can be placed in certain equivalence
        classes based on geometric attributes, and be compactly represented by a polynomial sized directed acyclic graph. 
	We show that the Subset Sum Problem has two aspects, namely  a \emph{combinatorial} aspect and a \emph{relational} aspect, and that it is the latter
	which is the primary determiner of complexity.
        We reveal a surprising connection between the size of the elements and their number, and the precise way
        in which they affect the complexity. In particular, we show that for all instances of the Subset Sum Problem, the complexity is
	independent of the size of elements, once the difference between consecutive elements exceeds $\myceil{7\log{n}}$ bits in size.

	We derive a metric to measure the degree of additive disorder (equivalently additive structure) in a given instance, based on some geometric translations, 
	and show how this metric determines the computational complexity. 
	We present examples of instances with varying degrees of additive structure to show this connection explicitly.
	We also discuss a relation between the geometric structure and the additive
	structure of the Subset Sum Problem. Along with our analysis technique, these results may be of general interest to \emph{Additive Combinatorics}.

	We provide some numerical examples to illustrate the algorithm, and also show how it can be used to estimate some difficult combinatorial
	quantities such as the number of restricted partitions.
\end{abstract}

\newpage
\tableofcontents

\section{Introduction}
\label{sec:intro}

Given a sequence of positive numbers $(a_1,a_2,\ldots,a_n)$, is there a subset that sums to a given target number $T$?
This is the famous \emph{Subset Sum Problem (SSP)} that is easily understood by anyone who knows how to add two numbers. 
In fact, young school children not only understand it, but even successfully solve it when it involves just a handful of small numbers.
All they have to do is try out various possible combinations, and check if
there is a match, i.e., perform an \emph{exhaustive search}. 
However, as $n$ increases and the size of numbers gets bigger, the problem becomes hard quickly, and the famousness turns to notoriety, particularly when one
fails to recognize any pattern. 
Could there be a pattern, in general, in the first place that obviates the need for an exhaustive search? 
We show that the rich geometric structure of the \emph{Subset Sum Problem}, indeed helps us to avoid exponential search.

\subsection{Previous approaches}

There has been only an incremental improvement to the \emph{exhaustive search} coming from the work of Horowitz and Sahni ~\citep{hs74}, 
with a time complexity of $O(n2^\frac{n}{2})$. 
It is a \emph{divide-and-conquer} type of algorithm and consists of: (1) breaking the problem into (roughly) two halves, 
(2) enumerating all the subset sums for each half in increasing
order and (3) checking for a combination of elements one from each half
such that their sum is $T$. In this so called {\it two-list} algorithm, the computational effort of enumeration and 
sorting is $O(n2^\frac{n}{2})$ while checking for solution
can be done in $O(2^\frac{n}{2})$. Therefore the overall complexity is $O(n2^\frac{n}{2})$. It is noteworthy that no better algorithm in terms
of \emph{worst-case} time complexity than this has been found since 1974. An improvement in terms of space was made in ~\citep{rs81}, where the worst case
space needs were shown to be $O(n2^\frac{n}{4})$, but the time complexity remained at $O(n2^{n/2})$. 

Following this idea of splitting the given set
into two parts, and with the goal of beating the $O(n2^{n/2})$ bound, Howgrave-Graham and Joux~\cite{hgj2010} obtained an improved bound of 
$\tilde{O}(2^{0.337n})$ (the symbol $\tilde{O}$ means that some polynomial factors are suppressed)
by splitting into four parts for instances of \emph{average case}
complexity. This was followed by another improvement by Becker, Coron and Joux~\cite{bcj2011}, who showed a complexity of $\tilde{O}(2^{0.291n})$.
These approaches being probabilistic in nature, do not always find the solution. More importantly, for instances with no solution, they cannot
certify that there is no solution. 

A pseudo-polynomial time algorithm with a time complexity of $O(n2^m)$ (where $m$ represents the maximum number of bits needed to represent the numbers),
based on \emph{dynamic programming} \cite{gj79,dpv06}, can easily be constructed.
Clearly, this is useful only when $m$ is very small.
Lagarias, Odlyzko and others ~\citep{lo85, cj92} looked at this problem as a {\it Lattice} problem, where \ssp\ee is reduced to a 
\emph{shortest vector problem}(SVP) in a lattice. 
These methods are conditioned on the availability of efficient algorithms to find the \emph{shortest vector} in a lattice.
The currently known polynomial time algorithms provide a \emph{short vector} but do not guarantee a \emph{shortest vector}. As a result the lattice based methods
do not always guarantee a solution, even when one exists.

Apart from these approaches, there is no known unconditionally deterministic algorithm that always solves the \emph{Subset Sum Problem}. 
In addition, there haven't been any known efforts to characterize the structure of the SSP.
Given this, there is no exaggeration in saying that nothing really is known about the structure of the SSP.

\subsection{An important question}

One might take some solace in the modest progress, by noting that the \emph{SSP} is $\mathcal{NP}$-\emph{complete}. 
In other words, it is the hardest problem in $\mathcal{NP}$; The class of problems
where given a solution, the verification is easy (i.e., can be done in polynomial time), but there
is no known efficient method to find a solution in the first place. 
The $\mathcal{NP}$-\emph{completeness} of the \emph{Subset Sum Problem} (along with several other combinatorial problems)
was first proven in a seminal paper by Karp~\cite{k72}, following a landmark result of Cook~\cite{c71}, who in 1971 formally defined the notion of
\emph{NP-completeness} for the first time, by showing that the Boolean satisfiability problem \emph{SAT} is \npc.
This idea of $\mathcal{NP}$-\emph{completeness} was also arrived at independently by Levin in 1973 (see for e.g., the translation~\cite{tr84}),
who called it a ``universal sequential search problem''~\cite{gj79}.
It turns out that any  \npc \ee problem can be
transformed into another, through the notion of \emph{reducibility} \cite{k72}. Using this idea of \emph{reducibility},
hundreds of problems were proved to be \npc, and collected in the form of a book by Garey and 
Johnson \cite{gj79}. The list of \npc\ee problems is growing with new problems being identified on an almost regular basis. 

On the other hand, ${\mathcal P}$ is the class of all problems that can be solved in polynomial time, i.e., in time $n^c$ where $c$ is a positive constant.

The ${\mathcal P}$ vs. ${\mathcal NP}$ problem asks whether these two classes are in fact equal. 
The famous logician Kurt G\"odel wondered about this very question (in a letter to John von Neumann), 
as far back as $1956$, even before these classes were formally defined \cite{jh89, ms92}.

The theory of computational complexity has rapidly grown since the $70$s, with the discovery of many new complexity classes
(apart from the main ${\mathcal P}$ and $\mathcal{NP}$) and continues to grow. 
Out of these, one particular class, namely $\#\mathcal{P}$ (sharp-P) is also relevant to our work. 
It refers to the class of all problems, called \emph{enumeration problems} where we are 
interested in determining the number of solutions to a given instance.
The $\#\mathcal{P}$ class was defined by Valiant~\cite{va77}, who showed that computing the permanent of a matrix
is a complete problem for this class.
Unlike the class $\mathcal{NP}$, where given a solution the verification
can be done in polynomial time, here in the class $\#\mathcal{P}$ given the \emph{number of solutions} to a 
given instance, this fact can be verified efficiently. While the class $\mathcal{NP}$ deals with \emph{decision problems} (with a \emph{yes/no} answer),
the class $\#\mathcal{P}$ deals with \emph{function problems} (the number of satisfying solutions).
Just as \emph{SSP} is a \emph{complete} problem for the class $\mathcal{NP}$,
it turns out that the counting version of \emph{SSP} is \emph{complete} for the $\#\mathcal{P}$ class\cite{pc1994}. 
If we knew the number of subsets that sum to the target number, then clearly we know the answer to the decision problem. Thus 
the counting version feels more \emph{harder} than decision version.
Just as $\#\mathcal{P}$ is the function counterpart to $\mathcal{NP}$, the counterpart to $\mathcal{P}$ is  $\mathcal{FP}$ 
which is the class of all functions that can be computed in polynomial time.

Like the ${\mathcal P}$ vs. ${\mathcal NP}$ question, the function version asks if $\mathcal{FP} = \#\mathcal{P}$?
Both these questions are answered in the affirmative in this paper.

Apart from the above brief remarks on the main complexity
classes of $\mathcal{P,NP,FP}$,  and $\#\mathcal{P}$, we will not further delve on it here.
We do make some remarks on the historical context of the $\mathcal{P}$ vs. $\mathcal{NP}$, and its importance  towards the end of the paper.
There are entire books devoted to it \cite[see for e.g.,][]{pc1994,ms13,ab09,mm15}, which can be consulted for more details.
The history behind the ${\mathcal P}$ vs. ${\mathcal NP}$ question, its importance to \emph{Mathematics} and \emph{Computer Science}
as well as to many other branches of science have been beautifully described in many works \cite{rl10,rl13,ms13,aw19,sa16,co2000}. A popular and
entertaining account of the importance of this problem is narrated in \cite{lf13}. 

To show that these classes are equal, one has to simply provide a polynomial time algorithm to \emph{any one} of the 
thousands of \npc\ee problems. A majority of the researchers in complexity theory, strongly believe that
these classes are unequal. This belief is articulated in the above cited references, as well as many online blogs.
One commonly cited reason for this belief is the \emph{failure} in finding an efficient algorithm for any of
the numerous \npc\ee problems over the past $50+$ years. We discuss more about this in~\cref{sec:epi}.

\subsection{About this work \& presentation style}

This work started with the primary goal to understand the \emph{structure} of the \ssp, with the view that it will provide
an understanding of the structure of \emph{all} $\mathcal{NP}$\emph{-complete problems}. 
We chose to approach this by considering  \emph{geometry of the solution space}, and fortunately it led to the discovery of a fascinating structure.
Having met the initial goal of capturing the solution space by an \emph{universal geometric
structure\footnote{A curious reader may take a sneak peek at~\cref{fig:mainexample} and~\ref{fig:thin_fat}, and return.}}, 
the idea to devise an algorithm came quite naturally. Some obstacles appeared while assessing the algorithm,
but they ended up revealing an \emph{important relation} between the size of the elements in the \ssp\ee and their number, and
how they determine the complexity.

Due to the simplicity of the problem statement, the \ssp\ee has a universal appeal from hobbyists to career mathematicians. 
This fact along with the importance attached 
to the ${\mathcal P}$ vs. $\mathcal{NP}$ question, as well as the original and novel ideas in this work,
prompted me to write the paper in an expository style, with a linear progression of ideas,
at a somewhat leisurely pace by including a lot of details and several examples. 
Due to this adopted style, some readers may find it tedious at some places, while others may see increased clarity.
However, the final judgment belongs to the reader if we have met the goals for clarity and correctness of claims.

\subsection{Ideas \& Outline of the paper}

The ideas used in the paper are elementary in nature consisting of simple geometric concepts, sequences, combinatorics,
paths and path lengths in a graph, shortest and longest paths, and a creative combination of them. 
Several proofs in the paper make use of the principle of
mathematical induction, while some others make use of combinatorics and counting in two ways.

The following is the brief outline of the rest of the paper:
Given that our main goal is to understand the \emph{structure} of the \ssp, roughly the first half of the paper
(\cref{sec:prob,sec:FBM,sec:chains,sec:allchains}) is devoted to the construction of a \emph{universal geometrical structure} comprised of a family of
non-decreasing paths in the Cartesian plane that captures the complete solution
space of any instance. We show two different aspects of each non-decreasing path - a sequence of points and a sequence of \emph{links} called as a \emph{chain}
- and their equivalence. 
We show that the non-decreasing paths admit two symmetries, namely a \emph{reflection symmetry} and a \emph{local
character symmetry}. 
We describe transforms which convert one non-decreasing path to another. Armed with these transforms, we provide a
systematic way to generate a whole family of non-decreasing paths starting with a single path.
In this geometric characterization we have a family of curves that start at the origin and have a common
end point. The target sum $T$ corresponds to a horizontal line (\emph{orbital line}) that intersects all the curves.

In~\cref{sec:poc} we analyze the intersection points with the \emph{orbital line}. For this we introduce \emph{local reference frames}
and \emph{local coordinates}. This is followed by treating each curve as a sequence of linear segments (called \emph{edges}) and
provide relations that translate an intersection point in one edge to another edge.
We make the \emph{edge representation} a central part of the characterization of the \emph{orbital line} and construct a 
directed acyclic graph $G_0$ (\cref{sec:G0}) with \emph{edges} as nodes, and \emph{translations} as arcs. We show that the SSP is equivalent to finding
\emph{zero paths} in $G_0$. We estimate the number of nodes and number of arcs in $G_0$, and show that it is polynomial sized.

We present an iterative algorithm in~\cref{sec:algihm} where we characterize the complexity in terms of a certain \emph{growth factor} $\eta_{peak}$.
In order to bound $\eta_{peak}$ we take a deeper dive into the SSP in~\cref{sec:bhairav} by showing that it has two distinct aspects: a \emph{combinatorial aspect}
and a \emph{relational aspect}. We construct a \emph{configuration graph} that captures these two aspects, and by a systematic analysis obtain a bound
for the maximum number of distinct paths through an edge. We also show that \emph{size doesn't matter beyond a critical value} dependent on $\log{n}$.
In~\cref{sec:guru} we prove the main claims of the paper. In~\cref{sec:adds} we consider a set of instances with varying \emph{degrees of additive structure}
and show how the algorithm performs on them. We also try to provide a connection between the \emph{geometric} and \emph{additive} structures. 
We finally conclude by discussing the historical antecedents of the problem and the significance of our results.
We present several numerical examples along with applications of the SSP algorithm to computing some difficult combinatorial quantities 
such as the number of restricted partitions in~\cref{sec:appendix}.

\section{Problem statement and notation}
\label{sec:prob}

We use the standard notation of $\mathbb{N, R, C}$ denoting the positive integers, reals and complex numbers respectively.
For a positive integer $k$,  the set $\{1,2,\ldots,k\}$ is denoted by $[k]$. For an integer $k$ lying in between integers $i$ and $j$, where $i < j$, 
we write $k \in [i,j]$.

Given $({\bf a},T)$, where ${\bf a} = (a_1,a_2,\ldots,a_n)$ is a sequence of $n$ positive integers and $T$ is another given positive integer,
we are interested in finding the \emph{number of subsets}, if any,
that sum to $T$. In contrast to the problem statement in~\cref{sec:intro}, which is a \emph{decision problem}, we are interested in the
\emph{counting version} here. We assume, without loss of generality that
\begin{equation*}
	a_1 \le a_2 \le \ldots \le a_{n-1} \le a_n,
\end{equation*}
and that $0 < T \le \sum_ja_j$.
We further assume that each of the numbers $a_j$ can be represented using $m \in {\mathbb N}$ digits in binary, where $m$ is independent of $n$.
In other words, $a_n < 2^m$.
Let $b_j := 2^{j-1}$, where $j \in [n]$, be the binary basis sequence ${\bf b} = (2^0,2^1,\ldots,2^{n-1})$.
\begin{definition}
	Let $c_j := b_j + \iota a_j$, where $\iota = \sqrt{-1}$, for each $j \in [n]$ be the $n$ complex numbers formed from the given set ${\bf a}$
	and the binary basis ${\bf b}$. We denote this sequence of $n$ complex numbers by ${\bf c}$, i.e.,
	\begin{equation*}
		{\bf c} = (c_1,c_2,\ldots,c_n).
	\end{equation*}
\end{definition}
We denote the real part by $\Re$ and imaginary part by $\Im$, so that $\Re(c_1) = b_1$ and $\Im(c_1) = a_1$.

\begin{definition}
	For a positive integer $k \in [n]$, we define 
	\begin{align*}
		B_k & := b_1 + b_2 + \ldots + b_k = 2^k-1, \\
		A_k & := a_1 + a_2 + \ldots + a_k, \\
		\text{and } C_k & := B_k + \iota A_k.
	\end{align*}
\end{definition}

\begin{definition}(Power Set)
	Given ${\bf c}$, we denote the power set by $S_n$ which is defined as the set of all subset sums of ${\bf c}$. Formally, we have
	\begin{equation*}
		S_n := \{x_1c_1+x_2c_2+\ldots+x_nc_n : {\bf x} \in \{0,1\}^n\},
	\end{equation*}
	where ${\bf x}$ is a binary vector of length $n$.
	It is clear that the cardinality of $S_n$ denoted by $|S_n| = 2^n$. These subset sums can be represented by $2^n$ points in the complex plane ${\mathbb C}$. 
        We also refer to the power set  sometimes as \emph{point set}. 
        Each point of the point set corresponds to a unique binary vector ${\bf x}$ of length $n$ so that the $index = {\bf b} \cdot {\bf x}$ and 
        the $sum = {\bf a} \cdot {\bf x}$ where $\cdot$ indicates dot product. 

	Since ${\mathbb C} \cong {\mathbb R}^2$, there should not be any confusion when we frequently treat the complex numbers as ordered pairs in ${\mathbb R}^2$.
	It should be emphasized that, we use the term \emph{subset sum} for both the ordinary integer sum as well as the complex
	sum that has index as the real part, and ordinary subset sum as the imaginary part. It should be clear from the context, what is being referred to.
\end{definition}

The power set $S_n$ is our universe, in this article, for studying the \ssp. 
For the given positive integer $T$, we can imagine drawing a line $y=T$ in 
the complex plane. We refer to this line as the \emph{orbital line}, abbreviated as OL.
Then the \ssp, is to determine the number of points of the \emph{point set} that lie on the OL.

\begin{definition}(Extreme Points)
	The points $0$ and $C_n$ which belong to the point set are referred to as the extreme points, since they correspond
	to the smallest and largest subset sums respectively.
	These two extreme points correspond to the binary vectors
\begin{align*}
        \bfz & := (0,0,0,\ldots,0,0) \\
        \text{and } \bfo & := (1,1,1,\ldots,1,1).
\end{align*}
\end{definition}

\begin{definition}
        For a given non-negative integer $r$ (also referred to as index), where $r < 2^n$, we have an associated subset sum $\siga(r)$, defined by
        \begin{equation*}
                \siga(r) := {\sum}_{u=1}^n k_ua_u, \quad \text{ with } k_u \in \{0,1\},
        \end{equation*}
        where $k_nk_{n-1}\ldots k_2k_1$ is the binary expansion of $r$. Note that when $r$ is a small number that doesn't require $n$
        digits, we still use $n$ digits, and the additional digits are treated as zeros.
\end{definition}

\subsection{A running example}
To illustrate the main ideas of the paper, in particular the geometric structure, we use a running example which hereafter will be
referred to as the \emph{main example}.
\begin{example}(Main Example)
This example is an instance of the \ssp\ee with $n = 9$ and $m = 18$, 
\begin{align*}
	{\bf a} & = (43196, 84912, 106109, 119107, 119758, 121761, 125743, 131627, 147792) 
\end{align*}
and  $T =  663708$.
This instance has an unique solution with index equal to $365$ whose binary form is
$[365]_{binary} = 1  0 1 1  0  1  1  0 1$,
and it can be verified that
	$663708 = 147792 + 125743 + 121761 + 119107 + 106109 + 43196$. While the algorithm for the \ssp\ee (to be discussed later) works for any positive $(n,m)$,
we cannot illustrate the notions of the paper for large $n$ without visual difficulty. We found that this is a \emph{good} sized
example to show the point set  and paths through them.
\end{example}

\subsection{Subspaces}
The point set $S_n$ is the complete solution space for SSP. However, it will be beneficial to look at the subspaces as well,
which are defined below.

\begin{definition}(Subspace)
        For the given instance ${\bf a}$, a subspace of dimension $k \in [n]$ corresponds to the sequence $(a_1,a_2,\ldots,a_k)$
	with associated point set $2^{{\bf c}_k}$, where ${\bf c}_k = \{c_1,\ldots,c_k\}$.
        We denote the point set of subspace $k$ by $S_k$.
        Thus, we have $n$ subspaces corresponding to the $n$ possible values of $k$.
        All points in a subspace $S_k$ lie in the interval box $[0, B_k] \times [0,  A_k]$ in the Cartesian plane.
\end{definition}

With the above definition, we can make unambiguous statements such as:
\begin{itemize}
	\item[(i)] The point set $2^{{\bf c}}$ can be seen to be identical to two copies of subspace $2^{{\bf c}_{n-1}}$;
                one located at the origin, and the second located at $c_n$.
	\item[(ii)] Similarly, the point set $2^{{\bf c}}$ is the union of $2^{n-k}$ translates of the subspace $2^{{\bf c}_k}$.
\end{itemize}

\subsection{Additional Definitions}
We frequently refer to some progressions, defined below, that are utilized as some special cases of the \emph{Subset Sum Problem}.

\begin{definition}(Arithmetic Progression (AP))
	An instance ${\bf a}$ is said to be a $n$ term arithmetic progression, denoted by ${\bf a} \in AP(n,k_1,k_2)$, or even simply by ${\bf a} \in AP$,
	if and only if there exist positive integer constants $k_1,k_2 \in {\mathbb N}$ such that
       \begin{align*}
               a_i =  k_1 + (i-1)k_2, \text{ for all } i \in [n].
       \end{align*}
	\label{def:AP}
\end{definition}

\begin{definition}(Constant Progression (CP))
	An instance ${\bf a}$ is said to be a $n$ term constant progression, denoted by ${\bf a} \in CP(n,k_1)$, or even simply by ${\bf a} \in CP$,
       if and only if there is a positive integer constant $k_1 \in {\mathbb N}$ such that
       \begin{align*}
               a_i =  k_1  \text{ for all } i \in [n].
       \end{align*}
	\label{def:CP}
\end{definition}

\begin{definition}(Geometric Progression (GP))
        An instance ${\bf a}$ is said to be a $n$ term geometric progression, denoted by ${\bf a} \in GP(n, r)$, or even simply by ${\bf a} \in GP$,
       if and only if there exists a positive integer constant $r > 1$ such that
       \begin{align*}
               a_i =  r^{i-1}, \text{ for all } i \in [n].
       \end{align*}
        \label{def:GP}
\end{definition}

\begin{definition} (Dissociated set)
	Given an instance of SSP $({\bf a},T)$ we say that ${\bf a}$  is a dissociated set if and only if all the subset sums $2^{\bf a}$ are distinct.
	\label{def:diss}
\end{definition}

\section{Filling boxes, and paths from \bfz \ee to \bfo}
\label{sec:FBM}

Given an instance ${\bf a}$ of size $n$, the point set forms some configuration in the Cartesian plane, and this configuration
changes with each instance. The configuration can take a wide range of shapes: for example, when
${\bf a = b}$, all the points will be on a straight line; when ${\bf a} \in CP$, i.e. all elements are equal to a fixed  positive integer, the points form $n+1$ 
distinct levels along $y$; and for a random instance they appear spread out in the whole box $[0,B_n] \times[0,A_n]$. 
This makes it harder to study and characterize the structure of the point set.

Furthermore, let $x_1,x_2 \in [B_n]$ be two random positive integers, and let $y_1 = \siga(x_1)$ and $y_2 = \siga(x_2)$ be the 
corresponding subset sums. Then the relative positions of $z_1 = (x_1,y_1)$ and $z_2 = (x_2,y_2)$ in the $y$ coordinate can vary with instance.
If $y_1 > y_2$ for one instance, it is possible to have $y_1 < y_2$ 
for another instance of the same size. Given these observations, consider the following question:

\begin{question}
	For a fixed $(n,m)$, it can be seen that there are at most $2^{mn}$ possible instances of ${\bf a}$. 
	Is there a universal structure that captures all these instances, and enable us to make (non-trivial) statements that are true for any instance?
\end{question}
The answer to this question, as the reader will see, is a \emph{resounding yes}!

While the relative ordering of the point set changes with instance, there exist points where a relative ordering is preserved for any instance.
There are two extreme points, namely $(0,0)$ and $(B_n,A_n)$ where the second point is always at least as large in $y$ coordinate as
the first point for all instances of size $n$. 
Recall that these two extreme points correspond to the binary vectors \bfz\ee and \bfo.
We characterize the structure of the solution space through the notion of non-decreasing paths between \bfz\ee and \bfo, which is described next.

\begin{definition}(Path)
	A path $\psi$ of length $N$, from \bfz\ee to \bfo\ee is a sequence of binary vectors $(\bfx_1,\bfx_2,\ldots,\bfx_N)$ where $\bfx_i \in \{0,1\}^n$ for $i \in [N]$,
	such that $\bfz = \bfx_1$,  $\bfo = \bfx_N$, and no vector is repeated. It is clear that $2 \le N \le 2^n$.
\end{definition}

\begin{definition}(Non-Decreasing Path (NDP))
	A path $\psi$ from \bfz\ee to \bfo\ee is said to be a non-decreasing path, or simply a NDP, if and only if the subset sums along the path,
	are non-decreasing for {\bf every} instance ${\bf a}$ of size $n$.
\end{definition}
Consider the following simple example, to see that characterization of NDPs is not obvious even for simple cases.
\begin{example}
	Consider the simple case when $n=3$. The sequence of indices for the power set in increasing order are $\Re(S_n) =  (0,1,2,\ldots,7)$.
        If we write the corresponding sums, we get
        \begin{align*}
                \Im(S_n) = \{0,a_1,a_2,a_1+a_2,a_3,a_3+a_1,a_3+a_2,a_3+a_2+a_1\}.
        \end{align*}
	Does $\Im(S_n)$ form a non-decreasing sequence of sums for every ${\bf a}$?
        A moment's thought reveals that the answer is \emph{no}, since $a_3$ need not be greater than or equal to $a_1+a_2$.
        In other words, the location of $a_3$ in the sequence is uncertain:
        \begin{align*}
                \{0,a_1,a_2,a_1+a_2,\boxed{a_3},a_3+a_1,a_3+a_2,a_3+a_2+a_1\} \quad \text{if } a_3 > a_1+a_2 \\
                \text{or  }  \{0,a_1,a_2,\boxed{a_3},a_1+a_2,a_3+a_1,a_3+a_2,a_3+a_2+a_1\}\quad \text{if } a_3 < a_1+a_2.
        \end{align*}
        For the special case, when $a_3 = a_1 + a_2$, both the sequences will be NDPs.
\end{example}

Why do we care about NDPs? Since NDPs are independent of the instance, they provide a basis for the 
\emph{universal geometric structure} that we seek.
By definition, a NDP $\psi$ remains non-decreasing for all instances of ${\bf a}$ that satisfy $a_1 \le a_2 \le \ldots \le a_n$.
This suggests that the ``non-decreasing property'' should be independent of the magnitude of elements of ${\bf a}$. Alternatively, the
non-decreasing property depends only on the binary vectors (indices) along the path. 
To be able to systematically characterize the NDPs, we consider a \emph{model for filling $n$ boxes with balls}, which is described next.

\subsection{A model for filling $n$ boxes}
We are given $n$ boxes, that abut each other with tops open and arranged in a line from left to right. We also have a bag of $n$ balls
to the immediate right of the rightmost box. Initially all the boxes are empty, and we associate this \emph{empty state} with \bfz.
We label the boxes as $n,n-1,\ldots,2,1$ from left to right. The bag containing the balls is thought of as being at location $0$.
We have a machine that follows a single rule: \emph{move a ball from location $k$ to location $k+1$ provided it is empty, where
$k \in [0, n-1]$}.
For example, in the first step, the machine can only take a ball from the bag and put in box $1$. 
In the second step, the machine picks up the ball in box $1$, and displaces it to box $2$. In the third step, the machine can either
displace the ball in box $2$ to box $3$, or pick a new ball from the bag and put in box $1$. At each step, there is a unit displacement
from one box to the box on left (provided it is empty), or from the bag to box $1$. Note that the displacements are always to the left by unit distance, and
never to the right. Also we cannot displace a ball to a position that is not empty.
This process is repeated until all the boxes are filled.  The state where all the $n$ boxes are filled is associated with \bfo.

The following proposition is straightforward.
\begin{proposition}
	To go from the empty state to the filled state, we need exactly $n(n+1)/2$ unit displacements.
\end{proposition}

\begin{definition}(state and sequence of states)
	We refer to the configuration of balls in the $n$ boxes at any time step as a state. 
	By associating a $0$ with an empty box, and a $1$ with a filled box, a state corresponds to a binary vector of length $n$.
	Let $\pi := (\pi_0,\pi_1,\pi_2,\ldots,\pi_{n(n+1)/2})$ be the sequence of states in filling the boxes, where $\pi_k \in \{0,1\}^n$
	for $k \in \{0,1,\ldots,n(n+1)/2\}$, $\pi_0$ is the empty state, and $\pi_{n(n+1)/2}$ is the filled state. 
\end{definition}

\begin{proposition}
	Any sequence of states obeying the rules for filling the boxes corresponds to a NDP.
\end{proposition}
\begin{proof}
	Since we always move a ball from a lower indexed box to a higher indexed box, it amounts to replacing $a_k$ by $a_{k+1}$
	for $k \in [n-1]$, or adding a $a_1$ to the subset sum. Since ${\bf a}$ satisfies $a_1 \le a_2 \le \ldots \le a_n$, 
	the subset sums corresponding to the states are non-decreasing, and the conclusion follows.
\end{proof}

\subsection{Two complementary ways of filling boxes}
It is clear that, as long as the balls keep moving to the left, one displacement at a time, any sequence of choices can be 
used to fill all the boxes, and the sequence of corresponding binary vectors will result in non-decreasing subset sums. Instead of 
randomly choosing which ball to move at a time, we will focus on two particular ways, temporarily labeled LB and HB:
\begin{itemize}
	\item[(LB)] At each time step, choose the ball at the lowest \emph{possible} bit position that \emph{can} be moved one unit to the left.
	\item[(HB)] At each time step, choose the ball at the highest \emph{possible} bit position that \emph{can} be moved one unit to the left.
\end{itemize}
An example will clarify these two ways of filling all boxes.
\begin{example}
	Let $n=4$. The sequence obtained using LB is on the left, and the sequence obtained using HB is in the center.
	We use $0$ to represent an empty box, and a $1$ for a filled box. The balls enter from the right, starting in the rightmost
	box. Notice that both sequences have a length of
	$1 + 4\times5/2 = 11$. The third table, labeled RC,  uses a random choice.
	\begin{equation*}
		\begin{array}{|c|c|c|c|}
			\hline
			& & LB & \\
			\hline
			0 & 0 & 0 & 0 \\
			\hline
			0 & 0 & 0 & 1 \\
			\hline
			0 & 0 & 1 & 0 \\
			\hline
			0 & 0 & 1 & 1 \\
			\hline
			0 & 1 & 0 & 1 \\
			\hline
			0 & 1 & 1 & 0 \\
			\hline
			0 & 1 & 1 & 1 \\
			\hline
			1 & 0 & 1 & 1 \\
			\hline
			1 & 1 & 0 & 1 \\
			\hline
			1 & 1 & 1 & 0 \\
			\hline
			1 & 1 & 1 & 1 \\
			\hline
		\end{array}
		\quad \quad
		\begin{array}{|c|c|c|c|}
			\hline
			& & HB & \\
			\hline
			0 & 0 & 0 & 0 \\
			\hline
			0 & 0 & 0 & 1 \\
			\hline
			0 & 0 & 1 & 0 \\
			\hline
			0 & 1 & 0 & 0 \\
			\hline
			1 & 0 & 0 & 0 \\
			\hline
			1 & 0 & 0 & 1 \\
			\hline
			1 & 0 & 1 & 0 \\
			\hline
			1 & 1 & 0 & 0 \\
			\hline
			1 & 1 & 0 & 1 \\
			\hline
			1 & 1 & 1 & 0 \\
			\hline
			1 & 1 & 1 & 1 \\
			\hline
		\end{array}
		\quad \quad
		\begin{array}{|c|c|c|c|}
			\hline
			& & RC & \\
			\hline
			0 & 0 & 0 & 0 \\
			\hline
			0 & 0 & 0 & 1 \\
			\hline
			0 & 0 & 1 & 0 \\
			\hline
			0 & 1 & 0 & 0 \\
			\hline
			0 & 1 & 0 & 1 \\
			\hline
			1 & 0 & 0 & 1 \\
			\hline
			1 & 0 & 1 & 0 \\
			\hline
			1 & 0 & 1 & 1 \\
			\hline
			1 & 1 & 0 & 1 \\
			\hline
			1 & 1 & 1 & 0 \\
			\hline
			1 & 1 & 1 & 1 \\
			\hline
		\end{array}
	\end{equation*}
\end{example}

We are interested in LB and HB, as they turn out to be important for the SSP. For an arbitrary $n$, we can generate the LB/HB sequence algorithmically, simply
by applying the rule for $n(n+1)/2$ steps, starting with the zero vector. 

\begin{lemma}
\label{lem:one}
Let rule LB be used to fill $k$ boxes. If we add another box on the left, add a new ball to the bag, and continue using rule LB,
then we get $k+1$ new elements into the sequence.
\end{lemma}
\begin{proof}
For the problem with size $k$, the last element in the sequence has the following binary form of length $k$.
\begin{equation*}
\begin{array}{ccccccc}
x_k & x_{k-1} & x_{k-2} & \ldots & x_3 & x_2 & x_1 \nonumber \\
1 & 1 &  1 & \ldots & 1 & 1 & 1 \nonumber \\
\end{array}
\end{equation*}
Now the next elements will be, using the rules defined above,
\begin{equation*}
\begin{array}{cccccccc}
x_{k+1} & x_k & x_{k-1} & x_{k-2} & \ldots & x_3 & x_2 & x_1 \nonumber \\
\fbox{0} & \fbox{1} & \fbox{1} & \fbox{1} & \ldots & \fbox{1} & \fbox{1} & \fbox{1} \nonumber \\
1 & 0 & 1 & 1 & \ldots & 1 & 1 & 1 \nonumber \\
1 & 1 & 0 & 1 & \ldots & 1 & 1 & 1 \nonumber \\
\vdots \\
1 & 1 & 1 & 1 & \ldots & 1 & 0 & 1 \nonumber \\
1 & 1 & 1 & 1 & \ldots & 1 & 1 & 0 \nonumber \\
1 & 1 & 1 & 1 & \ldots & 1 & 1 & 1 \nonumber \\
\end{array}
\end{equation*}
	Since there are $k$ moves of $1$s and an added $1$ (from bag) for the last element, we get $k+1$ new elements going from problem size $k$ to $k+1$.
\end{proof}

\subsection{Formulae to generate NDS}
It turns out that the non-decreasing sequences described algorithmically above can be generated by formulae.

\begin{definition}
	For a non-negative integer $k$, let
	\begin{equation*}
		\vartheta(k) := \lfloor \frac{-1 + \sqrt{1+8k}}{2} \rfloor
	\end{equation*}
	where $\lfloor x \rfloor$ denotes the largest integer less than or equal to $x$. Thus, $\vartheta(0)=0, \vartheta(1)=1, \vartheta(2)=1,\vartheta(3)=2$ and so on.
	\label{def:theta}
\end{definition}

\begin{theorem}
	The non-decreasing sequence corresponding to using rule LB, denoted by $\phi_n$, can be generated by the formula\footnote{While this formula was discovered independently,
	it is not new. In a different context, the sequence $\phi_n$ is known [see OEIS at https://oeis.org/A089633].}
	\begin{align} 
		{\phi}_n(k) := 2^{1+\vartheta(k)} - 2^{\vartheta(k)-k+\frac{\vartheta(k)(1+\vartheta(k))}{2}}-1,
	\end{align}
	where $k \in [0, \frac{n(n+1)}{2}]$.
	\label{thm:phiformula}
\end{theorem}
\begin{proof}
The proof is given by induction on $n$. \\
(Base Step) Consider first the case when $n=2$. We have a total of $\frac{2.3}{2}=3$ elements in the sequence. Using the formula for $\vartheta$ defined
above and letting $i=0,1,2,3$ we have \\
	\begin{equation*} 
		\vartheta(0) = 0, \vartheta(1) = 1, \vartheta(2) = 1, \text{and } \vartheta(3) =2.
	\end{equation*}
Now the $\phi$ values are
\begin{align*} 
	{\phi}_2(0) &= 2^{(1+0)} - 2^{(0-0+\frac{0.1}{2})} -1 = 0 \equiv [0,0]_{2,2} \\
        {\phi}_2(1) &= 2^{(1+1)} - 2^{(1-1+\frac{1.2}{2})} -1 = 1 \equiv [0,1]_{2,2} \\
        {\phi}_2(2) &= 2^{(1+1)} - 2^{(1-2+\frac{1.2}{2})} -1 = 2 \equiv [1,0]_{2,2} \\
        {\phi}_2(3) &= 2^{(1+2)} - 2^{(2-3+\frac{2.3}{2})} -1 = 3 \equiv [1,1]_{2,2} 
\end{align*}
	where the corresponding binary representation is shown at the right. (On the right, the first subscript means base $2$, and the second one means we are using
	$2$ digits.) So the assertion is true for $n=2$. \\
(Inductive Step) Now, assume that it is true for $n=k > 2$. For this sequence,
the last element has an index equal to $\frac{k(k+1)}{2}$. When we increment $n$ to be equal to $k+1$, we are interested in the range \\
	\begin{equation*} 
		\frac{k(k+1)}{2}+1, \frac{k(k+1)}{2}+2, \ldots, \frac{(k+1)(k+2)}{2}
	\end{equation*}
The corresponding $\vartheta$ values are
\begin{align*}
\vartheta(\frac{k(k+1)}{2}+j) &= \lfloor\frac{-1+\sqrt{(2k+1)^2+8j}}{2}\rfloor \\
&= \begin{cases}
k & \text{if}  \qquad j=1,2,\ldots,k \\
k+1 & \text{if} \qquad  j=k+1
\end{cases}
\end{align*}
Therefore,
\begin{align*}
{\phi}_{k+1}(\frac{k(k+1)}{2}+j) = \begin{cases}
2^{1+k} - 2^{k - (\frac{k(k+1)}{2}+j) + \frac{k(k+1)}{2}} - 1  \text{ if }  j \in \{1,2,\ldots,k\} \\
2^{2+k} - 2^{1+k} -1 \qquad \text{if} \qquad j=k+1
\end{cases}
\end{align*}
First consider the case $j \in [k]$:
it can be seen that $2^{1+k} - 2^{k - (\frac{k(k+1)}{2}+j) + \frac{k(k+1)}{2}} - 1 = 2^{1+k} - 2^{k-j} - 1$ after canceling some terms. For $j = 0$,
we get $2^{1+k} - 2^{k} - 1 = 2^k - 1$ whose binary form consists of $k$ $1$s. This is the final element for $n=k$.
When $j=1$, we get $2^{1+k} - 2^{k-1} - 1 = 2^k + 2^{k-1}-1$ as the $1$ in the $k$th place moves to $k+1$th place leaving a zero at the $k$th place.
As $j$ is increased further, the zero location propagates all the way to the $1$st position. This position gets filled when $j = k+1$.
Thus, the bit transitions happen as in~\cref{lem:one} and we have a non-decreasing sequence. Hence it is true for all $n$.
\end{proof}

\begin{corollary}
	The sequence corresponding to using rule HB, denoted by $\varphi_n$, can be described by the formula 
	\begin{align}
		{\varphi}_n(k) = {\phi}_n^{c}(n(n+1)/2-k), \text{ for } k = 0, 1, \ldots, n(n+1)/2,
	\end{align}
	where  $\phi_n^c(k)$ is the binary $1$s complement of $\phi_n(k)$, i.e.,  
	\begin{align}
		{\phi}_n^{c}(k) = 2^n - 1 - {\phi}_n(k) = 2^n - 2^{(1+\vartheta(k))} + 2^{{\vartheta}(k) - k + \frac{\vartheta(k)(\vartheta(k)+1)}{2}}.
	\end{align}
	\label{cor:psiformula}
\end{corollary}
\begin{proof}
	Since $\phi_n$ is non-decreasing, the reverse of this sequence denoted $REV(\phi_n)$ is non-increasing. Now consider the $1$s complement
	of $REV(\phi_n)$, denoted by $REV(\phi_n^c)$. Clearly, this is non-decreasing. Since the sequence $\phi_n$ corresponds to rule LB, where
	we move the ball in lowest possible bit position, when we apply the reversal and the $1$s complement, it becomes the highest possible bit position.
	As a result, it corresponds to filling boxes using rule HB.
\end{proof}

As a consequence of~\cref{thm:phiformula} and~\cref{cor:psiformula}, we have the following result, due to their complementary nature.

\begin{proposition}
	For any $j \in [n]$, and $k \in [0,j(j+1)/2]$, we have the identities
	\begin{itemize}
		\item[(i)] $\phi_j(k) + \varphi_j(\frac{j(j+1)}{2}-k) = B_j$, and 
		\item[(ii)] $\siga(\phi_j(k)) + \siga(\varphi_j(\frac{j(j+1)}{2}-k)) = A_j$. 
	\end{itemize}
\end{proposition}

The two families of complementary sequences are shown below for clarity.
The first family of sequences labeled ${\phi}_k, k=1,2,\ldots,n$ for varying problem sizes is
\begin{align*}
{\phi}_1 & = \{0, 1\} \\
{\phi}_2 & = \{0, 1 ,2 , 3\} \\
{\phi}_3 & = \{0, 1, 2, \fbox{3} , 5 , 6 , 7\} \\
{\phi}_4 & = \{0, 1, 2, \fbox{3 , 5 , 6 , 7, 11}, 13, 14 , 15\} \\
{\phi}_5 & = \{0, 1, 2, \fbox{3 , 5 , 6 , 7, 11, 13, 14, 15, 23, 27}, 29, 30, 31\} \\
{\phi}_6 & = \{0, 1, 2, \fbox{3 , 5 , 6 , 7, 11, 13, 14, 15, 23, 27, 29, 30, 31, 47, 55, 59}, 61, 62, 63\} \\
\ldots \\
{\phi}_n & = \{0,1 , 2, \boxed{3, 5, \ldots, 2^n-5}, 2^n-3, 2^n-2, 2^n-1\}
\end{align*}
The second family of sequences labeled ${\varphi}_k, k=1,2,\ldots,n$ is
\begin{align*}
{\varphi}_1 & = \{0, 1\} \\
{\varphi}_2 & = \{0, 1, 2, 3\} \\
{\varphi}_3 & = \{0, 1, 2, \fbox{4} , 5 , 6 , 7\} \\
{\varphi}_4 & = \{0, 1, 2, \fbox{4, 8 , 9 , 10, 12}, 13, 14 , 15\} \\
{\varphi}_5 & = \{0, 1, 2, \fbox{4, 8, 16, 17, 18, 20, 24, 25, 26, 28}, 29, 30, 31\} \\
{\varphi}_6 & = \{0, 1, 2, \fbox{4, 8, 16, 32, 33, 34, 36, 40, 48, 49, 50, 52, 56, 57, 58, 60}, 61, 62, 63\} \\
\ldots \\
{\varphi}_n & = \{0, 1, 2, \boxed{4, 8, \ldots, 2^n-4}, 2^n-3, 2^n-2, 2^n-1 \} 
\end{align*}
Note that while ${\phi}_1 = {\varphi}_1$ and ${\phi}_2 = {\varphi}_2$, other pairs are different in some elements. The differences are
\emph{boxed} to make them clear. 
The sequences  ${\phi}_n(i)$ and  ${\varphi}_{n}(i)$ for all $n > 2$ have exactly six points in common. 
These six points in decimal values are $0,1,2,2^n-3,2^n-2$, and $2^n-1$.
As a result, the pair $\{{\phi}_n,{\varphi}_n\}$ of sequences together contain $2(1+\frac{n(n+1)}{2})-6 = n(n+1)-4$ unique points for all $n > 2$.

\subsection{From sequences to curves}

\begin{definition}
	For a given positive integer $j \in [n]$, let $p_j$ be the sequence of complex numbers defined by
        \begin{equation}
		p_j(k) = \phi_j(k) + \iota \siga(\phi_j(k)) \quad \text{ for } k=0,1,2,\ldots,\frac{j(j+1)}{2}.
                \label{eq:vtx1}
        \end{equation}
        Similarly, let $q_j$ be the sequence  of complex numbers defined by
        \begin{equation}
		q_j(k) = \varphi_j(k) + \iota \siga(\varphi_j(k)) \quad \text{ for } k=0,1,2,\ldots,\frac{j(j+1)}{2}.
                \label{eq:vtx2}
        \end{equation}
        The real part corresponds to the index, and the imaginary part corresponds to the subset sum associated with the index,
        for both sequences.
	\label{def:pq_seq}
\end{definition}

By joining adjacent points of $p_j$ by straight lines, we obtain a continuous piecewise-linear curve $p_j(t)$ in the complex plane, with a natural 
parametrization, $t \in [0,\frac{j(j+1)}{2}]$.  Similarly for $q_j(t)$.

\begin{definition}(Vertex Points)
        The integral points of $p_j$ and $q_j$, defined by~\cref{eq:vtx1} and~\cref{eq:vtx2} respectively, are called \emph{vertex points} to distinguish them
        from other points on the piecewise-linear curves.
	In other words, a point $(x,y) \in p_j$ (or $q_j$) is said to be a vertex point if and only if it belongs to the point set $S_n$.
	Clearly, $p_j$ has exactly $1+\frac{j(j+1)}{2}$ vertex points, and so does $q_j$. 
\end{definition}

\subsection{Reflection Symmetry}
By the complementarity of the indices ($\phi_k$ and $\varphi_k$ in bit form) discussed above, it follows that for any $t \in [0,N]$ where $N = \frac{k(k+1)}{2}$,  we have
\begin{align}
        p_k(t) + q_k(N-t) & = C_k.
        \label{eq:rs}
\end{align}
This can be written as
\begin{align*}
        p_k(t)  - \frac{C_k}{2} = \frac{C_k}{2} - q_k(N-t)
\end{align*}
which has the form of an equation representing \emph{Point Reflection Symmetry} of these two curves about the point $\frac{C_k}{2}$.
Thus the complementary curves $p_k(t)$ and $q_k(t)$ satisfy a \emph{Point Reflection Symmetry}. An example is shown in~\cref{fig:illust_refsym}
below.

\begin{figure}[!htbp]
\begin{center}
\epsfxsize=5in \epsfysize=3in {\epsfbox{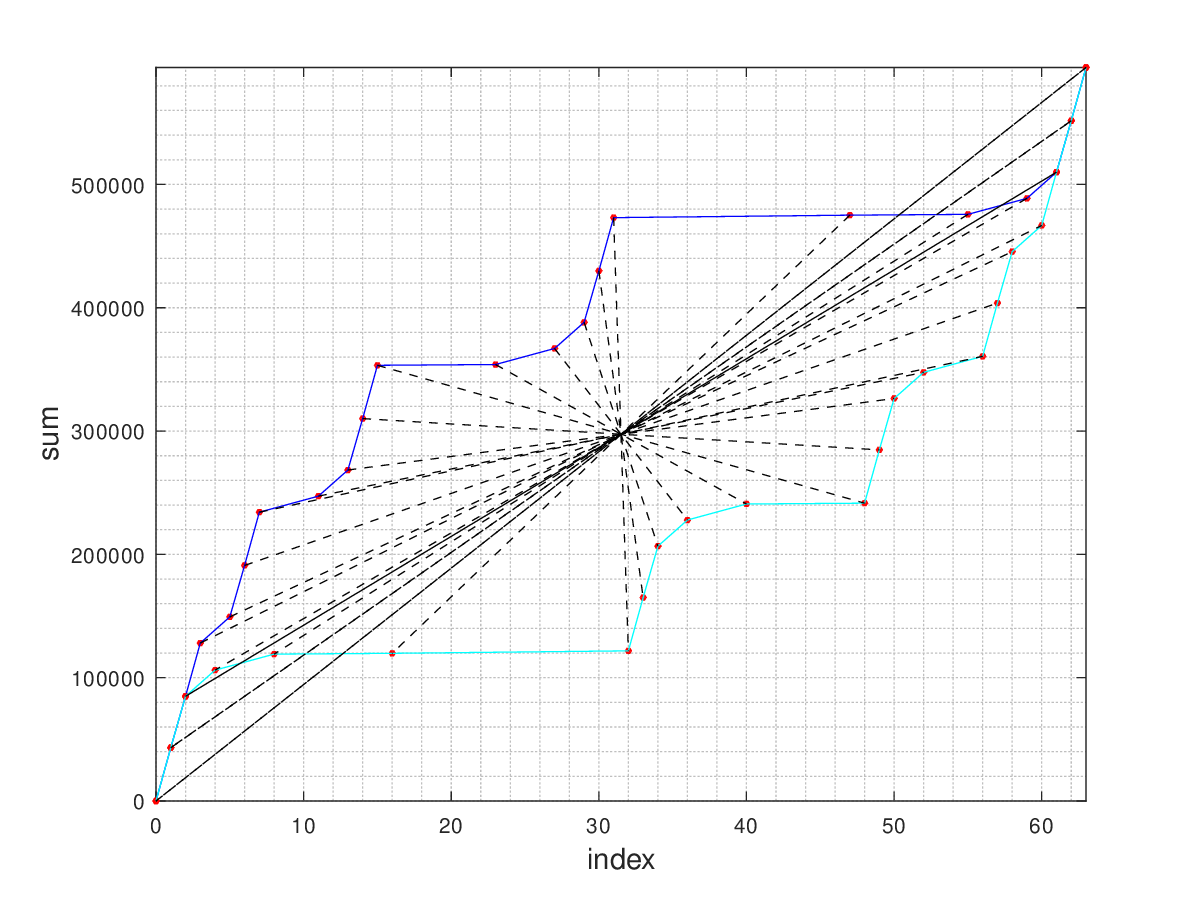}}
	\caption{Point Reflection Symmetry for complementary curves $p_6$ (blue) and $q_6$ (cyan) of the main example. The symmetric
	pairs of points are connected by a dotted line.}
\label{fig:illust_refsym}
\end{center}
\end{figure}

\subsection{The curves $p_k$ and $q_k$ and coverage} 
\begin{figure}[!htbp]
\begin{center}
\epsfxsize=6in \epsfysize=5in {\epsfbox{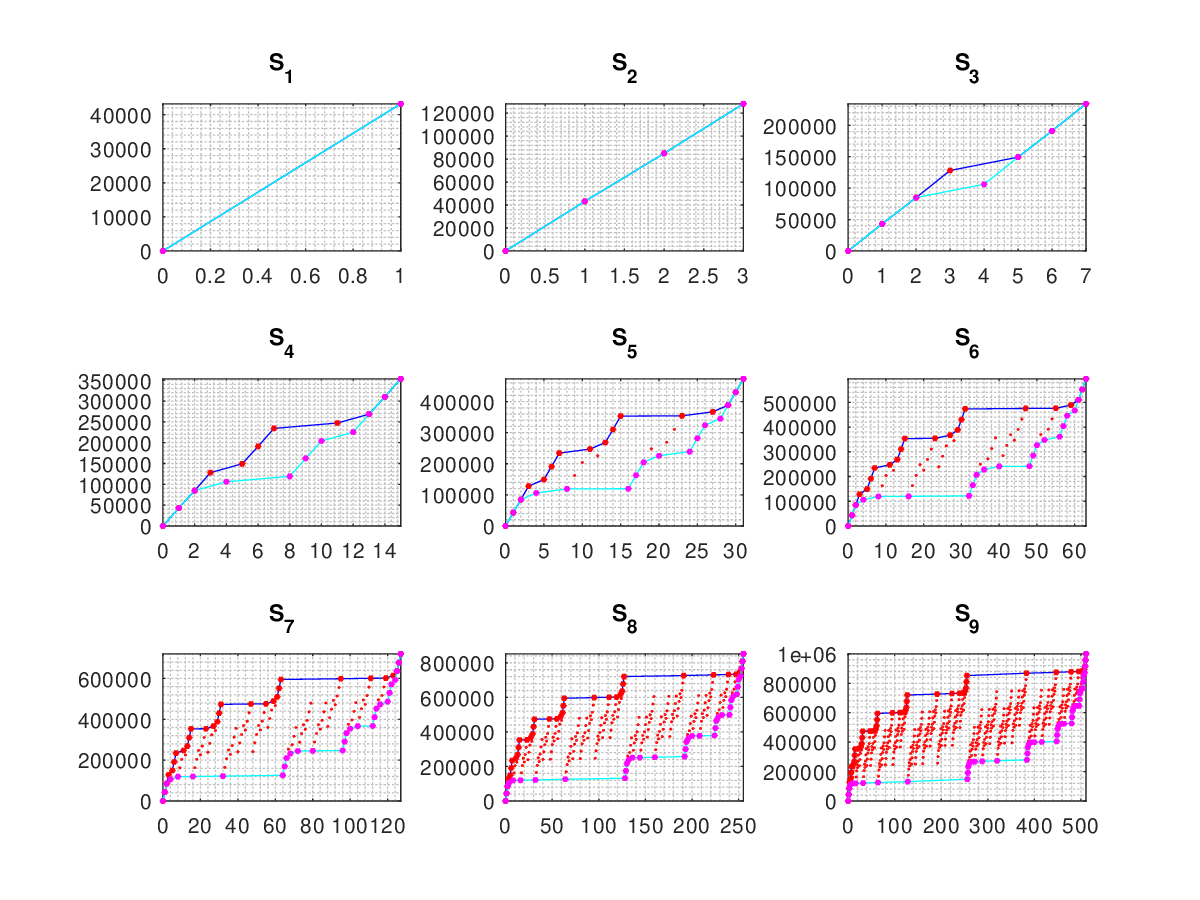}}
	\caption{Illustration of coverage of subspaces $S_1,S_2,\ldots,S_9$, by complementary curves $p_k$ and $q_k$. In each case the $p_k$ curve is in blue
	and the $q_k$ curve in cyan. In $S_1$ and $S_2$ the complementary curves are identical and overlap.
	In $S_3$, the curves $p_3$ and $q_3$ are distinct in one point, and together they cover
	the space. In $S_4$, the curves $p_4$ and $q_4$ are distinct in more points, and together they cover the space. In
	$S_5$, the curves $p_5$ and $q_5$ are distinct in several points but do not cover the space. There are several
	isolated points. This pattern continues for higher subspaces.}
\label{fig:illust2}
\end{center}
\end{figure}

Here, we provide the geometrical intuition, by showing the pair of curves $p_k,q_k$ and the subspace $S_k$ as $k$ ranges in $\{1,\ldots,9\}$ 
using the main example. 
These are shown in~\cref{fig:illust2}, where the $p_k$ curve is in blue, and the $q_k$ curve is in cyan.
Some observations follow:
\begin{itemize}
	\item{[Cases $n=1$ and $n=2$]} 
		For the case $n=1$ the solution space is $\{0,1\}$, and the sequences $\phi_1$ and $\varphi_1$ are both identical and equal to $\{0,1\}$.  
		For $n=2$, we have $4$ subset sums corresponding to indices $\{0,1,2,3\}$. The sequences, $\phi_2 = \{0,1,2,3\} = \varphi_2$ are identical. 
		In these trivial cases, just one sequence of indices covers the solution space.

	\item{[Case $n=3$]}  For $n=3$, the indices of the solution space are $\{0,1,2,3,4, 5,6,7\}$. The 
		sequences $\phi_3 = \{0,1,2,3,5,6,7\}$ and $\varphi_3 = \{0,1,2,4,5,6,7\}$ are distinct.  Note that the index $3$ is 
		in $\phi_3$ but not in $\varphi_3$. Similarly, the index $4$ is in $\varphi_3$ but not in $\phi_3$. However, together they cover the solution space. 

	\item{[Case $n=4$]} For $n=4$, we have $16$ elements in the solution space. It is easy to verify that the $16$ indices are 
		covered by ${\phi}_4 = \{0,1,2,3,5,6,7,11, 13,14,15\}$ and ${\varphi}_4 = \{0,1,2,4,8,9,10,12,13,14,15\}$ together,
		i.e., ${\phi}_4 \cup {\varphi}_4 = S_4$. 

	\item{[Case $n=5$]}  For $n=5$, the indices are $\{0,1,2,\ldots,31\}$. The sequences $\phi_5$ and $\varphi_5$ are 
		$\{0,1,2,3,5,6,7,11,13,14,15,23, 27,29,30,31\}$ and $\{0,1,2,4,8,16,17,18,20$, $24,25,26, 28,29,30,31\}$ respectively.  However, notice 
		that these two sequences, even together do not cover the solution space. In particular, the indices $\{9,10,12,19,21,22\}$ are missing. 

	\item{[Cases $n > 5$]} For $n > 5$, the number of elements of the subspace that are not covered by $p_n$ and $q_n$ will increase with $n$.  
		We will show later on that by including additional curves (formed by certain operations on the $p_n$ and $q_n$ curves) we can get complete coverage 
		of the solution space.  This will be discussed in detail in the next section, after developing the suitable framework.
\end{itemize}

\section{Links, Chains and Transforms}
\label{sec:chains}
In the previous section we used the Filling Boxes Model to characterize two particular NDPs namely $p_n(k)$ and $q_n(k)$ where $k \in [0, \frac{n(n+1)}{2}]$.
Here we show an alternate characterization
of these NDPs using the differences of adjacent elements of the sequence ${\bf c}$. Furthermore, both characterizations are related in a way analogous 
to \emph{differentiation} and \emph{integration}. 

\begin{definition}(Set of Differences $D$ and links) 
        Let $d_j := c_j - c_{j-1}$, for $j \in [n]$ where $c_0 := 0$, and let $D := \{d_1,d_2,\ldots,d_n\}$ be the set of complex numbers
	corresponding to the differences. 
	The complex number $d_j$ can also be viewed as a directed line segment from the origin to the point $(b_j-b_{j-1}) + \iota(a_j-a_{j-1})$
	for each $j \in [n]$. When referring to the line segment, we call it as a link but still denote it as $d_j$ for simplicity.
        \label{def:dseg}
\end{definition}

Since the maximum size of each element in the given instance is bounded by $2^m$, it follows that $\Im(d_j) < 2^m$ for all $j \in [n]$. 

For any positive integer $j \in [n]$, we can write $c_j$ as a telescoping sum:
\begin{align*}
	c_j = d_1 + \ldots + d_j. 
\end{align*}

\begin{proposition}
	Let $j \in [n]$ be a positive integer. Then for any non-negative integer $k \in [0, \frac{j(j+1)}{2})$ we have
	\begin{align*}
		p_j(k+1) - p_j(k) & = d_r, \\
		\text{and } q_j(k+1) - q_j(k) & = d_s
	\end{align*}
	for some $r,s \in [n]$.
\end{proposition}
\begin{proof}
	Recall that for any non-negative integer $k$, we have $p_j(k) = \phi_j(k) + \iota\siga(\phi_j(k))$.
        In the sequence $\phi_j$, as we go from one element to the next element, the Filling-Boxes model shows that
        we are displacing a ball from a lower bit position to  an immediate higher bit position. Thus, the difference
        between adjacent indices has the form $b_r - b_{r-1}$ for some suitable  $r \in [n]$. The corresponding change in the subset sum, is
	clearly $a_r - a_{r-1}$. Thus, from~\cref{def:pq_seq} it follows that $p_j(k+1) - p_j(k) = d_r$. Similarly for $q_j$ whose indices 
	are defined by the sequence $\varphi_j$.
\end{proof}

\begin{definition} (Chain)
        A chain is simply an ordered concatenation of two or more elements of $D$, where the operation of concatenation is denoted by $\oplus$ which separates the
        given numbers. A chain is indicated by the symbol $\pp$ preceding an alphabetical character.
        For example, the chain $\pp{L}$ which is the concatenation of  $d_i$ and $d_j$, for some $i,j \in [n]$, is written as $\pp{L} = d_i\oplus d_j$.
        We refer to the $k$th element of a chain by $\pp{L}(k)$.
        Similarly for $i, j \in [n]$ with $i < j$ the expression $\pp{L}[i,j]$ indicates the
        sequence of elements (subchain) starting at index $i$ and ending at index $j$.
\end{definition}

With these definitions, we can associate the curves $p_n$ and $q_n$ to chains as shown in the proposition below.
The elements in the chain are \emph{boxed} to disclose the pattern clearly.
\begin{proposition}
        The sequence of differences between adjacent points in the sequence $p_n$, corresponds to the chain $\partial{p_n}$, where
        \begin{align}
		\partial p_n & = \boxed{d_1} \oplus \boxed{d_2 \oplus d_1} \oplus\boxed{d_3 \oplus d_2 \oplus d_1} \oplus \ldots \oplus \boxed{d_n \oplus d_{n-1} \oplus \ldots \oplus d_1}.
		\label{eq:chain_pn}
        \end{align}
        The sequence of differences between adjacent points in the sequence $q_n$ corresponds to the chain $\partial{q_n}$, where
        \begin{align}
		\partial q_n & = \boxed{d_1 \oplus d_2 \oplus \ldots \oplus d_n} \oplus \boxed{d_1 \oplus \ldots \oplus d_{n-1}} \oplus \ldots \oplus \boxed{d_1 \oplus d_2} \oplus \boxed{d_1}.
		\label{eq:chain_qn}
        \end{align}
\end{proposition}
\begin{proof}
        We give the proof by induction on $n$. \\
	(Base Step) For $n=1$, the sequence of indices $\phi_1 = (0,1)$ and sums $\siga(\phi_1) = (0,a_1)$. The chain has only one element, i.e.,
        $(b_1+\iota a_1) - (b_0+\iota a_0) = 1 + \iota a_1$
        which is defined by $d_1$. Thus $\partial p_1 = (d_1)$. \\
        (Inductive Step) By hypothesis suppose it is true for $n=k$. So we have,
        \begin{align*}
		\partial p_k  = \boxed{d_1} \oplus \boxed{d_2 \oplus d_1} \oplus \boxed{d_3 \oplus d_2 \oplus d_1} \oplus \ldots \oplus \boxed{d_k \oplus d_{k-1} \oplus \ldots \oplus d_1}.
        \end{align*}
        For $n=k+1$, let's refer to the Filling-Boxes model.
        We have an empty box on the left and an extra ball in the bag. Also, all the $k$ boxes to the left of the
        bag are filled. The next sequence of displacements happen as indicated in~\cref{lem:one}, which correspond to
        \begin{align*}
                (d_{k+1},d_{k},\ldots,d_2,d_1).
        \end{align*}
        Thus we have
        \begin{align*}
                \partial p_{k+1} & = \partial p_k \oplus \boxed{d_{k+1} \oplus d_{k} \oplus \ldots \oplus d_1} \\
                & = \boxed{d_1} \oplus \boxed{d_2 \oplus d_1} \oplus \boxed{d_3 \oplus d_2 \oplus d_1} \oplus \ldots \oplus \boxed{d_{k+1} \oplus d_{k} \oplus \ldots \oplus d_1}.
        \end{align*}
        and the claim follows. By recalling that $\varphi_n$ is the bit complement of $\phi_n$, it follows that the chain corresponding
        to $q_n$ is the reverse of that of $p_n$.
\end{proof}

\begin{corollary}
	Let $N = \frac{n(n+1)}{2}$. Then for any $k \in [N]$ we have 
	\begin{align*} 
		\partial p_n(k) = \partial q_n(N+1-k).
	\end{align*}
\end{corollary}

\begin{proposition}
	To each real number $t \in [0,n(n+1)/2]$, the unique point $p_n(t)$ can be determined as
	\begin{align*}
		p_n(t) = \sum_{j=1}^{\myfloor{t}}\partial{p_n}(j) + (t-\myfloor{t})\partial{p_n}(\myfloor{t}+1)
	\end{align*}
	where, in the first term, the empty sum is treated as zero.  Similarly,
	\begin{align*}
		q_n(t) = \sum_{j=1}^{\myfloor{t}}\partial{q_n}(j) + (t-\myfloor{t})\partial{q_n}(\myfloor{t}+1).
	\end{align*}
\end{proposition}
\begin{proof}
	First suppose that $t \in {\mathbb N}$. Then $p_n(t)$ is simply the sum of the first $t$ elements of the chain
	$\partial p_n$ which is nothing but a telescoping sum. If $t$ is not an integer, then we get the extra term on
	the right corresponding to a partial traversal on a link,  since the curve $p_n(t)$ is a piecewise-linear curve.
\end{proof}

With  some abuse of notation, the above proposition can be written more elegantly as follows.
\begin{proposition}
	For any positive integer $j \in [n]$ and for any real $t \in [0, \frac{j(j+1)}{2}]$
	\begin{align*}
		p_j(t) & = {\int}_{\hspace*{-0.08in}0}^t \partial p_j(t')dt' \\
		\text{and } q_j(t) & = {\int}_{\hspace*{-0.08in}0}^t \partial q_j(t')dt'.
	\end{align*}
\end{proposition}

Thus we have two equivalent ways of representing the curve: the curve itself (i.e. a sequence of points), and a chain (i.e. a sequence 
of line segments) where to get the coordinates of a point we take a path integral.
Thus, informally, a chain is like the derivative of the curve with respect to parameter $t$, while the curve is like the
integral of chain.

\subsection{Elemental representation of chains}
A clear pattern is evident in~\cref{eq:chain_pn} and~\cref{eq:chain_qn}, as shown by \emph{boxing} the elements, which leads to the definition
of an \emph{elemental chain}. This will simplify the notation and make the description easy as we move forward.

\begin{definition} (Elemental Chains)
        We view the elements of $D$ as building blocks of elements of ${\bf c}$. In particular, to each $c_i \in {\bf c}$, we associate a
        chain $\pp{c_i}$ defined by
        \begin{align}
                \pp{c_i} = d_1 \oplus d_2 \oplus \ldots \oplus d_i.
        \end{align}
        We also define the reverse chain (with a hat symbol)
        \begin{align}
                \pp{\hat{c}_i} = d_i \oplus d_{i-1} \oplus \ldots \oplus d_1.
        \end{align}
        We refer to $\pp{c}_i$ and $\pp{\hat{c}}_i$ as elemental chains which are reverse of each other.
\end{definition}

With this definition, the chain $\partial p_n$ can be written compactly as
\begin{align}
	\partial p_n  = \phc_1 \oplus \phc_{2} \oplus \ldots \oplus \phc_n.
\end{align}

Similarly, the chain $\partial q_n$ can be written as,
\begin{align}
	\partial q_n  = \pc_n \oplus \pc_{n-1} \ldots \oplus \pc_1.
\end{align}

\begin{observation}
	The following observations can be made with regard to the elemental representation of chains $\pp{p}_n$ and $\pp{q}_n$.
	\begin{itemize}
		\item[(i)] In $\pp{p}_n$, if we consider only the first elemental chain, we see that it corresponds to $\partial p_1$. 
			In general, if we consider the first $k$ elemental chains, for any $k \in [n]$, the chain 
			corresponds to $\partial p_k$. Thus, the chain $\partial p_n$ contains all the 
			sub-chains $\partial p_1, \partial p_2, \ldots, \partial p_{n-1}$.  
		\item[(ii)] In $\pp{q}_n$, if we consider only the last elemental chain, we see that it corresponds to $\partial q_1$. 
			In general, if we consider the last $k$ elemental chains, for any $k \in [n]$, the chain corresponds 
			to $\partial q_k$. Thus, the chain $\partial q_n$ contains all the 
			sub-chains $\partial q_1, \partial q_2, \ldots, \partial q_{n-1}$.
	\end{itemize}
\end{observation}
The above observations can be formally stated as:
\begin{proposition}
	Let $k \in [n]$ be a positive integer. Then the chain $\pp{p}_k$ contains $\pp{p}_1, \ldots, \pp{p}_{k-1}$ as subchains. 
	Similarly, the chain $\pp{q}_k$ contains $\pp{q}_1, \ldots, \pp{q}_{k-1}$ as subchains. 
	\label{prop:contain}
\end{proposition}

\begin{definition} (Partial Sums of a Chain)
        For a positive integer $N$, let $\pp{L} = l_1 \oplus \ldots \oplus l_N$ be a chain of length $N$. We denote the sum of the first $i$ elements of $\pp{L}$ by
        $\sigma(\pp{L},i)$ and the sum of all elements of $\pp{L}$ simply by $\sigma(\pp{L})$. We also define the empty sum to be $0$,
        i.e., $\sigma(\pp{L},0) = 0$.
        Thus, $\sigma(\pp{L},2) = l_1 + l_2$ and $\sigma(\pp{L}) = l_1 + \ldots + l_N$.
\end{definition}

\begin{definition}(Valid Subset Sum)
        A complex number $s$ is said to be a valid subset sum if and only if $s \in S_n$, i.e., $s$ is an element of the power set.
\end{definition}

It is straight forward to see that, for any $k \in [n]$,  all the partial sums of both $\pp{p}_k$ and $\pp{q}_k$ are indeed valid subset sums.

\subsection{Transforms on chains}
As noted above, the pair of complementary sequences $p_n$ and $q_n$ cover the solution space only for $n \le 4$. For $n \ge 5$
we will have many missing points, and in order to cover them we need \emph{new} sequences/chains that are distinct from $\pp{p}_n$ and $\pp{q}_n$. 
In this section, we define certain transforms that enable us to form new chains from a given chain.

\begin{definition}(Active Region)
	Let $\pp{\psi}$ be a chain consisting of $n$ elemental chains. Then the \emph{active region} of $\pp{\psi}$, denoted by $\alpha(\pp{\psi})$,
	is defined as the longest interval of indices $[i,j] \subseteq [1,n]$ 
	satisfying either $\pp{\psi}[i,j] = \pp{p}_k$ or $\pp{\psi}[i,j] = \pp{q}_k$, where $k = j+1-i$.
	The non-active part of a chain is said to be fixed.
\end{definition}

Thus, it readily follows that $\alpha(\pp{p}_n) = \alpha(\pp{q}_n) = [1,n]$. 
Below we define some transformations on chains which operate \emph{only} on the active region of a given chain. 

\begin{definition}(Character set $\Lambda$)
	Given a chain $\pp{\psi}$ with active region specified by $[i,j]$, the character set of $\pp{\psi}$ denoted by $\Lambda(\pp{\psi})$, is defined as 
	\begin{align*}
		\Lambda(\pp{\psi}) = [j+1-i].
	\end{align*}
	Thus, the character set $\Lambda$ of a chain $\pp{\psi}$ tells us that there exist subchains of either type $\pp{p}_r$ or
	type $\pp{q}_r$ within $\pp{\psi}$ for all $r = 1,2,\ldots,j+1-i$.
\end{definition}

\begin{example}
	Let $\pp{\psi}$ be a chain consisting of  seven blocks, described by 
	\begin{align*}
		\pp{\psi} = \pc_5 \oplus \pc_3 \oplus \pc_2 \oplus \pc_1 \oplus \phc_4 \oplus \phc_6 \oplus \phc_7
	\end{align*}
	Then, from the above definitions it follows that $\alpha(\pp{\psi})  = [2,4]$ since these indices correspond to the longest subchain 
	$\pp{q}_3 = \pc_3 \oplus \pc_2 \oplus \pc_1$. Thus $\Lambda(\pp{\psi})  = [3]$ and we have the complete subchains $\pp{q}_1 = \pc_1$,
	$\pp{q}_2 = \pc_2 \oplus \pc_1$ and $\pp{q}_3 = \pc_3 \oplus \pc_2 \oplus \pc_1$.

	On the other hand if we consider the chain $\pp{q}_{10}$ we have $\alpha(\pp{q}_{10})  = [1,10]$, and $\Lambda(\pp{q}_{10})  = [10]$.
\end{example}

\begin{definition}(Local Character Transform $\lambda_k$)
	Let $\pp{\psi}$ be a chain composed of $n$ elemental chains, and active region specified by the interval $[i,j]$. Then given a positive integer
	$k \in [j+1-i]$, the local character transform denoted by $\lambda_k(\pp{\psi})$ changes the active region as follows:
	\begin{align*}
		\alpha(\lambda_k(\pp{\psi})) = \begin{cases}
			[i, i+k-1] & \text{ if $\pp{\psi}[i,j]$ is of the form $\pp{p}_{j+1-i}$} \\
			[j+1-k, j] & \text{ if $\pp{\psi}[i,j]$ is of the form $\pp{q}_{j+1-i}$}. 
		\end{cases}
	\end{align*}
	Informally, the local character transform picks up one of the available characters from the character set.
	Note that the local character transform doesn't change the chain; it simply changes the active region - the region that can undergo further
	transforms. Furthermore, the $\lambda_k$ operation affects the leftmost blocks of $p$-type active region, but the rightmost blocks of $q$-type active region.
\end{definition}

\begin{example}
	Consider the chain $\pp{q}_7 = \pc_7 \oplus \pc_6 \oplus \pc_5 \oplus \pc_4 \oplus \pc_3 \oplus \pc_2 \oplus \pc_1$. Clearly, the active region 
	is $[1,7]$. If we apply $\lambda_4$ transform on this chain, we get
	\begin{align*}
		\lambda_{4}(\pp{q}_7) = \pc_7 \oplus \pc_6 \oplus \pc_5 \oplus \boxed{\pc_4 \oplus \pc_3 \oplus \pc_2 \oplus \pc_1}
	\end{align*}
	where the new active region is $[4,7]$ whose elemental chains are boxed for clarity. 

	Similarly, consider the chain $\pp{p}_7 = \phc_1 \oplus \phc_2 \oplus \phc_3 \oplus \phc_4 \oplus \phc_5 \oplus \phc_6 \oplus \phc_7$. Clearly, the 
	active region is $[1,7]$. If we apply $\lambda_5$ transform on this chain, we get
	\begin{align*}
		\lambda_{5}(\pp{p}_7) = \boxed{\phc_1 \oplus \phc_2 \oplus \phc_3 \oplus \phc_4 \oplus \phc_5} \oplus \phc_6 \oplus \phc_7
	\end{align*}
	where the new active region is $[1,5]$ which is boxed for clarity.  
	\label{ex_lambda}
\end{example}

\begin{definition}(Reversal Transform $\rho_k$)
	Let $\pp{\psi}$ be a chain with $n$ blocks and active region specified by the interval $[i,j]$. Then given a positive integer
	$k =j+1-i$, the reversal transform denoted by $\rho_k(\pp{\psi})$ reverses the order of elemental chains in the active region 
	and reverses the order of elements in each elemental chain. Thus
	\begin{align*}
		\rho_k(\pp{\psi}[i,j]) = \begin{cases}
			\phc_1 \oplus \phc_2 \ldots \oplus \phc_k  \quad \text{ if   } \pp{\psi}[i,j] = \pc_k \oplus \pc_{k-1}, \ldots \oplus \pc_1 \\
			\pc_k \oplus \pc_{k-1} \ldots \oplus \pc_1 \quad \text{ if   } \pp{\psi}[i,j] = \phc_1 \oplus \phc_{2} \ldots \oplus \phc_k. 
		\end{cases}
	\end{align*}
	In words, if the active region of the chain $\pp{\psi}$ originally corresponded to $\pp{p}_k$ then after the $\rho_k$ transform
	this portion of the chain becomes $\pp{q}_k$, and vice versa.
\end{definition}

\begin{example}(Continuation of~\cref{ex_lambda})
	Suppose we apply a $\rho_4$ transform on $\lambda_4(\pp{p}_7)$, we get:
	\begin{align*}
		\rho_4(\lambda_{4}(\pp{q}_7)) & = \rho_4(\pc_7 \oplus \pc_6 \oplus \pc_5 \oplus \boxed{\pc_4 \oplus \pc_3 \oplus \pc_2 \oplus \pc_1}) \\
		& = \pc_7 \oplus \pc_6 \oplus \pc_5 \oplus \boxed{\phc_1 \oplus \phc_2 \oplus \phc_3 \oplus \phc_4}.
	\end{align*}
	Similarly if we apply a $\rho_5$ transform on $\lambda_5(\pp{q}_7)$ we get
	\begin{align*}
		\rho_5(\lambda_{5}(\pp{p}_7)) & = \rho_5(\boxed{\phc_1 \oplus \phc_2 \oplus \phc_3 \oplus \phc_4 \oplus \phc_5} \oplus \phc_6 \oplus \phc_7) \\
		& = \boxed{\pc_5 \oplus \pc_4 \oplus \pc_3 \oplus \pc_2 \oplus \pc_1} \oplus \phc_6 \oplus \phc_7.
	\end{align*}
\end{example}

We can define a composite $\tau_k$ transform equivalent to a local character transform followed by a reversal transform as follows.
\begin{definition}(Reflection Transform $\tau_k$)
	Let $\pp{\psi}$ be a chain with $n$ elemental chains and active region specified by the interval $[i,j]$. Then given a positive integer
	$k  \in [j+1-i]$, the reflection transform denoted by $\tau_k(\pp{\psi})$ is defined as
	\begin{align*}
		\tau_k(\pp{\psi}) := \rho_k(\lambda_k(\pp{\psi})).
	\end{align*}
	Note that the $\tau_k$ transform operates on the first $k$ elemental chains if the active region of $\pp{\psi}$ is of the type $\pp{p}_{j+1-i}$,
	and on the last $k$ elemental chains if the active region is of the type $\pp{q}_{j+1-i}$.
\end{definition}

\begin{example}
	With the composite reflection transform, the previous example can be written more simply as:
	\begin{align*}
		\tau_4(\pp{q}_7) & = \tau_4(\pc_7 \oplus \pc_6 \oplus \pc_5 \oplus \pc_4 \oplus \pc_3 \oplus \pc_2 \oplus \pc_1) \\
		& = \pc_7 \oplus \pc_6 \oplus \pc_5 \oplus \boxed{\phc_1 \oplus \phc_2 \oplus \phc_3 \oplus \phc_4} \\
		\tau_5(\pp{p}_7) & = \tau_5(\phc_1 \oplus \phc_2 \oplus \phc_3 \oplus \phc_4 \oplus \phc_5 \oplus \phc_6 \oplus \phc_7) \\
		 & = \boxed{\pc_5 \oplus \pc_4 \oplus \pc_3 \oplus \pc_2 \oplus \pc_1} \oplus \phc_6 \oplus \phc_7.
	\end{align*}
\end{example}

Previously, we noted that all the partial sums of $\pp{p}_n$ (as well as $\pp{q}_n$) are valid subset sums. It turns out that applying
a reflection transform preserves the validity of all partial sums of the new chain.

\begin{proposition}
	Let $\pp{\psi}$ be a given chain with $n$ elemental chains and an active region $[i, j] \subseteq [1,n]$. If all the partial sums
	of $\pp{\psi}$ are valid subset sums, then for any $k \in [j+1-i]$, all the partial sums of $\tau_k(\pp{\psi})$ are also valid
	subset sums.
\end{proposition}
The proof is straight forward. Suppose the active region of the given chain $\pp{\psi}$ corresponded to $\pp{p}_k$. After a $\tau_k$ transform,
this portion of the chain will correspond to $\pp{q}_k$. Since the low $k$ bits of the indices (binary vectors) of this part of the curve are complements
of those before the $\tau_k$ transform, it follows that they are valid subset sums too.

The above result enables us to generate new chains by repeated application of $\tau_k$ transforms (with decreasing indices) to generate new
chains with the guarantee that all partial sums being valid subset sums.

\subsection{$p_k \ra \{q_i\}_{i=1}^k$ and $q_k \ra \{p_i\}_{i=1}^k$ structures}
The $\tau$ transform defined above enables us to generate the whole family of curves $\{q_i\}_{i=1}^k$ from a single curve $p_k$, and the whole family 
of curves $\{p_i\}_{i=1}^k$ from $q_k$. 
For example, consider the $p_n$ curve with associated chain $\pp{p_n}$. Since this chain has the character set $[1,n]$ and
contains all the subchains $\pp{p_1},\pp{p_2},\ldots,\pp{p_{n-1}},\pp{p_n}$, we can apply $\tau_i$ on this chain
to get
\begin{align*}
	\tau_i(\pp{p_n}) & = \rho_i(\lambda_i(\pp{p_n})) = \pp{q}_i \oplus \phc_{i+1} \oplus \phc_{i+2} \oplus \ldots \oplus \phc_{n}, \\
	& = \pp{q}_i \oplus \pp{p}_n [i+1,n]
\end{align*}
for all $i \in [n]$. The last expression means that the last $n-i$ elemental chains are fixed as in $\pp{p}_n$, but the first $i$ elemental
chains have been reversed to form $\pp{q}_i$ with new active region $[1, i]$.

\begin{figure}[!htbp]
\begin{center}
\epsfxsize=5in \epsfysize=3in {\epsfbox{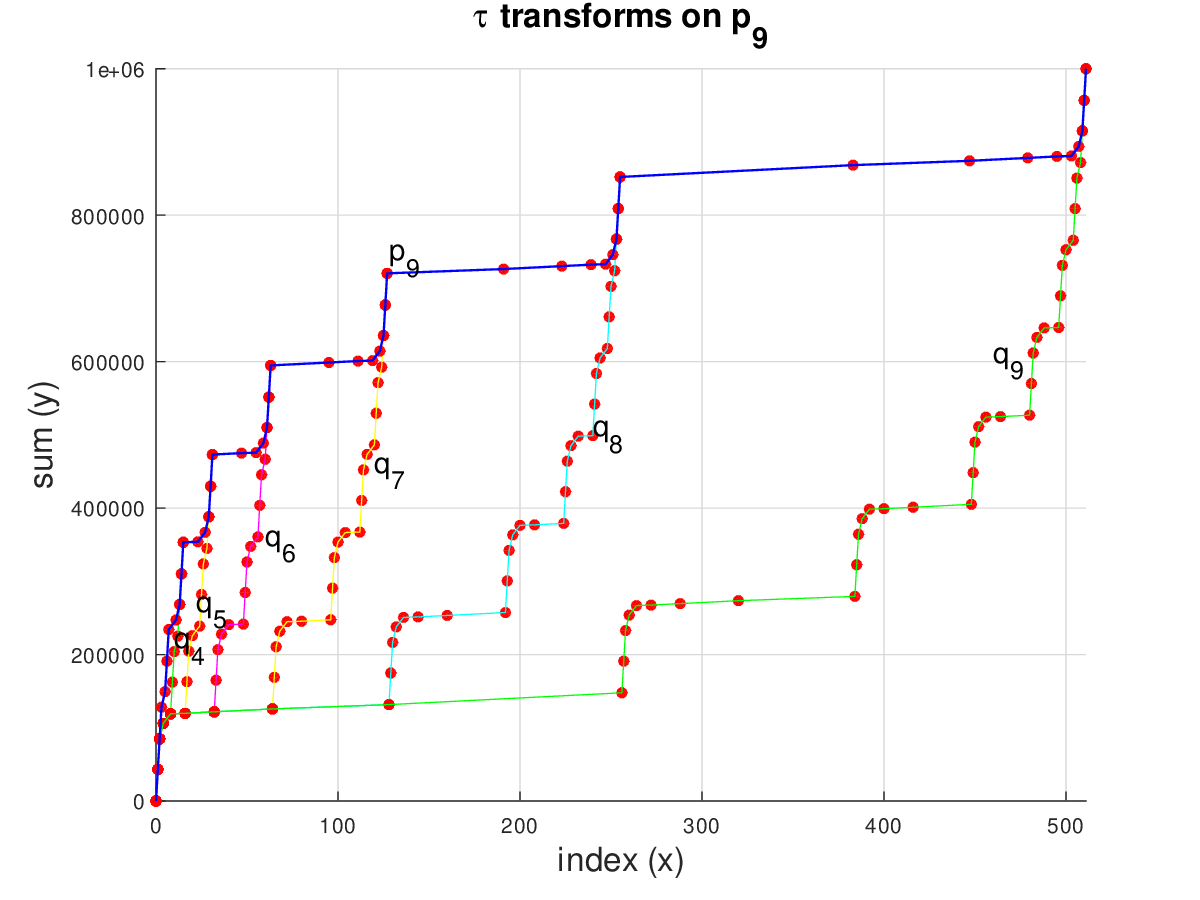}}
        \caption{Illustration of the application of $\{\tau\}_{k=4}^9$ on the $p_9$ curve of main example.
        The $p_9$ curve is shown in blue.}
\label{fig:illust_pQ}
\end{center}
\end{figure}

For the main example, the effect of the family of transforms $\{\tau\}_{k=4}^9$ on the
$p_9$ curve is shown in~\cref{fig:illust_pQ}.  Notice that different parts of the curve have differing number
of local characters. The lower part of the $p_9$ curve has more local characters, than the higher part.

\begin{figure}[!htbp]
\begin{center}
\epsfxsize=5in \epsfysize=3in {\epsfbox{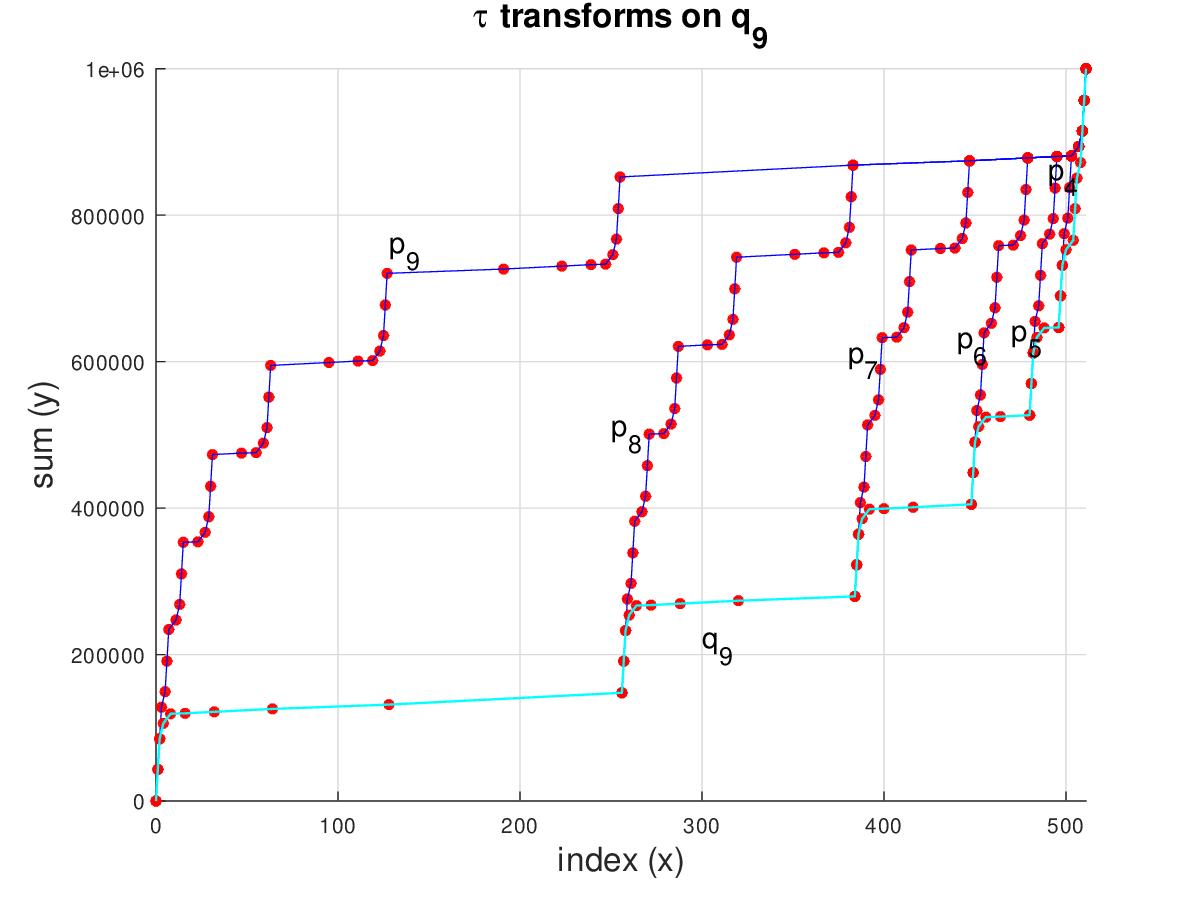}}
        \caption{Illustration of the application of $\{\tau\}_{k=4}^9$ on the $q_9$ curve of main example.
        The $q_9$ curve is shown in cyan.}
\label{fig:illust_qP}
\end{center}
\end{figure}

Similarly, we can apply $\tau_i$ transform on the chain $\pp{q}_n$ to get
\begin{align*}
	\tau_i(\pp{q_n}) & = \rho_i(\lambda_i(\pp{q_n})) = \pc_n \oplus \pc_{n-1} \oplus \ldots \oplus \pc_{i+1} \oplus \pp{p_i}, \\
	& = \pp{q}_n[1,n-i]\oplus \pp{p}_i
\end{align*}
for all $i \in [n]$. The last expression means that the first $n-i$ elemental chains are fixed as in $\pp{q}_n$, and the last $i$ elemental
chains have been reversed to form $\pp{p}_i$ with new active region $[n+1-i, n]$.
The effect of the family of transforms $\{\tau\}_{k=4}^9$ on the
$q_9$ curve is shown in~\cref{fig:illust_qP}.  In this case, the higher part of the $q_9$ curve has more local characters.

\subsection{Nested $\tau$ operations}
Here we show an example with two $\tau$ transforms applied one after other.
Consider the chain $\pp{q}_9$ and the nested operations $\tau_5(\tau_8(\pp{q}_9))$ using main example. Following the above discussion, 
applying the $\tau_8$ operation first we get
\begin{align*}
	\tau_8(\pp{q_9}) & = \pp{q}_9[1,1]\oplus \pp{p}_8.
\end{align*}
Now applying the $\tau_5$ operation on the result we get
\begin{align*}
	\tau_5(\tau_8(\pp{q_9})) & =  \tau_5(\pp{q}_9[1,1] \oplus \pp{p}_8), \\
	& = \pp{q}_9[1,1]\oplus \pp{q}_5 \oplus \pp{p}_8[6,8].
\end{align*}
It can be noticed that the $\tau$ transform operates only a complete chain $\pp{p_k}$ or $\pp{q_k}$ for some $k \in [n]$ and never on a partial chain.

\begin{figure}[!htbp]
\begin{center}
\epsfxsize=5in \epsfysize=3in {\epsfbox{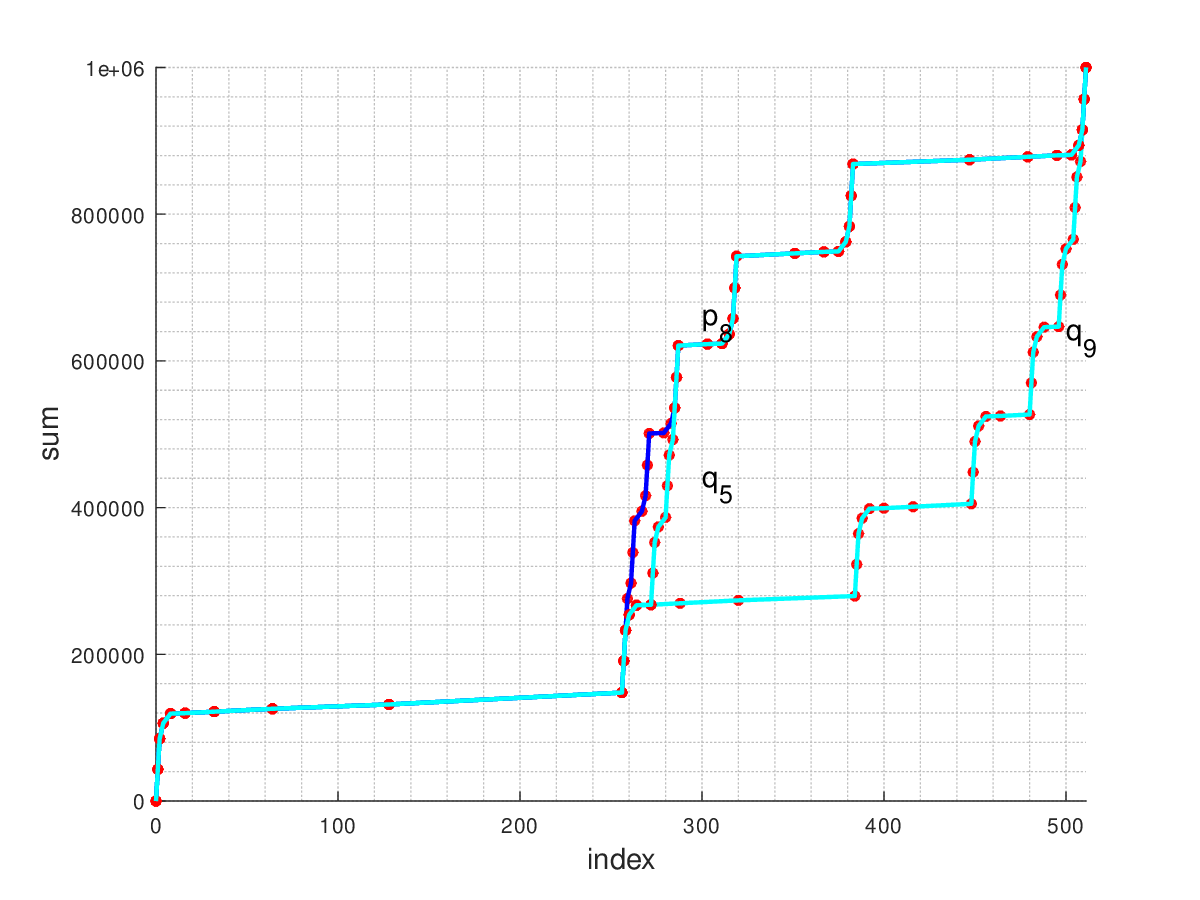}}
	\caption{Illustration of the application of two $\tau$ operations in sequence on the $q_9$ curve of main example.}
\label{fig:illust_twotau}
\end{center}
\end{figure}

\section{Generation of all NDPs}
\label{sec:allchains}

The above discussion suggests a natural generalization. By considering either $\pp{p}_n$ (or $\pp{q}_n$) and repeatedly applying the
$\tau$ transforms we can generate all possible chains. This process can be conveniently represented in the form of a layered directed acyclic graph denoted by $\hg$
and is defined next.

\subsection{A transformation graph $\hg$} 

\begin{definition}($\hg$ graph)
	Let $\hg = (V(\hg), E(\hg))$ be a layered directed graph with vertex set
	$V(\hg)$ consisting of chains $\pp{p}_i$ and $\pp{q}_i$ for $i \in [n]$, arc set $E(\hg)$ consisting
	the set of transformations $\tau_{j}$ with $j \in [n]$  between the curves. There is a root node 
	representing the chain $\pp p_{n}$ (or $\pp q_n$), and the graph $\hg$ has $n+1$ layers. 
\end{definition}

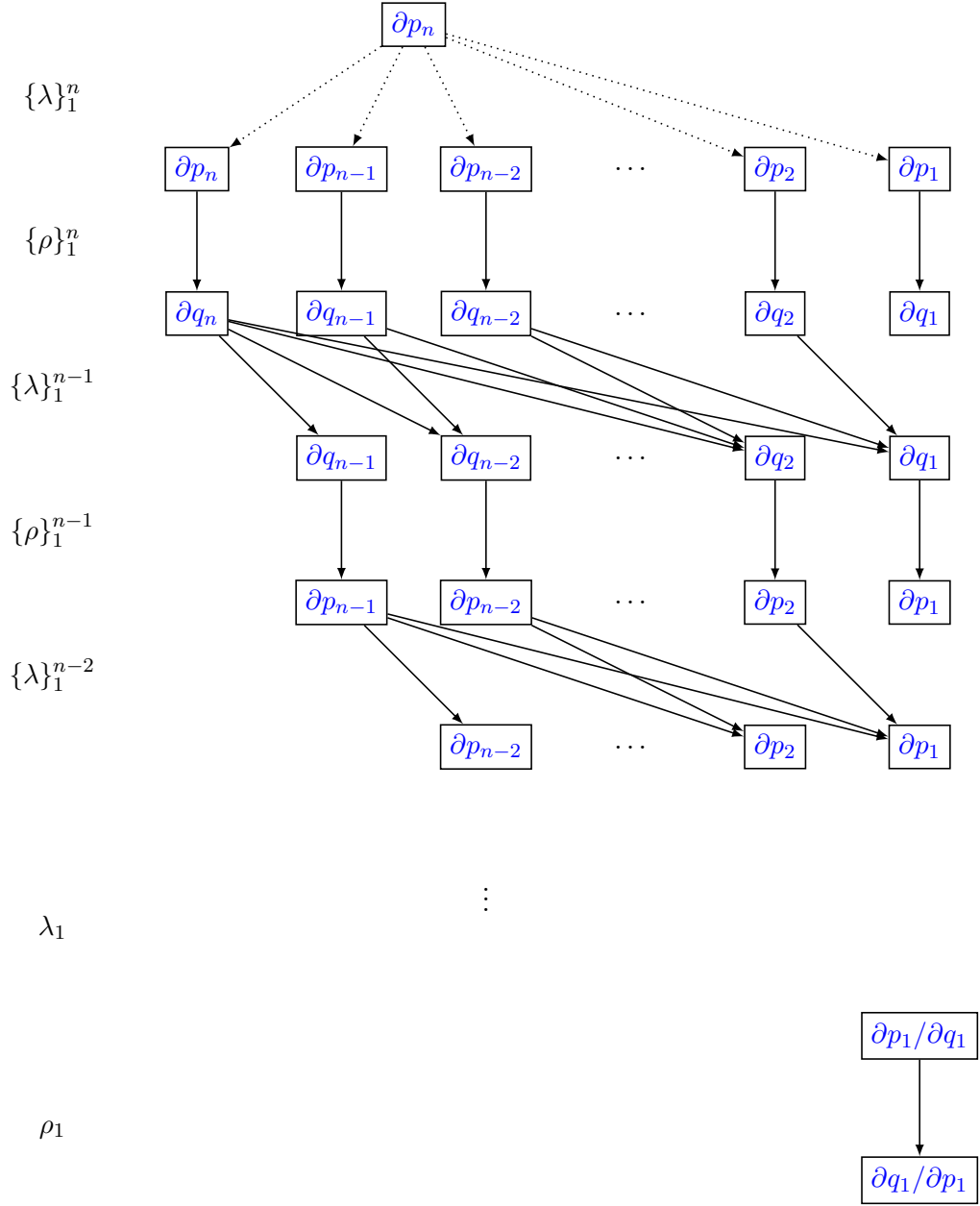
\begin{figure}[!htbp]
\begin {center}
\begin {tikzpicture}[-latex ,auto ,node distance =1cm and 1cm ,on grid ,
semithick ,
state/.style ={ draw,black, text=blue, minimum width =0.1 cm}]
        \node[state] (L0) at (4,14) {$\pp{p_n}$};

        \node[state] (L11) at (1,12) {$\pp{p_n}$};
        \node[state] (L12) at (3,12) {$\pp{p_{n-1}}$};
        \node[state] (L13) at (5,12) {$\pp{p_{n-2}}$};
        \node (L1x) at (7,12) {$\ldots$};
        \node (L1z) at (-1,13) {${\{\lambda\}_1^n}$};
        \node[state] (L15) at (9,12) {$\pp{p_2}$};
        \node[state] (L16) at (11,12) {$\pp{p_1}$};
        \foreach \from/\to in {L0/L11,L0/L12,L0/L13,L0/L15,L0/L16}
    \draw[dotted] (\from) -> (\to);

        \node[state] (L21) at (1,10) {$\pp{q_{n}}$};
        \node[state] (L22) at (3,10) {$\pp{q_{n-1}}$};
        \node[state] (L23) at (5,10) {$\pp{q_{n-2}}$};
        \node (L2x) at (7,10) {$\ldots$};
        \node (L2z) at (-1,11) {${\{\rho\}_1^n}$};
        \node[state] (L25) at (9,10) {$\pp{q_2}$};
        \node[state] (L26) at (11,10) {$\pp{q_1}$};
\foreach \from/\to in {L11/L21,L12/L22,L13/L23,L15/L25,L16/L26}
    \draw (\from) -> (\to);

        \node[state] (L32) at (3,8) {$\pp{q_{n-1}}$};
        \node[state] (L33) at (5,8) {$\pp{q_{n-2}}$};
        \node (L3x) at (7,8) {$\ldots$};
        \node (L3z) at (-1,9) {${\{\lambda\}_1^{n-1}}$};
        \node[state] (L34) at (9,8) {$\pp{q_2}$};
        \node[state] (L35) at (11,8) {$\pp{q_1}$};
\foreach \from/\to in {L21/L32,L21/L33,L21/L34,L21/L35,L22/L33,L22/L34,L23/L34,L23/L35,L25/L35}
    \draw (\from) -> (\to);

        \node[state] (L42) at (3,6) {$\pp{p_{n-1}}$};
        \node[state] (L43) at (5,6) {$\pp{p_{n-2}}$};
        \node (L4x) at (7,6) {$\ldots$};
        \node (L4z) at (-1,7) {${\{\rho\}_1^{n-1}} $};
        \node[state] (L44) at (9,6) {$\pp{p_2}$};
        \node[state] (L45) at (11,6) {$\pp{p_1}$};
\foreach \from/\to in {L32/L42,L33/L43,L34/L44,L35/L45}
    \draw (\from) -> (\to);

        \node[state] (L51) at (5,4) {$\pp{p_{n-2}}$};
        \node (L53) at (7,4) {$\ldots$};
        \node (L53) at (-1,5) {${\{\lambda\}_1^{n-2}}$};
        \node[state] (L54) at (9,4) {$\pp{p_2}$};
        \node[state] (L55) at (11,4) {$\pp{p_1}$};
\foreach \from/\to in {L42/L51,L42/L54,L42/L55,L43/L54,L43/L55,L44/L55}
    \draw (\from) -> (\to);

        \node  at (5,2)  {$\vdots$};
        \node  (Lxx) at (4,2.5)  {};
        \node  (Lyy) at (-1,1.5)  {${\lambda_1}$};
        \node  (Lzz) at (-1,-1.3)  {${\rho_1} $};

\node[state] (L45) at (11,0) {$\pp{p_1}/\pp{q_1}$};
\node[state] (L55) at (11,-2) {$\pp{q_1}/\pp{p_1}$};
\foreach \from/\to in {L45/L55}
    \draw (\from) -> (\to);
\end{tikzpicture}
\end{center}
\caption{Chains resulting from $\pp{p_n}$ by repeated alternating $\lambda$ and $\rho$ transforms. Each node represents only the active region of the chain.
This directed acyclic graph has $2n+1$ levels.}
\label{fig:pchain}
\end{figure}

Since only the active region of a chain can undergo a $\tau$ transform, the nodes of the graph represent \emph{only} the active regions,
 and the arcs represent $\tau$ transforms. 
First, we show the graph with the alternating $\lambda$ and $\rho$ transforms for clarity, and later we will show the graph using only the 
composite $\tau$ transforms.  The root node is the starting chain $\pp{p}_n$ as shown in~\cref{fig:pchain}. 

Some remarks on the graph are given:
\begin{itemize}
        \item[(i)] Except for the first layer, where we allow $\lambda_n(\pp{p_n})$, in all other layers of \emph{local character transformation},
                the resulting chain (active region) has a lower numbered local character. The active region monotonically decreases
                as we go down the graph.
        \item[(ii)] The nodes represent the active part of the chain only; the rest of the chain is fixed according to the path traversed from
                the root. Note that all the chains have a length equal to $n(n+1)/2$. Since each chain corresponds to a curve, we have as many
                curves as the number of paths in the graph.
        \item[(iii)] Once the local character reaches $1$, we cannot go down any further. Actually, we don't even need to go below $3$ because
                since $\pp{p_2} = \pp{q_2}$, and $\pp{p_1} = \pp{q_1}$.
        \item[(iv)] The root node is the chain $\pp{p_n}$, but we could start with $\pp{q_n}$ as well. The graph with both $\lambda$ and
		$\rho$ transforms has $2n+1$ levels including the root level.
\end{itemize}

\begin{figure}[!htbp]
\begin {center}
\begin {tikzpicture}[-latex ,auto ,node distance =1cm and 1cm ,on grid ,
semithick ,
state/.style ={ draw,black,  text=blue, minimum width =0.1 cm}]
	\node[state] (L0) at (4,14) {$p_{n}$};

        \node[state] (L11) at (1,12) {${q_n}$};
        \node[state] (L12) at (3,12) {${q_{n-1}}$};
        \node[state] (L13) at (5,12) {${q_{n-2}}$};
        \node (L1x) at (7,12) {$\ldots$};
	\node (L1z) at (-1,13) {${\{\tau\}_1^n}$};
        \node[state] (L15) at (9,12) {${q_2}$};
        \node[state] (L16) at (11,12) {${q_1}$};
        \foreach \from/\to in {L0/L11,L0/L12,L0/L13,L0/L15,L0/L16}
    \draw[dotted] (\from) -> (\to);

        \node[state] (L32) at (3,10) {${p_{n-1}}$};
        \node[state] (L33) at (5,10) {${p_{n-2}}$};
        \node (L3x) at (7,10) {$\ldots$};
	\node (L3z) at (-1,11) {${\{\tau\}_1^{n-1}}$};
        \node[state] (L34) at (9,10) {${p_2}$};
        \node[state] (L35) at (11,10) {${p_1}$};
\foreach \from/\to in {L11/L32,L11/L33,L11/L34,L11/L35,L12/L33,L12/L34,L13/L34,L13/L35,L15/L35}
    \draw (\from) -> (\to);

        \node[state] (L51) at (5,8) {${q_{n-2}}$};
        \node (L53) at (7,8) {$\ldots$};
	\node (L53) at (-1,9) {${\{\tau\}_1^{n-2}}$};
        \node[state] (L54) at (9,8) {${q_2}$};
        \node[state] (L55) at (11,8) {${q_1}$};
\foreach \from/\to in {L32/L51,L32/L54,L32/L55,L33/L54,L33/L55,L34/L55}
    \draw (\from) -> (\to);

        \node  at (5,6)  {$\vdots$};
        \node  (Lxx) at (4,6)  {};
        \node  (Lyy) at (-1,3)  {${\tau_1}$};

\node[state] (L45) at (9,4) {${p_2}/{q_2}$};
\node[state] (L55) at (11,2) {${q_1}/{p_1}$};
\foreach \from/\to in {L45/L55}
    \draw (\from) -> (\to);
\end{tikzpicture}
\end{center}
\caption{Transformation Graph of all paths generated from the starting curve $p_{n}$ by repeated $\tau$ transforms. Only the active regions
of the paths are shown at each level. The size of the active region decreases monotonically with depth. 
This directed acyclic graph structure has $n+1$ levels.}
\label{fig:pqgraph}
\end{figure}
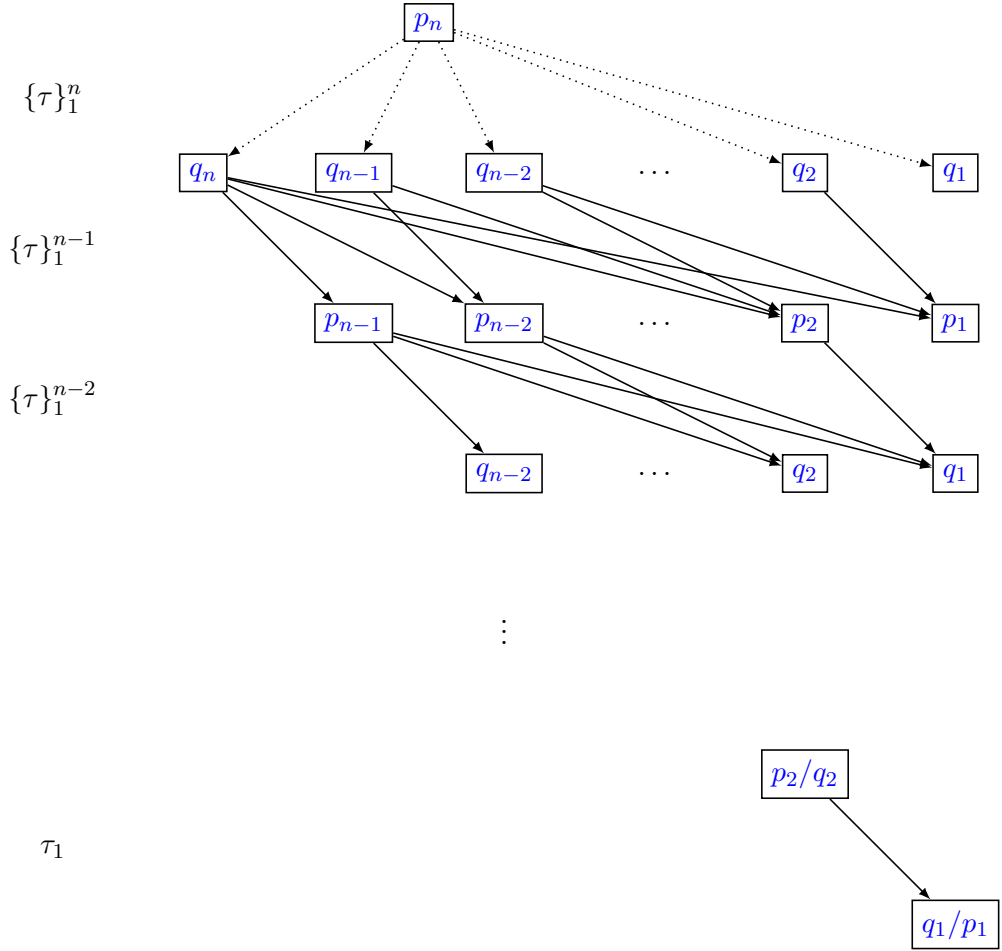

Since chains and curves are essentially equivalent objects, the same graph structure carries over to curves. For the sake of
clarity we again show the graph in~\cref{fig:pqgraph}, with curves instead of chains but with $\tau$ transforms which reduce the 
number of layers from $2n+1$ to $n+1$.

\subsection{Counting paths in $\hg$}
The graph $\hg$ can be conveniently represented by an upper triangular matrix of size $n\times n$ with elements being the 
curves $p_i/q_i$ for $i \in [n]$. The initial curve $p_{n}$ can be thought of in location $(0,0)$. The first row consists 
of curves $q_n,q_{n-1},\ldots,q_1$ comprising $Q_n$. The second row consists of $p_{n-1},p_{n-2},\ldots,p_1$ comprising $P_{n-1}$, 
and so on. The final row consists of either $p_1 = P_1$ or $q_1 = Q_1$, depending on $n$ being even or odd respectively.
The $\hg$ graph in the form of an upper triangular matrix in the general case is shown below. In each cell of the matrix, the
number of incoming paths from $p_n$, which is a binomial coefficient (proof is given below) is shown in red. The last column shows the total number
of paths into each row.

\begin{equation*}
\begin{array}{|c||c|c|c|c|c|c|c|c|c|c|c|}
\hline
p_{n}^{\rt{1}} & 1 & 2 & 3 & 4 & 5 & \ldots & n-3 & n-2 & n-1 & n & \#paths\\
\hline
\hline
	1 & q_{n}^{\rt{\binom{0}{0}}} & q_{n-1}^{\rt{\binom{1}{0}}} & q_{n-2}^{\rt{\binom{2}{0}}} & q_{n-3}^{\rt{\binom{3}{0}}} & q_{n-4}^{\rt{\binom{4}{0}}} & \ldots & q_4^{\rt{\binom{n-4}{0}}} & q_3^{\rt{\binom{n-3}{0}}} & q_2^{\rt{\binom{n-2}{0}}} & q_1^{\rt{\binom{n-1}{0}}} & \rt{\binom{n}{1}} \\
\hline
	2 &                 & p_{n-1}^{\rt{\binom{1}{1}}} & p_{n-2}^{\rt{\binom{2}{1}}} & p_{n-3}^{\rt{\binom{3}{1}}} & p_{n-4}^{\rt{\binom{4}{1}}} & \ldots & p_4^{\rt{\binom{n-4}{1}}} & p_3^{\rt{\binom{n-3}{1}}} & p_2^{\rt{\binom{n-2}{1}}} & p_1^{\rt{\binom{n-1}{1}}} & \rt{\binom{n}{2}} \\
\hline
	3 & &             & q_{n-2}^{\rt{\binom{2}{2}}} & q_{n-3}^{\rt{\binom{3}{2}}} & q_{n-4}^{\rt{\binom{4}{2}}} & \ldots & q_4^{\rt{\binom{n-4}{2}}} & q_3^{\rt{\binom{n-3}{2}}} & q_2^{\rt{\binom{n-2}{2}}} & q_1^{\rt{\binom{n-1}{2}}} & \rt{\binom{n}{3}} \\
\hline
	4 & &            &             & p_{n-3}^{\rt{\binom{3}{3}}} & p_{n-4}^{\rt{\binom{4}{3}}} & \ldots & p_4^{\rt{\binom{n-4}{3}}} & p_3^{\rt{\binom{n-3}{3}}} & p_2^{\rt{\binom{n-2}{3}}} & p_1^{\rt{\binom{n-1}{3}}} & \rt{\binom{n}{4}} \\
\hline
	5 & &            &             &             & q_{n-4}^{\rt{\binom{4}{4}}} & \ldots & q_4^{\rt{\binom{n-4}{4}}} & q_3^{\rt{\binom{n-3}{4}}} & q_2^{\rt{\binom{n-2}{4}}} & q_1^{\rt{\binom{n-1}{4}}} & \rt{\binom{n}{5}} \\
\hline
	\vdots & &            &             &             &             & \ddots & \vdots & \vdots & \vdots & \vdots & \vdots \\
\hline
	n-3 & &            &             &             &             &             & q_4^{\rt{\binom{n-4}{n-4}}} & q_3^{\rt{\binom{n-3}{n-4}}} & q_2^{\rt{\binom{n-2}{n-4}}} & q_1^{\rt{\binom{n-1}{n-4}}} & \rt{\binom{n}{n-3}} \\
\hline
	n-2 & &            &             &             &             &             &             & p_3^{\rt{\binom{n-3}{n-3}}} & p_2^{\rt{\binom{n-2}{n-3}}} & p_1^{\rt{\binom{n-1}{n-3}}} & \rt{\binom{n}{n-2}} \\
\hline
	n-1 & &            &             &             &             &             &             &             & q_2^{\rt{\binom{n-2}{n-2}}} & q_1^{\rt{\binom{n-1}{n-2}}} & \rt{\binom{n}{n-1}} \\
\hline
	n & &            &             &             &             &             &             &             &             & p_1^{\rt{\binom{n-1}{n-1}}} & \rt{\binom{n}{n}} \\
\hline
\end{array}
\end{equation*}

Each element of the first row has a single transformation arc from $p_{n}$. In other words $q_{n-i} = \tau_{n+1-i}(p_{n})$, for $i = 1,\ldots,n$.
\begin{definition}(paths)
	For an element $(i,j) \in [n]\times[n]$ with $i \le j$, the number of paths into it is denoted by $\wp(i,j)$,
	and it equals the sum of all paths into its parent curves in row $i-1$. Formally
	\begin{align}
		\wp(i,j) = \sum_{k=i-1}^{j-1}\wp(i-1,k).
	\end{align}
	We define $\wp(0,0) = 1$, and $\wp(0,j) = 0$ for all $j \in [n]$.
\end{definition}

From the definition, we have $\wp(1,j) = 1$ for all $j \in [n]$ since all of these curves have only one parent curve $p_{n}$.
For the second row, we have 
\begin{align*}
	\wp(2,2) & = \wp(1,1) = 1, \\
	\wp(2,3) & = \wp(1,1) + \wp(1,2) = 2, \\
	\wp(2,j) & = \wp(1,1) + \wp(1,2) + \ldots + \wp(1,j-1) = j-1, \text{for all } j = 3,\ldots,n.
\end{align*}

We use the following well known identity (sometimes called as the hockey-stick identity) in the proofs below. 
\begin{identity}
	For any positive integer $r \in [n]$, we have
	\begin{align*}
		\binom{n}{r} = \binom{n-1}{r-1} + \binom{n-2}{r-1} + \ldots + \binom{r-1}{r-1}.
	\end{align*}
	\label{id:good}
\end{identity}
This identity can be easily proved by repeatedly applying Pascal's identity: $\binom{n}{r} = \binom{n-1}{r-1} + \binom{n-1}{r}$.
\begin{proposition}
	The number of paths into an element $(i,j)$ with $i \le j$ is
	\begin{align*}
		\wp(i,j) = \sum_{k=i-1}^{j-1}\wp(i-1,k) = \binom{j-1}{i-1}.
	\end{align*}
\end{proposition}
\begin{proof}
	We give the proof by induction. It is true for the first row trivially since $\binom{j}{0} = 1$ for all $j \in [n]$.
	By induction hypothesis, suppose it is true for row $i$ for some $i \in [n]$. Now consider row $i+1$ and a column $j$.
	The parent curve indices for the element $(i+1,j)$ are $(i,i), (i,i+1), \ldots, (i,j-1)$. By definition of paths,
	\begin{align*}
		\wp(i+1,j) & = \wp(i,i) + \wp(i,i+1) + \ldots + \wp(i,j-1), \\
		& = \binom{i-1}{i-1} + \binom{i}{i-1} + \ldots + \binom{j-2}{i-1}, \\
		& = \binom{j-1}{i}, \text{(by~\cref{id:good})}
	\end{align*}
	and the claim follows.
\end{proof}

\begin{definition}
	The total number of paths into level $r \in [n]$ in $\hg$ is the sum of all paths into elements of row $r$. We denote this
	quantity by $\beta(r)$.
\end{definition}

\begin{theorem}
	The number of paths $\beta(r)$ into level $r \in [n]$ follows a Binomial distribution.
	\label{thm:binomial}
\end{theorem}
\begin{proof}
From the definition above, we have
\begin{align*}
	\beta(r) & = \sum_{j=r}^n\wp(r,j), \\
	& = \binom{r-1}{r-1} + \binom{r}{r-1} + \binom{r+1}{r-1} \ldots + \binom{n-1}{r-1}, \\
	& = \binom{n}{r}, \text{ (by~\cref{id:good})},
\end{align*}
and the claim follows.
\end{proof}

\begin{proposition} 
	Let ${\psi}_0 = p_{n}$. Then the total number of paths into all levels $r \in [n]$ is $2^n$,
	and the maximum number of distinct descendant paths of $\psi_0$ including $\psi_0$ itself is $2^n/8$.
\end{proposition}
\begin{proof} 
	Since the number of paths into level $r$ is $\binom{n}{r}$, the total number of paths into all levels is easily seen to be $2^n$.
	However, some of these NDPs are not distinct since $\pp{p}_1 = \pp{q}_1$ and $\pp{p}_2 = \pp{q}_2$. Furthermore, since the pair $(p_4,q_4)$ covers $S_4$, it
	follows that we can skip $\pp{p}_3$ and $\pp{q}_3$ as well. 
	Thus we can remove the last three columns and last three rows in the matrix form of $\hg$ to obtain $2^n/8$ distinct NDPs.
\end{proof}

\subsection{A covering by NDPs}

Starting with the $p_n$ curve (or $q_n$ curve), we have seen that many new curves can be generated
by applying the $\tau$ transforms repeatedly. 
A natural question arises: does this collection of curves cover
the solution space $S_n$? Next we prove this in the affirmative.

\begin{theorem}
The complete set of paths obtained from ${\psi}_0$ cover the point set $S_n$.
	\label{theorem:cover}
\end{theorem}
\begin{proof} 
	We give the proof by induction.\\ 
	\emph{(Base Step)} Consider the base case $n=4$. Set ${\psi}_0 = p_4$ with \emph{active region} being the whole curve. We have only 
	one possibility for descendant paths, namely ${\psi}_1 = {\tau}_4({\psi}_0)$ which is a $q_4$ curve. Since the set $\{p_4,q_4\}$ exhausts 
	all the $16$ points of the solution space, the hypothesis is true. \\ 
	\emph{(Inductive Step)} Suppose the hypothesis is true for $n=k$ where $k>4$. In other words, if $\psi_0 = p_k$, then the set of all descendant paths
	of $\psi_0$ cover $S_4$. When we go from $k$ to $k+1$ we get two clusters $S_k$ 
	of point sets; one located at $0$ and one at $c_{k+1}$. We have ${\psi}_0 = p_{k+1}$ with active region being whole of $p_{k+1}$. 
	By performing ${\tau}_{k+1}({\psi}_0)$, we obtain $q_{k+1}$. Now applying $\tau_k(q_{k+1})$, we obtain
	$p_k$ which along with $q_{k+1}$ (locally $q_k$)encloses the second cluster $S_k$ located at $c_{k+1}$. All points within this cluster $S_k$ 
	can be reached by hypothesis. 

	Now we need to show that all points in the first cluster can be reached from ${\psi}_0$ as well. 
	This is easy since $p_{k+1}$ has multiple local characters of $\{p_4,p_5,\ldots,p_k,p_{k+1}\}$. By choosing the local character $p_k$,
	i.e., by applying $\tau_k(p_{k+1})$ we obtain the $q_k$ curve in the first cluster located at $0$.
	This $q_k$ curve along with $p_{k+1}$ (locally $p_k$) encloses all the points of the first cluster.
	By hypothesis all points in this cluster $S_k$ can be reached.  Hence it is true that all points in $S_n$ are covered by paths 
	obtained from ${\psi}_0$.
\end{proof}

\begin{figure}[!htbp]
\begin{center}
\epsfysize=4in \epsfxsize=6in {\epsfbox{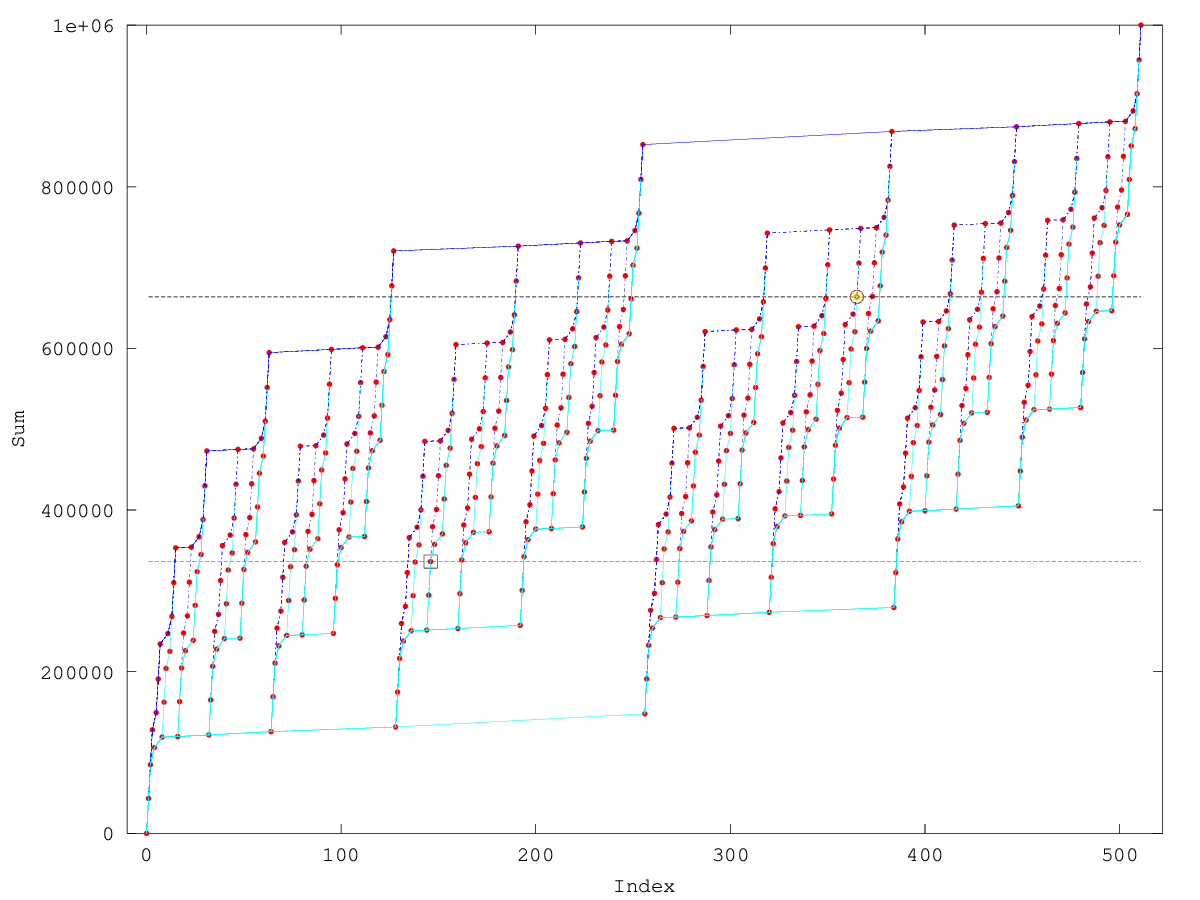}}
\caption{Illustration of the set of all NDPs $\Psi$ for the main example. The $p$-curves are in blue, and the $q$-curves
are in cyan. The symmetries of this family of curves and the subspaces are clearly evident. 
The dotted black line is the OL $y=T$, while the dotted red line is the complement $y = A_n -T$.
The highlighted circle corresponds to the solution with target value T, while the highlighted square corresponds to the complement, i.e., $A_n-T$.
After all, the solution space of the SSP has a rich and beautiful structure! 
}
\label{fig:mainexample}
\end{center}
\end{figure}

\begin{definition}
	Let $\Psi$ denote the set of all descendant paths, obtained from $\psi_0$, including $\psi_0$ itself.
\end{definition}

Recall that the NDPs are constructed to be independent of the magnitude of elements in ${\bf a}$. 
Thus the family of paths $\Psi$ will remain non-decreasing for all instances of size $n$. Thus, we have proved our first claim:
\begin{theorem}
	The set of non-decreasing paths $\Psi$ is a universal geometric structure for all instances of size $n$.
\end{theorem}

The complete set of paths, obtained starting from $p_n$ (or equivalently from $q_n$), for the \emph{main example} are shown in~\cref{fig:mainexample}. 
The point set that would appear like a random set of points in the Cartesian plane without the paths, once we overlay the paths suddenly 
reveals a hidden order and symmetry. 
The fascinating symmetries of paths and subspaces, as well as coverage of the point set is evident in the figure. 
While the configuration of points will change with each instance, the family $\Psi$ will track them.
All the $2^{9-3} = 64$ paths between $0$ and $C_9$ can be seen (in fact be counted visually).  
The path containing the solution (marked by a circle) corresponds to the following sequence of transformations:
	\begin{align*}
		\pp{\psi_{sol}} & = \tau_5(\tau_7(\tau_8(\pp{q_9}))), \quad \text{(if starting from } \pp{q}_9\text{)} \\
		\text{or } \pp{\psi_{sol}} & = \tau_5(\tau_7(\tau_8(\tau_9(\pp{p_9}))))  \quad \text{(if starting from } \pp{p}_9\text{)}.
	\end{align*}

\subsection{Segments and Path multiplicity}
\label{subsec:seg_mult}
\begin{definition} (segment)
        Let $s_i = [z_i^-, z_i^+]$ be an interval  with $z_i^-, z_i^+ \in S_n$. We call $s_i$ a segment
        if and only if it is on at least one NDP $\psi \in \Psi$.
\end{definition}

We have $|\Psi| = \frac{2^n}{8}$ distinct NDPs in the geometry, and each NDP has $\frac{n(n+1)}{2}$ edges.
Thus there are $\frac{2^n n(n+1)}{16}$ total segments in the geometry. It can proved that the number of unique segments is only $2^n+2$
for all $n > 3$, and that for $n=3$ it is $2^3$.

\begin{definition}(path multiplicity)
To each unique segment  $s_i, i \in [2^n+2]$ in the geometry, we associate a positive integer $\nu_i$, which is the number of NDPs that go through
$s_i$. We call $\nu_i$ the path multiplicity of segment $s_i$.
\end{definition}

For example, for the first segment $s_1 = [0, b_1+\iota a_1]$, $\nu_1 = \frac{2^n}{8}$, since all the NDPs go through it.
Similarly the last segment with end points $[(B_n-b_1) +\iota(A_n-a_1), B_n +\iota A_n]$ also has a path multiplicity of $\frac{2^n}{8}$.
However, the segment $[b_2+\iota a_2, b_1+b_2 + \iota(a_1+a_2)]$ has only one NDP going through it. In general, the path multiplicity of an arbitrary segment
in an NDP $\psi \in \Psi$ lies in the interval $[1, \frac{2^n}{8}]$. 

Next, consider the segments of~\cref{fig:mainexample} that intersect the OL. With not too much difficulty, one can see that there are $25$ distinct
segments that intersect the OL with multiplicities $(16, 8, 4, 2, 2, 1, 1, 8, 4, 2, 1, 1, 1, 1, 1, 1, 1, 1, 1, 1, 1, 1, 1, 1, 2)$ from left to right
respectively. Also note that the sum of these multiplicities is equal to $64 = 2^{9-3}$, the total number of distinct NDPs in the \emph{main example}.
The first intersection point is that of $p_7$ which contains all the subspaces leading to a multiplicity of $16$. This same
pattern continues for larger $n$ with many intersection points having an exponential multiplicity. The main thing to note is
that many of $2^n/8$ curves do not directly intersect the OL; they only do through higher curves (subspaces) in which they are
contained.  When all the elements of ${\bf a}$ are distinct, there is a gradual rise of the graph as we go from left to right. This staggering
causes many intersection points to have high multiplicities. We will refer to this discussion again in~\cref{sec:bhairav}.

\subsection{A transformation vector}
We have shown that the set of NDPs $\Psi$ derived from $\psi_0$ covers the solution space. 

Given a binary vector ${\bf x}$ what is the NDP $\psi$ that contains it? The answer is given by the following theorem.

\begin{theorem}
	Given any binary vector ${\bf x} \in \{0,1\}^n$, and $\psi_0 = q_n$ (or $p_n$), there is at least one NDP $\psi = K(\psi_0)$ that contains it,
where $K = (k_l,k_{l-1},\ldots,k_2,k_1)$ is the sequence of transformations $\tau_{k_i}$ on $\psi_0$.
Furthermore, the sequence of transformations encoded as a binary vector $K$, that produce $\psi$ from $\psi_0$ 
	are encoded in the vector $\bfx =(x_1,x_2,\ldots,x_n)$ itself as 
	\begin{align*} 
		\text{index}(K) = {\sum}_{i=1}^{n} |x_{i+1} - x_{i}| b_{i},
	\end{align*}
	where 
	\begin{align*}
		x_{n+1} = \begin{cases}
			1 \quad \text{if } \psi_0 = q_n \\
			0 \quad \text{if } \psi_0 = p_n. 
		\end{cases}
	\end{align*}
\end{theorem}

\begin{proof}
	Let ${\bf x}$ be the given binary vector with coordinates given by $z = ({\bf b}\cdot {\bf x}, {\bf a} \cdot {\bf x})$. 
	Recall that the space $S_n$ comprises of two copies of $S_{n-1}$, one at $0$ and the other
	at $c_n$. We refer to these as the left ($L$) and right ($R$) clusters. If $x_n = 1$ then it is clear that the point $z$ in the cluster $R$,
	whereas if $x_n =0$ then it belongs to cluster $L$. We pick the cluster that contains $z$ and descend into $S_{n-1}$. Then we check if
	$x_{n-1}$ is $1$ or $0$ and descend into the appropriate cluster $S_{n-2}$. We keep repeating this following a binary tree 
	until $x_1$ at which point we have located the point.
	If $\psi_0 = q_n$ and $x_n=1$, then the point is in the cluster $R$ we don't need to do anything. On the other hand if $x_n = 0$ we need to go to cluster $L$,
	which requires the transformation $\pp{p}_n = \tau_n(\pp{q}_n)$. It is straight forward to see that a $\tau_k$ operation is needed whenever 
	$x_{k+1} = 1$ and $x_k =0$, or $x_{k+1} = 0$ and $x_k = 1$. Alternatively, we need a $\tau_k$ transform when there is a change in consecutive bits $1 \ra 0$ or
	$0 \ra 1$. When a $\tau_k$ operation is used we set the $k$th  bit of $index(K)$ to $1$, otherwise
	set it to $0$. Doing this we obtain the formula for $index(K)$ and the claim follows.
\end{proof}

\begin{example} 
	Consider the solution vector ${\bf x} = (1,0,1,1,0,1,1,0,1)$ of the main example, with bit positions $9,8,\ldots,1$ respectively. Then 
	using the above theorem, we can compute the index of the $K$ vector as
	\begin{align*}
		index(K) = b_8 + b_7 + b_5 + b_4 + b_2 + b_1 = 219
	\end{align*}
	with the sequence of $\tau$ transforms given by
	\begin{align*} 
		\pp\psi = \tau_1(\tau_2(\tau_4(\tau_5(\tau_7(\tau_8(\pp{q}_9)))))).
	\end{align*}
	Note that in this example, we can in fact stop at $\tau_5$ since the point corresponding to the solution vector becomes part of a fixed chain
	after this.
\end{example}

\begin{theorem}
	The \ssp\ee has a solution, if and only if, at least one path $\psi \in \Psi$, intersects the orbital line at a vertex point.
\end{theorem}
How do we find all the points on the OL?
A brute force way is to exhaustively search all the paths in $\Psi$, where for each path we check all the points of each curve 
if any equals the target value. Clearly this is infeasible!

By understanding the relationship between the NDPs in $\Psi$ and using that relationship, we can indeed obtain a better algorithm.
This will need a few more ideas which are covered in the subsequent sections.

\subsection{Wormhole paths of $\hg$}
We have seen that for any path in $\hg$, the active region is a monotonically decreasing function of path length (number of transformations). In this section we 
examine the lower and upper boundaries of the active region at any point in the path and how they create \emph{wormhole} paths.

Consider the chain is $\pp{p_{n}}$. If we apply the $\tau_{j}$ operation on it where $j < n$
we get
\begin{align*}
	\tau_{j}(\pp{p_{n}}) = \rho_j(\lambda_j(\pp{p_{n}})) = \pp{q_j}\oplus\pp{p_n}[j+1,n].
\end{align*}
The active region of the initial chain is the interval $[1,n]$ which corresponds to the whole curve $p_n(t)$, i.e., $[(0,0), (B_n,A_n)]$. 
After the transformation, the new active region corresponds to $[(0,0),(B_j,A_j)]$. 
Now consider a further transformation on $\pp{q_j}$ given by $\tau_{k}(\pp{q_j}\oplus\pp{p_n}[j+1,n])  = \pp{q_j}[1,j-k] \oplus \pp{p_k}\oplus\pp{p_n}[j+1,n]$.
Now the active region of the new curve is $[(B_j-B_k,A_j-A_k), (B_j,A_j)]$. We notice from this example that when we go from a $p$-curve to a $q$-curve,
the upper end of the active region is lowered, and when we go from a $q$-curve to a $p$-curve the lower end point is raised. In either transformation,
the length of the active region (measured by $y$ range) decreases. For any path in $\hg$, the active region monotonically decreases with
number of transformations, by alternatingly lowering the upper end point or raising the lower end point.

\begin{example}
	In the main example, we have seen previously that the curve containing the solution point can be obtained from $p_{9}$ via the
	sequence of transformations: $(\tau_9,\tau_8,\tau_7,\tau_5,\tau_4)$. In~\cref{fig:paths_H0_main} we show the intervals ($y$ coordinate only) containing
	the curves after each successive $\tau$ transformation. The final curve is that of $q_4$ after transformation $\tau_4$. The OL
	is shown as a blue line.  
	We notice that the $y$-interval $[395296, 748620] \in p_{9}$ corresponds to the projection of $q_4$ onto $p_{9}$.

        \begin{figure}[!htbp]
        \begin{center}
        \epsfxsize=5in \epsfysize=3in {\epsfbox{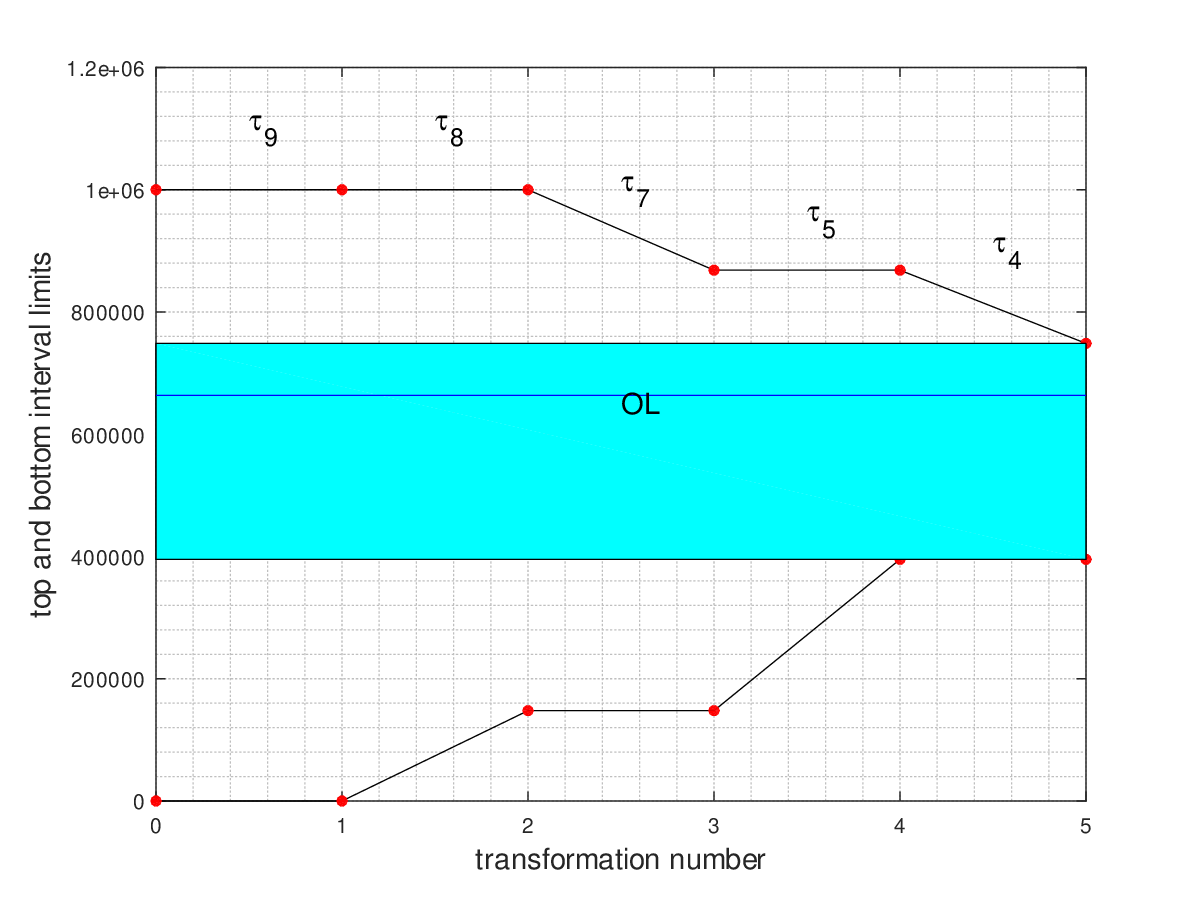}}
        	\caption{Illustration of the path $\pi = (\tau_9,\tau_8,\tau_7,\tau_5,\tau_4)$ for the main example.
                The top and bottom lines show the limits of the curve after each transformation. The OL is shown in
        	blue. 
        	}
        \label{fig:paths_H0_main}
        \end{center}
        \end{figure}
\end{example}

\begin{definition}
	For an arbitrary path $\pi$ of length $r$ in $\hg$, represented by
	\begin{align*}
		\pi = (k_1, k_2, \ldots, k_r)
	\end{align*}
	where $k_1 > k_2 > \ldots > k_r$ and $k_i \in [n]$ for all $i \in [n]$,
	let $\ell := (\ell_1, \ell_2, \ldots, \ell_r)$ be the sequence of 
	$y$ values for the lower curve, and let $\hbar := (\hbar_1,\hbar_2,\ldots,\hbar_r)$ be the sequence of $y$ values
	for the upper curve. Note that for any $i \in [r]$, the pair $(\ell_i,\hbar_i)$ corresponds to a complete 
	$p_j/q_j$ curve for some $j \in [n]$.
\end{definition}

\begin{proposition}
	For a path $\pi$ of length $r$ in $\hg$, represented by
        \begin{align*}
                \pi = (k_1, k_2, \ldots, k_r)
        \end{align*}
        where $k_1 > k_2 > \ldots > k_r$ and $k_i \in [n]$ for all $i \in [r]$,
        we have
	\begin{align*}
		\ell_i = \begin{cases}
			\sum_{j=1}^i (-1)^{(j+1)}A_{k_j} \quad \text{if $i$ is even}, \\
			\sum_{j=1}^{i-1} (-1)^{(j+1)}A_{k_j} \quad \text{if $i$ is odd}, 
		\end{cases}
	\end{align*}
	where the empty sum is treated as zero, and 
	\begin{align*}
		\hbar_i = \alpha_i + A_{k_i}.
	\end{align*}
\end{proposition}
The proof is omitted as it is simply a consequence of the transformation of chains.
The sequences $\ell$ and $\hbar$ given by the above proposition have the form
\begin{align*}
	\ell & = (0, (A_{k_1}-A_{k_2}), (A_{k_1}-A_{k_2}), (A_{k_1}-A_{k_2}) + (A_{k_3}-A_{k_4}), \\
	& (A_{k_1}-A_{k_2}) + (A_{k_3}-A_{k_4}), \ldots), \\
	\hbar & = (A_{k_1}, (A_{k_1}-A_{k_2})+A_{k_2}, (A_{k_1}-A_{k_2})+A_{k_3}, (A_{k_1}-A_{k_2}) + (A_{k_3}-A_{k_4})+A_{k_4}, \\
	& (A_{k_1}-A_{k_2})+(A_{k_3}-A_{k_4})+A_{k_5}, \ldots)
\end{align*}
as can be verified. It can be seen that $\ell$ is alternatingly staying the same and increasing, while $\hbar$ is alternatingly decreasing and staying the
same. 
\begin{definition} (wormhole)
	For any transformation path $\pi \in \hg$, let $\ell^{\pi}$ and $\hbar^{\pi}$ be the associated lower and upper curves respectively.
	We then call the region enclosed by the curves $\ell^{\pi}$ and $\hbar^{\pi}$
	as a wormhole path or simply a wormhole due to its shape.
\end{definition}

\begin{definition}(Valid Wormhole for a given OL)
	Given any path $\pi$ from $p_{n}$ in $\hg$, and a given $T$ we say that the path $\pi$ is valid if and only if the associated
	wormhole contains $T$. Alternatively, a path $\pi$ of length $k$ is valid if and only if
	\begin{align*}
		\ell_i^{\pi} \le T \le \hbar_i^{\pi}, 
	\end{align*}
	for all $i \in [k]$. 
\end{definition}

Next we give a larger example illustrating the fact that for a given OL, not every path in $\hg$ need contain it. 
\begin{example}
	This is a random example with $n=30, m=40$, with some path $(30,27,25,24,21,17,16,12,9,7,5)$ in $\hg$. The corresponding intervals are shown along with the OL
	in blue. 
        \begin{figure}[!htbp]
        \begin{center}
        \epsfxsize=5in \epsfysize=3in {\epsfbox{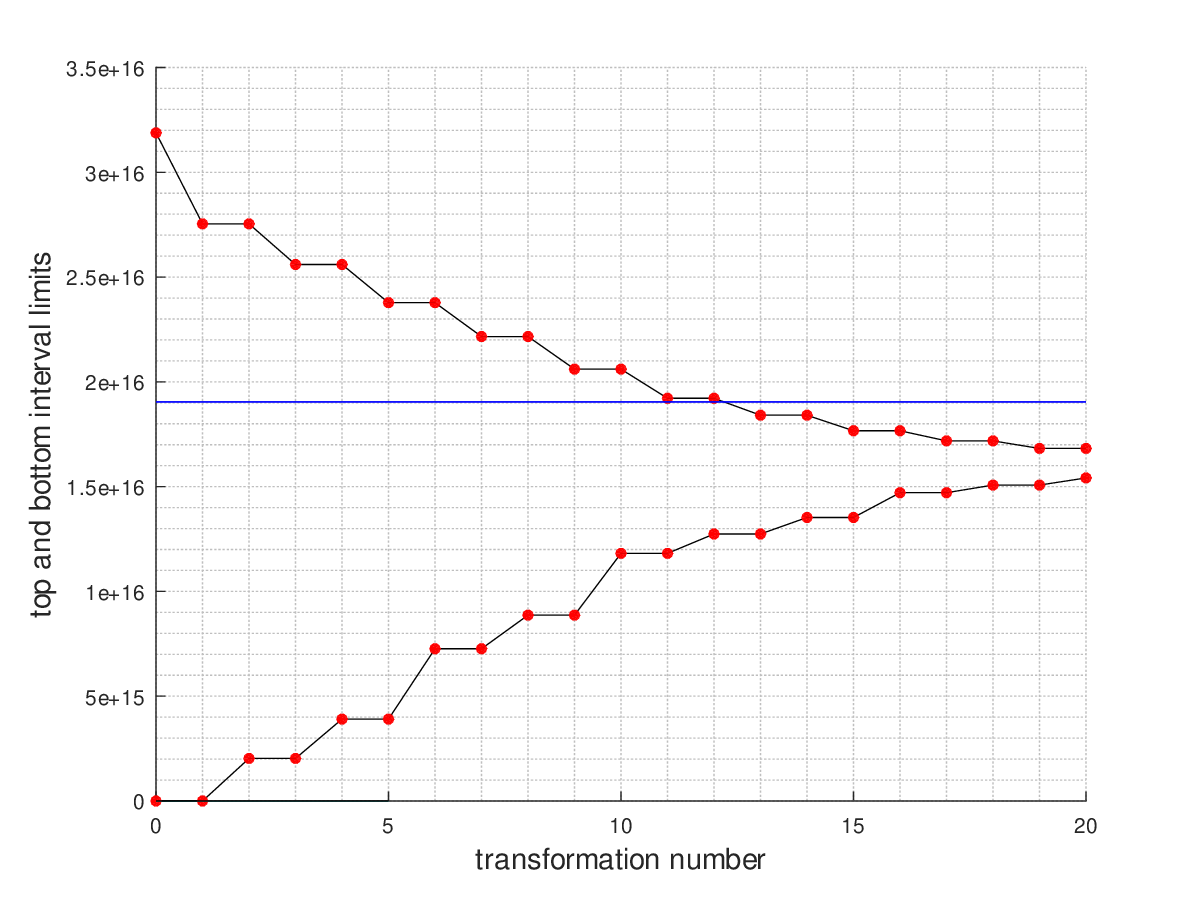}}
        	\caption{Illustration of an arbitrary path to level $r=20$ for a $n=30$ example.
		The OL (blue line) is contained in the initial part of the path but later misses it.
		This shows that not every path in $\hg$ contains a given OL $y=T$.
        	}
        \label{fig:paths_H0}
        \end{center}
\end{figure}
\end{example}

\subsection{Number of OLs into each curve/level for a given $T \in p_{n}$}
We have seen above that not all wormholes associated with the paths in $\hg$ contain a given OL $y=T$. 
To provide a clear picture for the upcoming analysis in the subsequent sections, we provide some definitions below:
\begin{definition}
	\begin{align*}
		\text{Let } \beta(r) := & \hs \text{\# transformation paths into level $r$ in $\hg$}, \\
		\beta(r,i) := & \hs \text{\# transformation paths into curve $p_i/q_i$ in level $r$ in $\hg$}, \\
		\beta_{T}(r,i) := & \hs \text{\# valid OLs into curve $p_i/q_i$ in level $r$ in $\hg$ for a given $T \in p_{n}$}, \\
		\beta_{T}(r) := & \hs \text{\# valid OLs into all curves of level $r$ in $\hg$ for a given $T \in p_{n}$}, \\
		\hat{\beta}_{T}(r,i) := & \hs \text{\# distinct valid OLs into curve $p_i/q_i$ in level $r$ in $\hg$ for a given $T \in p_{n}$}, \\
		\text{and } \hat{\beta}_{T}(r) := & \hs \text{\# distinct valid OLs into all curves of level $r$ in $\hg$ for a given $T \in p_{n}$}.
	\end{align*}
	We have proved earlier that $\beta(r,i)  = \binom{n+1-i}{r-1}$ and $\beta(r)  = \binom{n}{r}$.
\end{definition}

The following proposition follows directly from the above defintion.
\begin{proposition}
	For any given instance $({\bf a}, T)$ it is true that
	\begin{align*}
		\hat{\beta}_T(r,i) \le \beta_T(r,i) \le \beta(r,i) = \binom{n+1-i}{r-1}
	\end{align*}
	and
	\begin{align*}
		\hat{\beta}_T(r) \le \beta_T(r) \le \beta(r) = \binom{n}{r}.
	\end{align*}
\end{proposition}

\section{Intersection Points and Local Coordinates}
\label{sec:poc}
We have seen that there are at most $2^n/8$ distinct NDPs from $0$ to $C_n$. For a given orbital line $y=T$, all these
paths intersect the OL. We are interested in understanding the relation between these intersection points and this section is devoted
for this task.

\subsection{Edges and Local Reference Frames}
Let $\psi$ be an NDP with associated chain $\pp{\psi}$. 
The chain $\pp{\psi}$ comprises of $\frac{n(n+1)}{2}$ links, and the corresponding curve $\psi$ comprises of the same number
of line segments which are referred to as \emph{edges} (formally defined below). Recall that the points on the curve $\psi$ are 
given by
\begin{align*}
	\psi(i) = \sigma(\pp{\psi}, i), \quad \text{for all } i \in [\frac{n(n+1)}{2}]
\end{align*}
where $\psi(0) = 0$. 

\begin{definition}(Edge)
	Let $\psi$ be a NDP. The line segments of $\psi$ are called edges and denoted by $e_i$, where
	$i \in [\frac{n(n+1)}{2}]$. The lower end point of edge $e_i$ is denoted by $z_i^{-}$ and the upper end point by $z_i^{+}$.
	Thus,
	\begin{align*}
		z_i^{-} & = x_i^- + \iota y_i^- = \sigma(\pp{\psi},i-1), \\
		\text{and } z_i^{+} & = x_i^+ + \iota y_i^+ = \sigma(\pp{\psi},i). 
	\end{align*}
	We associate the half open interval $[z_i^{-}, z_i^{+})$ to each edge $e_i$. Note that the point $z_i^+$ itself does not belong to $e_i$, 
	to avoid double counting of vertex points.
	Since each edge corresponds to a link in the chain, for an edge $e_i$ we have an associated link $d_j$ for some $j \in [n]$.
	We denote the length of the edge $e_i$ by $|e_i|$ and define it (using only the $y$ coordinates) as
	\begin{align*}
		|e_i| = y_i^+ - y_i^- = \Im(d_j).
	\end{align*}
	While the Euclidean length of $e_i$ is strictly positive (since $\Re(d_j) > 0$ and $\Im(d_j) \ge 0$), the length $|e_i|$ however can be zero if $a_j  = a_{j-1}$.
	\label{def:edge}
\end{definition}
Comparing the above definition to~\cref{def:dseg}, one can see that an \emph{edge} is nothing but a translated \emph{link}.
Another way of saying this is that a \emph{chain} is a sequence of \emph{links}, whereas a NDP is a sequence of \emph{edges}.

\begin{definition}(Local Reference Frame of an edge $e_i$)
	Let $e_i$ be an edge on a NDP $\psi$, with end points $z_i^-$ and $z_i^+$ corresponding to a difference $d_j$ for some $j \in [n]$.
	We imagine a Cartesian coordinate system located at $z_i^-$, and call this the local
	reference frame (LRF) of the edge $e_i$.  We call the point $z_i^-$ as the origin of the edge $e_i$. 
	An arbitrary point $z \in {\mathbb C}$ has a local value $z_i^{loc}$ in the LRF of $e_i$, given by
	\begin{align*} 
		z_i^{loc} = z - z_i^-.
	\end{align*}
	The point $z$ in global coordinates is simply $z = z_i^- + z_i^{loc}$. 
	\label{def:lrf}
\end{definition}

\begin{definition}(Valid points on an edge)
	A point $z$ is said to be valid point of edge $e_i$, if it lies in between the end points of the edge.
	For a valid point $z$, there exists a rational $\delta \in [0,1)$, such that
	$z = z_i^- + \delta d_j$.  
	Then the coordinates of $z$ in the LRF of the edge $e_i$, denoted $z_i^{loc}$, are
	\begin{align*}
		z_i^{loc} = x_i^{loc} + \iota y_i^{loc}  = \delta d_j. 
	\end{align*}
	For a valid point $z$ in the LRF of $e_i$, we simply write $z \in e_i$.
\end{definition}

\subsection{Local coordinates of intersection points}
\label{sec:loc_coords}
Next we look at the intersection point of a NDP $\psi$ with a horizontal line $y = y_0$.
\begin{proposition}
	Given a positive integer $y_0 \in [0,A_n]$, the number of points belonging to NDP $\psi$, that have
	the $y$-coordinate equal to $y_0$, is either $1$ or $\infty$. 
	\label{prop:oneinf}
\end{proposition}
\begin{proof}
	If all the elements of ${\bf a}$ are distinct, then it follows that $\Im(d_j) > 0$ for all $j \in [n]$. 
	Since $0 < y_0 \le A_n$, there is a unique edge $e_i \in \psi$, such that $y_i^- \le y_0 < y_i^+$, and there is a unique intersection point of line $y=y_0$ with 
	the edge $e_i$. In the case one or more elements of $D$ have $\Im(d_j) = 0$, it is possible for one or
	more edges $e_i$ to coincide with the line $y = y_0$. In this case, we have infinite number of points
	belonging to $\psi$ that have $y$ coordinate equal to $y_0$.
\end{proof}

For the case when all elements of ${\bf a}$ are distinct, let the intersection point of $\psi$ with $y = y_0$ line be 
denoted $z_{\psi} = x_{\psi} + \iota y_{\psi}$ where $y_{\psi} = y_0$. Furthermore, let $e_i \in \psi$
(with associated link $d_j$) contain the intersection point $z_{\psi}$. From the above definitions, we can write $z_{\psi} = z_i^{-} + z_i^{loc}$ 
where $z_i^{loc} = \delta_{\psi}d_j$ and $\delta_{\psi} = \frac{y_i^{loc}}{|e_j|}$.  

For the case where the edge $e_i$ coincides with the $y=y_0$ line, we use the following prescription for local coordinates where we set
\begin{align}
	\delta_{\psi} := \begin{cases}
		0 & \text{ if } |e_j| = 0 \\
		\frac{y_{\psi}^{loc}}{|e_j|} & \text{ otherwise}.
	\end{cases}
\label{subsec:prescription}
\end{align}
With this, the local $x_i^{loc}$ coordinate is unambiguously determined as
\begin{align*}
	x_i^{loc} = \delta_{\psi} \Re(d_j).
\end{align*}
Thus, eventhough there are infinite number of intersection points (global coordinates) we have only a single local coordinate!

\begin{remark} 
	If there are $h$ edges of $\psi$, that are all horizontal and next to each other, and coincide with the $y=y_0$ line, then we would get $h$ 
	local coordinates (one for each edge) rather than an infinite number of them using the above definition for local coordinates.
	\label{rem:lv}
\end{remark}

\subsection{A Bit Of Relativity} 

Consider a pair of complementary curves $p_k$ and $q_k$, and a horizontal line $y = y_0$ where $y_0 \in [0, A_k]$. 
Since both $p_k$ and $q_k$ curves start at $0$ and end at $C_k$, both curves will intersect the $y=y_0$ line. Let the
intersection points be denoted $z_p = x_p + \iota y_p$  and $z_q = x_q + \iota y_q$ respectively. It is clear that
$y_p = y_0 = y_q$.

These points  are in \emph{global} coordinates (relative to the origin of subspace translate $S_k$ that contains the $p_k$ and $q_k$ curves). 
For simplicity we assume that each intersection point lies in a unique edge. (If multiple edges intersect the line, we will show how it can
be handled easily later on).  Let $z_p$ belong to some edge $e_j \in p_k$ with lower end point at $z_j^{-}$. Similarly, let $z_q$ belong to
some edge $e_i \in q_k$, with lower vertex at $z_i^-$. 
Using the LRF of each edge we can write this as
\begin{align*}
	z_q^{loc} & = z_q - z_i^-  \\
	z_p^{loc} & = z_p - z_j^-  = (z_q - z_i^-) + (z_p - z_q) + (z_i^- - z_j^-). 
\end{align*}
Thus
\begin{align}
	z_p^{loc} =  z_q^{loc} + (z_p - z_q) + (z_i^- - z_j^-). 
	\label{eq:zloc}
\end{align}
Since both $z_i^-$ and $z_j^-$ are coordinates relative to the origin of space $S_k$, the translation from $z_i^-$ to $z_j^-$ 
is simply $z_j^- - z_i^-$. We denote this translation by $w_{ij} := z_j^- - z_i^-$,
and can view it as a directed arrow from $z_i^-$ to $z_j^-$. 

If we consider only the $y$ coordinate, it can be readily seen that
\begin{align}
	y_p^{loc} & =  y_q^{loc} +  (y_i^- - y_j^-) = y_q^{loc} + \Im(\overline{w}_{ij})
	\label{eq:yloc}
\end{align}
where $\overline{w}_{ij}$ is complex conjugate of $w_{ij}$. 
This tells us that if we know $y_q^{loc}$ in the $q_k$ curve and the translation $w_{ij}$ we can obtain the local coordinate $y_p^{loc}$ in the
$p_k$ curve, and vice versa.

In the general case, we can have multiple edges $\{e_{i_1},e_{i_2},\ldots,e_{i_r}\}$ where $r < k$, intersecting the OL instead of a single edge $e_i$ 
of curve $q_k$. Similarly, we could have $\{e_{j_1},e_{j_2},\ldots,e_{j_s}\}$ where $s < k$, intersecting the OL instead of a single
edge $e_j$ of curve $p_k$. In view of~\cref{rem:lv}, we have
$r$ local coordinates on the $q_k$ curve and $s$ local coordinates on the $p_k$ curve. Why are $r,s < k$? A brief thought reveals
that if $r \ge k$ then all the elements of $\{d_1,d_2,\ldots,d_k\}$ have imaginary part equal to zero, which further implies that all 
$a_i = 0, \forall i \in [k]$.
Since we assumed that ${\bf a}$ consists of positive numbers, this situation cannot happen.
The above arguments are summarized in the following proposition.
\begin{proposition}
	Given a point on a edge of  $p_k$ (in local coordinates), we can find at most $k-1$ points, in $k-1$ edges of the complementary 
	curve $q_k$, such that all these points fall on the same OL. Furthermore, all the points on $q_k$  can be obtained by translations
	from the given point on $p_k$. The same is true when $p_k$ is replaced by $q_k$.
	\label{prop:lock}
\end{proposition}

Since the NDPs form a rigid structure, for any fixed edges $e_1 \in p_k$ and $e_2 \in q_k$ in the same subspace, their relative translation remains
invariant, as we go from one instance of subspace $S_k$ to another. Since each curve has $k(k+1)/2$ edges, we can compute all the
$(k(k+1)/2)^2$ translations. These translations remain the same for any translate of $S_k$ in the point set.
Using these translations, given any point in any edge of $p_k$, we can determine the image point (again
in local coordinates) in a suitable edge in $q_k$, such that the global values (relative to origin of $S_k$) have the same $y$
coordinate. We will build on this in the next section.

\subsection{Interacting edges}
\begin{definition}(Edge interaction)
	Let $e_r$ be an edge of the curve $p_k$, defined by the interval $[z_r^-, z_r^+)$. Similarly let $e_s$
	be an edge of the curve $q_k$, defined by the interval $[z_s^-,z_s^+)$. We say that $e_r$ and $e_s$ interact,
	if they have non-empty overlap in the $y$-coordinate, i.e.,
	\begin{equation}
		[y_r^-,y_r^+) \cap [y_s^-,y_s^+) \neq \emptyset.
	\end{equation}
	Otherwise, the edges are said to be non-interacting. 
	We write $e_r \cap e_s \neq \emptyset$, for simplicity,  if they are interacting.
\end{definition}

In view of this definition, it can be seen that an edge $e_r$ on curve $p_k$ can interact with \emph{more than one} edge
of curve $q_k$ and vice versa. For each of the $k(k+1)/2$ edges of $p_k$, we can determine which edges of $q_k$ interact with it.

\begin{definition}(Interaction length)
	The length of the interval over which two edges $e_r$ and $e_s$ interact is called the
	interaction length, which is denoted by $|e_r \cap e_s|$ and is easily seen to be 
	\begin{align*}
		|e_r \cap e_s| = \min(y_r^+,y_s^+) - \max(y_r^-,y_s^-).
	\end{align*}
	The edges $e_r$ and $e_s$ are said to be interacting if and only if $|e_r \cap e_s| \ge 0$.
\end{definition}

\begin{example}
	Let $e_p \in p_k$ and $e_q \in q_k$, such that $|e_p \cap e_q| \ge 0$. Then given $z_p^{loc} \in e_p$, 
	we have $y_q^{loc} = y_p^{loc} + \Im(\overline{w}_{pq})$. We have three cases 
	illustrated in~\cref{fig:edgeInter}.
	\begin{itemize}
		\item[(i)] The local $y$ values are related by $y_q^{loc} = y_p^{loc} + (y_p^- - y_q^-)$. Here, the 
			coordinates in both edges are valid.
		\item[(ii)] Given a valid $y_q'^{loc} \in e_q$, we have $y_p'^{loc} = y_q'^{loc} + (y_q^- -y_p^-) < 0$, 
			which is an invalid point.  
		\item[(iii)] Similarly, given a valid $y_p'' \in e_p$, we have $y_q''^{loc} = y_p''^{loc} + (y_p^- - y_q^-) > |e_q|$, 
			which is above the edge. Again, this is an invalid point.
	\end{itemize}
        \begin{figure}[!htbp]
        \begin{center}
        \begin{tikzpicture}
        %
        %
        \node[outer sep=0pt,circle, fill,inner sep=1.5pt,label={[fill=white]left:$z_p^-$}] (E1a) at (1,-1) {};
        \node[outer sep=0pt,circle, fill,inner sep=1.5pt, label={[fill=white]right:$z_p^+$}] (E1b) at (2,2) {};
        \node (N1) at (2.3,1) {$e_p$};
        \draw[] (E1a) -- (E1b);
        
        \node[outer sep=0pt,circle, inner sep=0.5pt] (P1) at (1.35,0) {$\circ$};
        \node[outer sep=0pt,circle, inner sep=0.5pt] (P2) at (5.4,0) {$\circ$};
        \draw[dotted] (P1) -- (P2);
        	\node (X1) at (0.9,0.3) {$z_p$};
        	\node (X1) at (5.8,0.3) {$z_q$};
        \node[outer sep=0pt,circle, blue, inner sep=0.5pt] (PB1) at (0.9,-1.5) {$\circ$};
        \node[outer sep=0pt,circle, blue, inner sep=0.5pt] (PB2) at (5.1,-1.5) {$\circ$};
        \draw[dotted,blue] (PB1) -- (PB2);
        	\node (B1) at (0.7,-1.5) {$z_p'$};
        	\node (B1) at (5.8,-1.5) {$z_q'$};
        \node[outer sep=0pt,circle, red, inner sep=0.5pt] (PB1) at (1.85,1.5) {$\circ$};
        \node[outer sep=0pt,circle, red, inner sep=0.5pt] (PB2) at (5.7,1.5) {$\circ$};
        \draw[dotted,red] (PB1) -- (PB2);
        	\node (B1) at (1.0,1.6) {$z_p''$};
        	\node (B1) at (6.0,1.6) {$z_q''$};
        %
        %
        \node[outer sep=0pt,circle, fill,inner sep=1.5pt,label={[fill=white]left:$z_q^-$}] (E2a) at (5,-2) {};
        \node[outer sep=0pt,circle, fill,inner sep=1.5pt, label={[fill=white]right:$z_q^+$}] (E2b) at (5.6,1) {};
        \node (N2) at (6.5,-0.5) {$e_q$};
        \draw[] (E2a) -- (E2b);
        \end{tikzpicture}
        \end{center}
		\caption{Illustration of two interacting edges. All coordinates shown are relative to the origin of $S_k$ that contains
		the curves $p_k$ and $q_k$.}
        \label{fig:edgeInter}
        \end{figure}
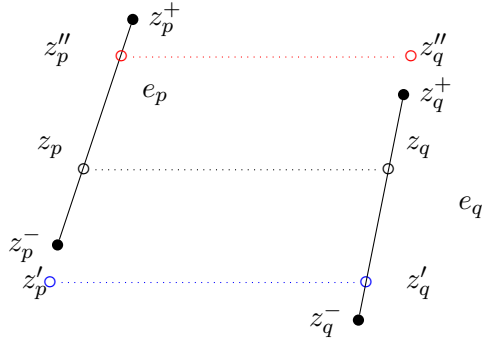
\end{example}

The above ideas can be summarized by the following proposition.
\begin{proposition}
	Let $\psi_1$ be a NDP with active region corresponding to $p_j$ for some $j \in [n]$. Let $\psi_2$
	be a derived path of $\psi_1$, given by the relation $\pp\psi_2 = \tau_k(\pp\psi_1)$, where
	$k < j$.
	Let $e_p \in \psi_1$ and $e_q \in \psi_2$, such that $|e_p \cap e_q| \ge 0$. Then given a point
	$z_p^{loc} \in e_p$, we have $y_q^{loc} = y_p^{loc} + \Im(\overline{w}_{pq})$.  
	\label{prop:lamrho}
\end{proposition}

If we look at~\cref{fig:mainexample}, in particular the OL, we see many edges intersecting it as we go from left to right.
All these intersection points have local coordinates in the reference frames of corresponding edges. With the work done
in previous section and this, we have a way to hop from one edge to another. As the reader may guess,
a repeated application of~\cref{prop:lamrho}, starting with an edge on $p_n$,  can enable us to visit all the intersection points
without leaving the OL.

\subsection{An unique representation of an edge}
\label{sec:5t}
In order to hop on the OL from one edge to another, we need a way to refer to the edges, as there are an exponential number of edges. 
Fortunately, the graph structure in~\cref{fig:pqgraph} itself provides part of the solution, as the curves are arranged in levels, i.e., equivalence classes. 
The chain $\pp{q_j}$ consists of $j(j+1)/2$ links, but only $j$ unique links, i.e., $d_1,d_2,\ldots,d_j$,
and the links repeat as we go along the chain. Thus, for the curve $q_j$, the edges which correspond to links, repeat.
As a result, we need a way to distinguish two edges of the same kind at different locations on the curve.

To handle this, we provide a unique $5$-tuple representation of an edge, defined below.

\begin{definition}($5$-tuple representation)
        For an edge on a path $\psi$, we associate a $5$-tuple: (level number, curve type, edge number, curve index, instance number) which
        are defined as
        \begin{align*}
		\text{level } & \in [n] : \text{ number of $\tau$ operations needed on $\pp{p}_n$ to obtain $\pp\psi$}, \\
                \text{curve type } & \in \{p-type,q-type\} : \text{ the type of the active region in $\pp\psi$} \\
                \text{edge number } & \in [n] : \text{ $j$ if the edge corresponds to $d_j$} \\
                \text{curve index } & \in \{1,\ldots,n\} : \text{ $j$ if active region corresponds to $p_j$ or $q_j$} \\
                \text{instance number } & \in [n] : \text{ $k$ if it is the $k$th copy of edge $d_j$}.
        \end{align*}
	Thus, for each edge $e_i$ we can associate a unique $5$-tuple $\digamma(e_i) = (f_1,f_2,f_3,f_4,f_5)$, where
	$f_1$ is the level number, $f_2$ is the curve type, $f_3$ is the edge number, $f_4$ is the curve index and $f_5$ is 
	the instance number. We will show later that the maximum number of unique $5$-tuples is $\frac{n(n+1)(n+2)(n+3)}{24}$.
        \label{def:3t}
\end{definition}

\subsection{Number of interacting edge pairs}

Consider a pair of complementary curves $p_k$ and $q_k$ for some $k \in [n]$.  Since there are $k(k+1)/2$ edges in each curve, the maximum number of pairs is 
trivially $(\frac{k(k+1)}{2})^2$. However, unless ${\bf a} = \bfz$, not all pairs will be interacting. 
Here, we provide bounds on the minimum and maximum number of interacting edges between two complementary curves.
Recall that two edges are said to be interacting if their interaction length is non-negative, i.e.,  even if their intersection is a single point. 

\begin{definition}(Number of interacting edges $J_k$)
	Let $k \in [n]$ be a positive integer. Then we define $J_k({\bf a})$ to be the number of interacting edge pairs between curves $p_k$ and $q_k$.
\end{definition}

\begin{proposition}
	Let ${\bf a}_{diss}$ be a dissociated set and let $3 \le k \le n$. Then $J_k({\bf a}_{diss}) = k^2+k-5$.
	\label{prop:jmin}
\end{proposition}
\begin{proof} 
	Since ${\bf a}_{diss}$ is dissociated, the subset sums of $p_k$ and $q_k$ (except for the $6$ common points) will all be distinct.
	Thus, the number of unique points between both curves is $k(k+1) - 4$.  If we use this set of 
	points to interpolate both curves, we get $k(k+1)-5$ disjoint segments where each segment on $q_k$ interacts 
	with only one segment on $p_k$. Thus, the total number of interactions between $q_k$ 
	and $p_k$ is $J_k({\bf a}_{diss}) = k^2 +k -5$.  
\end{proof}

To maximize the total number of interacting edge pairs between two complementary curves, with the requirement that ${\bf a} \neq \bfz$, 
a momentary thought shows that we can take ${\bf a} \in CP$ where all elements of ${\bf a}$ are equal. 

\begin{proposition}
	If ${\bf a} \in CP$, the case when all elements are equal to a positive integer constant (say) $c$, 
	then $J_k({\bf a}_{CP}) = \frac{k^3+6k^2-k}{6}$.
	\label{prop:constJn}
\end{proposition}
\begin{proof}
	Note that the number of unique sums in the curve $p_k$ (as well as $q_k$) is $k+1$. 
	Consider the curve $p_k$ and let's go from the first edge, all the way to the $k(k+1)/2$th edge one by one.
	For the first edge of $p_k$ (sum going from $0$ to $c$), the number of interactions is $1$. 
	For each edge of $p_k$ leading from sum $jc$ to $(j+1)c$, where $1 \le j \le k-1$, the number of edges of $q_k$ that interact with it can 
	be seen to be $k+1-j$ (It will be helpful to consider the Filling Boxes Model of~\cref{sec:FBM}).
	Meanwhile, there are $(j+1)$ edges of $p_k$ that have lower vertex sum to be $jc$, all of which interact with the $k+1-j$
	edges of $q_k$.
	Therefore, the total number of interactions is
	\begin{align*}
		J_k({\bf a}_{CP}) & = 1 + {\sum}_{j=1}^{k-1} (j+1)(k+1-j) \\
		& = 1 + \frac{k^3 + 6k^2 - k - 6}{6}, \\
		& =  \frac{k^3 + 6k^2 - k}{6}. 
	\end{align*}
\end{proof}

\begin{corollary}
	For a random nonzero ${\bf a}$, and $k \in [n]$, we have
	\begin{align*}
		k^2 + k -5 \le J_k({\bf a}) \le \frac{k^3 + 6k^2 - k}{6}.
	\end{align*}
\end{corollary}

\section{The Orbital Graph $G_0$}
\label{sec:G0}
In this section, we describe the details of a layered directed acyclic graph, referred to as the \emph{Orbital Graph}
that allows us to visit all the intersection points without leaving the OL.
\begin{definition}
	Let $P_k := \{p_1,p_2,\ldots,p_k\}$ be the set of the first $k$ curves of $p$-type, and $Q_k := \{q_1,q_2,\ldots,q_k\}$
	be the set of the first $k$ curves of $q$-type. The number of edges in $P_k$ is denoted by $|P_k|$. Similarly for $Q_k$.
\end{definition}
The starting point for the construction of the layered directed acyclic graph is the graph structure in~\cref{fig:pqgraph}
which can be written compactly as:
\begin{align}
	p_{n} \xra{\{\tau\}} Q_n \xra{\{\tau\}} P_{n-1} \xra{\{\tau\}} Q_{n-2} \xra{\{\tau\}} \ldots \xra{\{\tau\}} P_2 /\ Q_2 \xra{\{\tau\}} Q_1 /\ P_1,
	\label{eq:compactlinegraph}
\end{align}
where each arrow represents multiple $\tau$ transforms. 
This graph structure has $n+1$ layers, where each layer corresponds to a family of curves of either $q$-type
or $p$-type but not both, and they alternate. While the object of study in the graph will be edges (nodes),
we can still use the curves as a packaging of edges.
Thus, we view each curve $p_j$ (or $q_j$), not as a curve but a collection of $\frac{j(j+1)}{2}$ 
disjoint edges(nodes). Note that we will be using this new compact graph going forward.

Also, note that one could use the alternate graph,
\begin{align*}
	q_{n} \xra{\{\tau\}} P_n \xra{\{\tau\}} Q_{n-1} \xra{\{\tau\}} P_{n-2} \xra{\{\tau\}} \ldots \xra{\{\tau\}} Q_2 /\ P_2 \xra{\{\tau\}} P_1 /\ Q_1,
	\label{eq:compactlinegraphq}
\end{align*}
instead where we start with $q_{n}$. We can use either form, and we use~\cref{eq:compactlinegraph}.

The number of edges in each layer is given by the following proposition.
\begin{proposition}
	The sets $P_k$ and $Q_k$ have $\frac{k(k+1)(k+2)}{6}$ edges each.
	\label{prop:Qsize}
\end{proposition}
\begin{proof}
	We know that the curve $p_j$ has $\frac{j(j+1)}{2}$ edges. Summing over $j$ as it ranges in $[k]$, we have
	\begin{align*}
		|P_k| = {\sum}_{j=1}^k \frac{j(j+1)}{2} = \frac{k(k+1)(k+2)}{6}.
	\end{align*}
	Similarly for $|Q_k|$.
\end{proof}

Next we estimate the number of arcs between two adjacent levels, say $Q_k$ and $P_{k-1}$ (or equivalently $P_k$ and $Q_{k-1}$)  for some $4 \le k  \le n$ 
in the worst case.
\begin{proposition}
        Let $4 \le k \le n$. For any given ${\bf a}$, the maximum number of interacting edges between all the edges of $Q_k$ and all the edges of $P_{k-1}$, in the worst case
        is $\approx \frac{k^5}{120}$.
        \label{prop:PQarcs}
\end{proposition}

\begin{proof}
        To compute the total number of interacting edges between $Q_{k}$ and $P_{k-1}$, we note that for each $q_i \in Q_{k}$, we
        have the transformation structure given by
        \begin{equation*}
                q_i \xra{\{\tau\}_{k=4}^i} \{p_i,p_{i-1},p_{i-2},\ldots,p_4\}.
        \end{equation*}
        Since $Q_{k}$ is composed of curves $(q_{k},q_{k-1},\ldots,q_4)$, we have $(k-i)$ copies of transformations $q_i \ra p_i$.
        Therefore,
        \begin{align*}
                \text{\# interacting edges } & \le  {\sum}_{i=4}^{k-1} (k-i) J_i({\bf a}_{CP}) \quad \text{(Using~\cref{prop:constJn})} \\
                 & =  {\sum}_{i=4}^{k-1} (k-i) (\frac{(i^3+6i^2-i)}{6})  \\
                & = \frac{k^5+10k^4-5k^3-10k^2-2276k+6000}{120} \\
                & \approx \frac{k^5}{120} \text{ for large $k$}.
        \end{align*}
\end{proof}

\subsection{Determining the root node $e_0$}

To determine the root node, we search the edges of curve $p_n$ for the edge that contains the given target value $T$. Since $p_n$ is a NDP, and we have a formula
for the indices, we could use binary search to locate the edge(s) containing the intersection point with $y=T$. This can be done in logarithmic time.
We have two possible cases as shown by~\cref{prop:oneinf}:
\begin{itemize}
        \item[(1)] Only one intersection point; in this case we can identify a unique edge $e_j \in p_n$, that contains the point.
                Furthermore, we can obtain the local $y$-coordinate as $y_0 = T - y_j^-$, where $y_j^-$ is the $y$ component of the
                coordinates of the lower vertex. We can find $x_0$ as $x_0 = \frac{y_0}{|e_j|}\Re(d_j)$.
        \item[(2)] An infinite number of intersection points; In this case, by~\cref{prop:lock}
		we have only at most $n-1$ points (in local coordinates), one each in the intersecting edges of $p_n$. Let
                the edges that lie on the OL be $e_{j_1},e_{j_2},\ldots,e_{j_h}$, where $h \le n-1$. In this case, we have
                \begin{align*}
                        y_{j_1}^- = \ldots = y_{j_h}^- = T,
                \end{align*}
                so that $y_0 = 0$ in local coordinates for all these $h$ edges. We also have $x_0 =0$ for all these edges. Thus, in this
                case we have $h$ root nodes.
\end{itemize}

\begin{remark} 
	In order to capture both these cases uniformly, one could create a \emph{dummy} node $e_0$ and connect 
	it to the edges of $p_n$ that intersect the OL, by suitable arc weights. Thus, we can assume, without loss of generality, that there is 
	a single node $e_0 \in p_n$ that contains the intersection point whose local coordinates are $z_0 = x_0 + \iota y_0$.
\end{remark}

\subsection{The Graph $G_0$}
We let $G_0 := (V_0,E_0)$ be the layered directed acyclic graph corresponding to~\cref{eq:compactlinegraph}, where $V_0$ is
the set of nodes, and $E_0$ the set of arcs. We denote the number of nodes by $|V_0|$, and the
number of arcs by $|E_0|$. The layers of the graph correspond to levels $0,1,2,\ldots,n$.
The distinguished node $e_0 \in V_0$ is called the \emph{root node} which is in level zero.

\begin{definition}(Node in graph $G_0$)
	A node in the graph $G_0$, is an edge $e_j \in p_k (\text{or } q_k)$ for some $k \in [n]$, with lower vertex $z_j^-$ and  higher vertex $z_j^+$ 
	measured relative to the origin of subspace $S_k$ containing $p_k$ and $q_k$. 
\end{definition}

\begin{definition}(Arc in graph $G_0$)
	An arc in the graph $G_0$ connects a pair of edges $e_i$ and $e_j$ if and only if they satisfy 
	the following conditions:
	\begin{itemize}
		\item[(i)] $e_i \in p_k$ and $e_j \in q_k$, or vice versa, for some $k \in [n]$,
		\item[(ii)] they are on consecutive layers, i.e., $level(e_j) = level(e_i) \pm 1$, and
		\item[(iii)] and they interact, i.e., $|e_i \cap e_j| \ge 0$.
	\end{itemize}
	Furthermore, we associate the complex number $\overline{w}_{ij}$ to the arc, where
	$w_{ij} = z_j^- - z_i^-$, and call it the arc weight.
\end{definition}

\subsection{Computation of arcs between complementary curves}
If we look at~\cref{fig:pqgraph} we see arcs going from $p$ type curves to $q$ type curves and vice versa alternatingly.
Every arc in $G_0$ is an arc going from some $e_i \in p_k$ to some $e_j \in q_k$ or vice versa, where $k \in [n]$. Thus, if we compute
all the arcs corresponding to interacting edges in all the $n$ pairs of complementary curves, we can copy them into 
the various layers, and populate the graph.  Let edge $e_i \in p_k$
and $e_j \in q_k$, and if $|e_i \cap e_j| \ge 0$, then the arc weight from $e_i$ to $e_j$ is simply $w_{ij} = z_j^- - z_i^-$. 
Similarly, the arc weight from $e_j$ to $e_i$ is simply the negation, i.e., $w_{ji} = -w_{ij}$.
For the purpose of copying, it is useful to represent the arcs by an \emph{adjacency list}, defined below.
\begin{definition}(Adjacency List for an edge $e_i \in p_k$)
	Let $k \in [n]$ and $i \in [\frac{k(k+1)}{2}]$. To each edge $e_i \in p_k$, we associate a list of pairs, denoted by $\mc{A}(p,k,i)$ and defined as
	\begin{align*}
		\mc{A}(p,k,i) :=\{(r_1,w_{i{r_1}}),(r_2,w_{i{r_2}}),\ldots\},
	\end{align*}
	called the adjacency list, where the first element of each pair represents the edge in $q_k$ that interacts with $e_i$, and 
	the second element represents the complex arc weight. Similarly for an edge  $e_i \in q_k$, the adjacency list is
	denoted by $\mc{A}(q,k,i)$.
\end{definition}
By checking for interactions between all edges of a pair of complementary curves $p_k$ and $q_k$, and for all $k \in [n]$,
we can compute all the $|P_n| + |Q_n|$ adjacency lists.
The complexity of this operation is at most $O(n^4)$, since there are $n$ pairs of complementary curves, and for each pair we 
need to check at most $\frac{n^3+6n^2-n}{6}$ interactions by~\cref{prop:jmin}.

\subsection{Taking character sets into account: appending adjacency lists}
For an edge $e_i \in q_k$ (or $p_k$), we have seen that it can have at most $k$ local characters (including itself). For example, an edge $e_i \in q_k$,
can be congruent to edges $e_{i_1} \in q_{1}, \ldots, e_{i_{k-1}} \in q_{k-1}$. In our compact graph~\cref{eq:compactlinegraph}, since we collapsed
all the $\lambda$ transforms and have only $\tau$ transformations,  we need to append the adjacency lists of the congruent edges on lower indexed
curves to the adjacency list of $e_i$.
\begin{definition}(Collated adjacency lists)
	Let $e_i \in p_k$ for some $k \in [n]$ with adjacency list $\mc{A}(p,k,i)$. Let the character set of $e_i$ be $\Lambda(e_i) = [r]$ where
	$r \le k$. This means that the edges $e_{i_1} \in p_{1}, \ldots, e_{i_{r}} \in p_{r}$ are congruent to $e_i$ by $\lambda$ transforms.
	We append the adjacency lists of these edges to that of $e_i$ to get
	\begin{equation*}
		\mc{\bar{A}}(p,k,i) := \underset{\substack{e_{i_u} \in p_u \\ u \in [r], e_{i_u} \cong e_i}}{\bigcup} \mc{A}(p,u,i_u)
	\end{equation*}
	to get a collated adjacency list $\mc{\bar{A}}(p,k,i)$. 
\end{definition}
After computing the initial adjacency lists for all edges in $\{P_n,Q_n\}$, we obtain the collated adjacency lists by taking into account
the local characters of each edge. The complexity of this operation is simply $O(|P_n|+|Q_n|) = O(n^3)$.

\subsection{Computation of nodes}
The nodes can be computed easily since they are simply edges of the curves $\{P_n,Q_n\}$.
For each edge $e_i \in p_k$, where $k \in [n]$ and $i \in [k(k+1)/2]$, recall that 
\begin{align*}
	z_i^- = \phi_k(i) + \iota \siga(\phi_k(i)), \\
	\text{and } z_i^+ =\phi_k(i+1)  + \iota \siga(\phi_k(i+1)).
\end{align*}
Similarly for edges on $q_k$ curve which make use of $\varphi_k$ sequence. We copy these edges in layers as in~\cref{eq:compactlinegraph}
and assign an unique index to each node. Then for each node in $G_0$, we connect to nodes in the next layer
using the arcs given by the collated adjacency lists $\mc{\bar{A}}$. This completes the construction of graph $G_0$. 

Using the above description, we get a directed acyclic graph $G_0 = (V_0,E_0)$, where $V_0 = V(G_0)$ is the set of nodes,
and $E_0 = E(G_0)$ is the set of arcs.  We have designated root node $e_0 \in V_0$, and a start point $z_0 \in e_0$ given in local coordinates.

\subsection{Size of $G_0$}

The total number of nodes in the graph $G_0$, (including the root node $e_0 \in p_{n+1}$) can be calculated
(see~\cref{eq:compactlinegraph}) as
\begin{align*}
	|V_0| & = 1 + \sum_{j=1}^{n} |Q_j| \\
	& = 1 + \sum_{j=1}^n \frac{j(j+1)(j+2)}{6} \text{ (by~\cref{prop:Qsize}) } \\ 
	& = 1 + \frac{n(n+1)(n+2)(n+3)}{24} \approx \frac{n^4}{24}, \text{ for large enough $n$}.
\end{align*}

To compute the arcs consider one layer $Q_i \xra{\tau} P_{i-1}$ first. We know from~\cref{prop:PQarcs} that 
\begin{align*}
	\# arcs \text{ in $Q_i \xra{\tau} P_{i-1}$} & \approx \frac{i^5}{120}.
\end{align*}
Now summing over all the layers, and adding the first layer arcs between the root node in $p_{n}$ and $Q_n$ (the second term
in the expression below), we have
\begin{align*}
	|E_0| & \le {\sum}_{i=3}^n \frac{i^5}{120} \text{(by~\cref{prop:PQarcs})}\\
	& + \frac{n^3}{6} \quad \text{(by connecting root node to all nodes of first layer)}\\
	& \approx \frac{n^6}{720}, \text{ for $n$ large enough}.
\end{align*}
For a random instance,  if we use the estimate of~\cref{prop:jmin} (which is more realistic), 
then the total number of arcs will be $\approx n^5/60$, which is a factor of $n/12$ less than the above estimate.

For later reference, we have the definition of a \emph{leaf node} given below.
\begin{definition}(Leaf node)
	A node $e_k \in G_0$, is said to be a leaf node if and only if there are no outgoing arcs from $e_k$ into
	the next level of the graph. Since each level of the graph corresponds to the family of curves $Q_j$ or $P_j$
	for some $j \in [n]$, and recalling that nodes in $p_4$ and $q_4$ curves cannot have any children since 
	$4$ is the smallest local character allowed, it follows that leaf nodes can exist in all levels of the graph $G_0$.
\end{definition}

\subsection{Paths in $G_0$}
Given an instance $({\bf a},T)$, the previous section showed in detail the construction of a directed acyclic graph
$G_0 = (V_0,E_0)$, with a start node $e_0$, and a start value $z_0 = x_0 + \iota y_0 \in e_0$.

In this section we discuss paths in $G_0$, their geometrical meaning and their classification.

\begin{definition} (Path in $G_0$)
	A path in $G_0$ is an ordered sequence of distinct nodes $(e_{k_1},e_{k_2},\ldots,e_{k_l})$, such that every pair of adjacent
	nodes $e_{k_i}$ and $e_{k_{i+1}}$ are on adjacent levels, and connected by a directed arc with weight $w_{{k_i}k_{i+1}}$, 
	for all $1 \le i < l$. We denote the path by $\pi(e_{k_1},e_{k_l})$, identifying it by the first and last nodes.  
	Sometimes we simply refer to a path by $\pi$ when the start and end nodes are clear.
	The path length of the path $\pi(e_{k_1},e_{k_l})$, denoted by $|\pi(e_{k_1},e_{k_l})|$ is given by
	\begin{align*}
		|\pi(e_{k_1},e_{k_l})| = \sum_{i=1}^{l-1} \overline{w}_{{k_i}k_{i+1}}.
	\end{align*}
\end{definition}

\begin{definition}(Image point)
	Let $\pi(e_{k_1},e_{k_l})$ be a path from $e_{k_1}$ to $e_{k_l}$. 
	Given a point $z_1 \in e_{k_1}$, the image of the point in any edge $e_{k_j}$, where $j \in [l]$, along the path is obtained by
	traversing the path up to the node, and is given by
	\begin{align*}
		z_j = z_1 + |\pi(e_{k_1},e_{k_j})| = z_1 + \sum_{i=1}^{j-1} \overline{w}_{{k_i}k_{i+1}}.
	\end{align*}
	Note that if $z_1 = 0$, then the image point is simply the path length.
\end{definition}

\begin{definition}(Valid path through $e_i$)
	Let $\pi(e_0,e_i)$ be a path in $G_0$, with start point $z_0 \in e_0$. The image point in $e_i$ is $z_i = z_0 + |\pi(e_0,e_i)|$. 
	We say $\pi$ is a valid path into $e_i$, if and only if $y_i \in [0, |e_i|)$.
	Otherwise, it is called an invalid path.
\end{definition}

\begin{definition}(Valid path)
	The path $\pi$ is said to be a valid path if and only if it is valid at each of the nodes along the path.
\end{definition}

The geometrical meaning of a path $\pi$ in $G_0$ is fairly intuitive and as follows:
Consider a path into a node $e_i$ from $z_0 \in e_0$. If the path length $y_i < 0$, then geometrically this means that 
$e_i$ does not not intersect the OL, and the lower vertex of $e_i$ is \emph{above} the OL. Similarly, if $y_i > |e_i|$, 
then $e_i$ as a whole, is \emph{below} the OL. If $y_i \in [0,|e_i|)$, then the OL intersects this instance of edge $e_i$.

\begin{definition}(Zero Path)
	Let $\pi(e_0,e_k)$ be a path from $e_0$ to $e_k$, and let $e_i$ be an intermediate node along the path.
	For a given point $z_0 \in e_0$, we say that the path from $e_0$ to $e_i$, i.e., $\pi(e_0,e_i)$ is a 
	zero path if and only if it is a valid path, and  $y_i = 0$ or $y_j = 0$ for some $j \in [i+1,k]$. 
	In the former case, the OL passes through the lower vertex of $e_i$. In the latter case, although the path 
	may not pass through the lower vertex of $e_i$, it is still called a zero path, since it leads to a zero path length 
	further along the path.
\end{definition}

\begin{definition}(Distinct Paths through $e_i$)
	Let $e_i \in V_0\setminus{\{e_0\}}$ be an arbitrary node. Let $\pi$ and $\pi'$ be two paths through $e_i$
	coming from the start point $z_0 \in e_0$. Let the image points in $e_i$ corresponding to the two paths be  $y_i$ and $y_i'$ respectively.
	We say that the two paths $\pi$ and $\pi'$ are distinct if and only if $y_i \neq y_i'$. 
\end{definition}

\begin{definition}(Distinct Valid Paths and Distinct Zero Paths)
	Let $e_i \in V_0\setminus{\{e_0\}}$ be an arbitrary node.  If $\pi$ and $\pi'$ are two distinct paths through $e_i$, 
	and furthermore they both are valid, then we call them \emph{distinct valid paths}.  Similarly, if $\pi$ and $\pi'$ 
	are two distinct paths through $e_i$, and furthermore they both are zero paths, then we call them \emph{distinct zero paths}.
\end{definition}

\subsection{Classification of paths through a node}
\label{subsec:pathclass}
Consider $G_0$ with $z_0 \in e_0$ as the starting point, and let $e_i$ be an arbitrary node at level $l > 0$ from $e_0$. 
If we compute all paths in $G_0$, in general, we will have many paths through $e_i$.  
We have the following quantities of interest: 
\begin{itemize}
        \item[(i)] $DZP_i := $  the set of all distinct zero paths through $e_i$. There could be many paths through
                $e_i$ that go to a destination node (including $e_i$ itself) and meet it at the vertex point. Out of
                these we only consider those which have a unique intersection in $e_i$. In other words, if more than
                one zero path goes through $e_i$ at the same point in $e_i$, we will consider it as one. The cardinality
		of this set is denoted by $|DZP_i|$.
        \item[(ii)] $ZP_i := $  the set of all zero paths through $e_i$. Same as $DZP_i$ but with multiplicities counted.
		It is clear that
		\begin{align}
			|DZP_i| \le |ZP_i|.
		\end{align}
        \item[(iii)] $DVP_i := $  the set of all distinct valid paths through $e_i$, i.e., paths in $[0,|e_i|)$,
                and are distinct. These include both zero and non-zero paths, and thus
		\begin{align}
			|DZP_i| \le |DVP_i|.
		\end{align}
		If $|e_i| > 0$ then it is clear that $|DVP_i| \le |e_i|$. In the special case of $|e_i| = 0$, we have $|DVP_i| \le 1$ since
		there could be a path through the lower vertex.
\end{itemize}

With the above definitions of paths in $G_0$, the \emph{decision version} of \ssp\ee reduces to deciding the existence of zero
paths in $G_0$. Formally, we have the following theorem.

\begin{theorem}
	Let $({\bf a},T)$ be a given instance of SSP and let the OL intersect some node $e_0 \in p_n$ with local coordinates $(x_0,y_0)$.
	Then the given instance has a solution if and only if there is at least one zero path $\pi$ from $e_0$ to some $e_k \in V_0$.
\end{theorem}

\section{Hopping On The Orbital Line: An Algorithm}
\label{sec:algihm}
In this section we present an iterative algorithm to determine the number of \emph{zero paths}, or alternatively the number of \emph{vertex points}
on the OL. The algorithm consists of repeatedly applying the alternating steps of \emph{refining} and \emph{filtering} the graph, 
which are described below. At the end, we have a refined graph $G_m$, which if not empty, has the property that \emph{every} path is a \emph{zero path}.
In the case where there is no solution to the given instance of the \ssp, the final graph will be empty, i.e.,  $G_m = \emptyset$.

First, let's consider briefly on how one might find zero paths. A naive approach would be to compute all paths from $z_0 \in e_0$ to all
other nodes. 
However, this would be wasteful since zero paths are expected to be a small subset, if any at all, of the set of all valid paths. Furthermore, it is
not clear what the complexity of finding all valid paths would be (we will later reveal this complexity too in next section).
Instead, we do this:
in each iteration of the algorithm, we make use of a routine called \emph{Single Source Shortest Paths} abbreviated as SSSP, to find the
shortest paths to all nodes in the graph. While it is not the same as finding zero paths, when combined with the refining and filtering
steps, it suffices. So we first review this well-known graph algorithm~\cite{dpv06,mm15} which has a linear running time in the size of the graph. 

\subsection{Single Source Shortest and Longest Paths}
It is a well-known fact that every directed acyclic graph admits a \emph{topological ordering} of its nodes. Given a
directed acyclic graph $G = (V,E)$, the topological ordering of the graph is simply a linear ordering of the nodes
on a horizontal line so that all the arcs go from left to right. Furthermore, the topological ordering can be accomplished 
in linear time in the size of the graph, i.e., in time $O(|V| + |E|)$, using a \emph{depth first search}~\cite{dpv06}.  

The SSSP algorithm is slightly modified so that we also have the \emph{longest} path to each node.
The algorithm for computing the \emph{shortest} and the \emph{longest} paths from the root node to all other nodes, 
in a directed acyclic graph, is given below (~\cref{alg:SSSP}). Here, we assume that we have already carried out the topological ordering
and that the ordered nodes are stored in an array labeled $TS$. 

\begin{algorithm}
	\caption{Pseudo-code to compute the shortest and longest paths to all nodes from root, given the topologically sorted order of nodes ($TS$),
	and initial value $z_0 \in e_0$ in the root node.}\label{alg:sssp}
	\label{alg:SSSP}
\begin{algorithmic}[1]
	\Procedure{SSSP}{G = (V,E), TS, $e_0$, $z_0$, ${\bf \alpha^{\ast}, \alpha^{\square}}$}
	\State ${\mathbb \alpha_i}^{\ast}\gets \infty$ for all $i \in [|V|]$ \Comment{initialize all shortest paths to $\infty$}
	\State ${\mathbb \alpha_i}^{\square} \gets -\infty$ for all $i \in [|V|]$ \Comment{initialize all longest paths to $-\infty$}
	\State ${\mathbb \alpha_0}^{\ast}\gets y_0$ \Comment{set root shortest path to $y_0$}
	\State ${\mathbb \alpha_0}^{\square}\gets y_0$ \Comment{set root longest path to $y_0$}
\For{$j=2$ to $|V|$}
\State $u\gets TS[j]$ \Comment{for each node in the topological order}
\For{$k \in P(u)$}\Comment{For each parent of u}
	\If{$\alpha_k^{\ast} + \overline{w}_{ku} < \alpha_u^{\ast} $}
	\State $\alpha_u^{\ast} \gets \alpha_k^{\ast} + \overline{w}_{ku}$
\EndIf
	\If{$\alpha_k^{\square} + \overline{w}_{ku} > \alpha_u^{\square} $}
	\State $\alpha_u^{\square} \gets \alpha_k^{\square} + \overline{w}_{ku}$
\EndIf
\EndFor
\EndFor
	\State \textbf{return} ${\bf \alpha^{\ast}, \alpha^{\square}}$\Comment{Return the shortest and longest paths}
\EndProcedure
\end{algorithmic}
\end{algorithm}

Given that the arc weights (in the $y$ component) in the graph can be negative, in general, this algorithm may output negative shortest paths,
and large positive paths (greater than the edge length) to many nodes.  
\begin{notation}
For any node $e_i \in V_0$, we denote the shortest
path from $e_0$ by $\shp{i}$ and the longest path by $\lop{i}$, where $\shp{i},\lop{i} \in {\mathbb Z}$.
\end{notation}

\subsection{Refining Nodes and Arcs}
In this subsection, we define the notion of \emph{refining} of nodes and arcs, and the generation of a refined graph.
\begin{definition}(Refine operation on a node)
	Let $e_i$ be a node with the associated interval $[z_i^{-}, z_i^{+})$, and with length $|e_i| = y_i^+ - y_i^- > 1$ where $i \in [|V_0|]$.
	Let $z_i^{h} = \myfloor{\frac{(z_i^- + z_i^+)}{2}}$ be the midpoint of the interval.
	A refine operation on $e_i$, replaces it with two smaller disjoint nodes $e_i'$ and $e_i''$ whose union is $e_i$.
	In particular, $e_i'$ corresponds to the interval $[z_i^-,z_i^h)$, and $e_i''$ corresponds to $[z_i^h,z_i^+)$.
\end{definition}
\begin{remark}
	For a node $e_i$, if the length $|e_i| \le 1$, we do not split it into two nodes. We simply
	leave it as it is. 
	The floor operation ensures that the end points of nodes 
	(in $y$ coordinate) before and after refining remain as integers.
\end{remark}

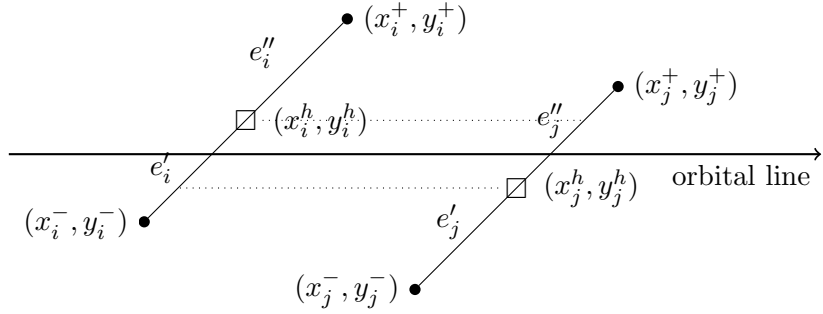
\begin{figure}[!htbp]
\begin{center}
	\begin{tikzpicture}[scale=0.9]
\draw[thick,->] (-1,0) -- (11,0) node[anchor=north east]{ orbital line};
\node[outer sep=0pt,circle, fill,inner sep=1.5pt,label={[fill=white]left:$(x_i^-,y_i^-)$}] (E1a) at (1,-1) {};
\node[outer sep=0pt,circle, fill,inner sep=1.5pt, label={[fill=white]right:$(x_i^+,y_i^+)$}] (E1b) at (4,2) {};
\node (N1) at (2.75,1.5) {$e_{i}''$};
\node (N1) at (1.25,-0.2) {$e_{i}'$};
\draw[] (E1a) -- (E1b);

\node[outer sep=0pt,circle, inner sep=0.5pt, label={right:$(x_{i}^h,y_{i}^h)$}] (P1) at (2.5,0.5) {$\square$};
\node[outer sep=0pt,circle, inner sep=0.5pt, label={right:$$}] (P1a) at (1.5,-0.5) {$$};
\node[outer sep=0pt,circle, inner sep=0.5pt, label={right:$$}] (P2a) at (7.5,0.5) {$$};
\node[outer sep=0pt,circle, inner sep=0.5pt, label={right:$(x_{j}^h,y_{j}^h)$}] (P2) at (6.5,-0.5) {$\square$};
\draw[dotted] (P1a) -- (P2);
\draw[dotted] (P1) -- (P2a);

\node[outer sep=0pt,circle, fill,inner sep=1.5pt,label={[fill=white]left:$(x_j^-,y_j^-)$}] (E2a) at (5,-2) {};
\node[outer sep=0pt,circle, fill,inner sep=1.5pt, label={[fill=white]right:$(x_j^+,y_j^+)$}] (E2b) at (8,1) {};
\node (N2) at (7,0.5) {$e_{j}''$};
\node (N2) at (5.5,-1.0) {$e_{j}'$};
\draw[] (E2a) -- (E2b);
\end{tikzpicture}
\end{center}
\caption{Illustration of refining edges $e_i$ and $e_j$ connected by an arc. Only three pairs of new edges interact; 
	edges $e_{j}'$ and $e_{i}''$ do not interact since their $y$ intervals do not overlap.}
\label{fig:edgesplit}
\end{figure}

Consider a pair of nodes $e_i$ and $e_j$ connected by an arc of weight $w_{ij}$. Recall that the arc weight is
determined by the lower vertices of the two nodes, i.e., $w_{ij} = z_j^- - z_i^-$. 
When we \emph{simultaneously} refine the nodes $e_i$ and $e_j$, we have four new nodes, and have potentially four possible arcs between
them.  See~\cref{fig:edgesplit} below for an illustration.

The refine operation on $e_i$ leads to new edges $e_{i}'$ and $e_{i}''$. Similarly, $e_{j}'$ and $e_{j}''$ are generated 
from $e_j$. The end points associated with these new edges are as follows:
\begin{align*}
	e_{i}' \Leftrightarrow [z_i^-, \myfloor{\frac{z_i^-+z_i^+}{2}}), \\
	e_{i}'' \Leftrightarrow [\myfloor{\frac{z_i^-+z_i^+}{2}}, z_1^+), 
\end{align*}
\begin{align*}
	e_{j}' \Leftrightarrow [z_j^-, \myfloor{\frac{z_j^-+z_j^+}{2}}), \\
	\text{and } e_{j}'' \Leftrightarrow [\myfloor{\frac{z_j^-+z_j^+}{2}}, z_2^+).
\end{align*}
We have four possible arcs between the new nodes, with arc weights as follows: 
\begin{align*} 
	(e_{i}',e_{j}') & \Leftrightarrow w_{ij}, \\ 
	(e_{i}',e_{j}'') & \Leftrightarrow w_{ij} + \myfloor{\frac{z_j^+-z_j^-}{2}}, \\ 
	(e_{i}'',e_{j}') & \Leftrightarrow w_{ij} - \myfloor{\frac{z_i^+-z_i^-}{2}}, \\ 
	\text{and } (e_{i}'',e_{j}'') & \Leftrightarrow w_{ij} + \myfloor{\frac{z_j^+-z_j^-}{2}} - \myfloor{\frac{z_i^+-z_i^-}{2}}.  
\end{align*} 
However, the proposition below shows that we will have at most three interacting pairs of edges and never four.

\begin{proposition} 
	Let $e_i$ and $e_j$ be two nodes with length greater than unity each, and connected by an arc with weight $w_{ij}$. A simultaneous 
	refine operation on $e_i$ and $e_j$ will generate
	at most three pairs of interacting edges between $\{e_{i}',e_{i}''\}$ and $\{e_{j}', e_{j}''\}$.
\end{proposition}
We omit the proof as it is a simple exercise to prove it by considering various cases.

\begin{corollary}
	Let $G_0 = (V_0,E_0)$ be the initial graph. Then after simultaneous refinement of all nodes, and connecting interacting
	pairs of nodes by arcs, the refined graph $G_1 = (V_1,E_1)$ satisfies:
	\begin{itemize}
		\item[(i)] the node with maximum length, say $e_i$, has $|e_i| \le 2^{m-1}$,
		\item[(ii)] $|V_1| \le 2|V_0|$,
		\item[(iii)] and $|E_1| \le 3|E_0|$.
	\end{itemize}
\end{corollary}

The complexity of the \emph{refine} algorithm is clearly $O(|V| + |E|)$ since it involves one loop over nodes and one loop over arcs. 
The check for interactions can be done in constant time.  
Since the source 
node $e_0$ is also split into two halves, the original point $y_0 \in e_0$ can be in either $e_{0}'$ or $e_{0}''$ but not both. In
case it is in $e_{0}'$, the (local) value of $y_0$ does not change. Otherwise, we update the local value in the
reference frame of $e_{0}''$ as $y_{0} := y_{0} - \frac{|e_0|}{2}$. The half containing the start point is kept and relabeled $e_0$, and
the other half is removed. We can find all the nodes in the refined graph that are \emph{reachable} from $e_0$ by doing a simple
\emph{Depth First Search (DFS)}~\cite{dpv06}.
This step will get rid of  some unreachable nodes and arcs of the refined graph.

The pseudo-code for the \emph{refine} operation is shown below (~\cref{alg:refine}). The complexity is clearly linear in the
size of the input graph, namely $O(|V| + |E|)$.
\begin{algorithm}
	\caption{(Pseudo-code for the Refine operation)}
	\label{alg:refine}
\begin{algorithmic}[1]
\Procedure{Refine}{$G = (V,E)$, $e_0$, $z_0$, $G'$}
	\State ${G' \gets \emptyset}$ \Comment{Initial graph $G'$ to $\emptyset$}
\For{$i=1$ to $|V|$}
	\State $e_i \ra \{e_i',e_i''\}$ \Comment{Apply the refine operation}
	\State Add $\{e_i',e_i''\}$ to $V'$
\EndFor
\For{$(i,j) \in E$}
	\For{$r \in \{', ''\}$}
	\For{$s \in \{',''\}$}
	\If{$e_{i}^r \cap e_{j}^s \neq \emptyset$} \Comment{Check if edges interact}
\State Add arc $(e_i^r,e_j^s)$ to $E'$
\EndIf
\EndFor
\EndFor
\EndFor
\If{$z_0 \in e_{0}'$} 
\State $e_0 \gets e_{0}'$
\Else
	\State $e_0 \gets e_{0}''$ \Comment{update new root node}
	\State $z_0 \gets z_0 - \myfloor{\frac{z_0^-+z_0^+}{2}}$ \Comment{Update value of $z_0$ in new reference frame}
\EndIf
\State $G' = REACH(G',e_0)$ \Comment{Get Reachable DAG from $e_0$ by DFS}
\State \textbf{return} ${G'}$\Comment{Return the refined graph $G'$}
\EndProcedure
\end{algorithmic}
\end{algorithm}

\subsection{A destination node $e_{\infty}$}
In the graph $G_0$ we have a distinguished \emph{root node} labeled $e_0$.
In order to facilitate the \emph{filtering} operation it is useful to have a destination node $e_{\infty}$ defined below.

\begin{definition}(TRUE and FALSE nodes)
	Let $e_i \in G_0$ be a node, with end points $z_i^-$ and $z_i^+$. The node $e_i$ is said to be a
	\emph{TRUE node} if and only if the lower vertex belongs to the point set, i.e., $z_i^- \in S_n$. Otherwise
	it is called a \emph{FALSE node}.
	In the initial graph $G_0$, all nodes are TRUE nodes. However, after applying a refine operation,
	only the lower-half nodes will be TRUE nodes. The upper-half nodes will be FALSE since their lower
	vertices are midpoints of the segments which are not part of the point set.
	\label{def:TF}
\end{definition}

\begin{definition}(destination node)
	We include a dummy node labeled as $e_{\infty}$, and attach it to all TRUE nodes of $V_0 \setminus \{e_0\}$
	by arcs with arc weight equal to zero. As a result of this operation, the number of arcs in the
	modified graph is equal to $|E_0| + |V_0| - 1$.  Note that while the rest of the nodes in the graph 
	have a geometrical interpretation, the dummy node $e_{\infty}$ does not have any. 
\end{definition}

The following proposition is straightforward, given the above definitions.
\begin{proposition}
	Given $G_0$, and a $z_0 \in e_0$, there is a zero path to some node $e_k \in V_0$,  if and only if 
	there is a zero path to $e_{\infty}$.
\end{proposition}
\begin{corollary}
	Given $G_0$, and a $z_0 \in e_0$, if there are $N$ zero paths to various nodes, then there will be
	$N$ zero paths to $e_{\infty}$.
\end{corollary}
In a sense, the node $e_{\infty}$ acts like a collector of all zero paths.
\begin{definition}(Reverse graph ${\hat G_0}$)
	Let ${\hat G_0}$ be the graph obtained by reversing all arcs, with $e_{\infty}$ as the root node,
	and $e_0$ as the destination node. Note that each arc $w_{ij}$ of the original graph,
	becomes $-w_{ij}$ in the reverse graph. We can start with $z_{\infty} = 0$ as the start point in $e_{\infty}$,
	and propagate paths back to $e_0$. Let $\bshp{k}$ and $\blop{k}$ denote the shortest and longest paths to a
	node $e_k$ in the reverse graph.
\end{definition}

\subsection{Filtering of nodes and arcs}
The main idea in \emph{filtering} is to remove nodes and arcs of the graph, that \emph{cannot support} any zero paths.
In other words, if no zero paths can go through a node (or an arc), they can be safely removed.
We have already seen above that, if all nodes of $G_0$ are refined, then for every pair of nodes connected by an arc
in the original graph, we will have only at most three arcs instead of four in the refined graph. 
In a sense, the refine operation also filters some arcs. If we view the nodes connected by arcs as \emph{channels},
the width is reduced, and they become more \emph{aligned}.
Here, we describe some tests which will enable us to remove more nodes and arcs. 

Given $z_0 \in e_0$, first we compute the \emph{shortest} and \emph{longest} paths to all other nodes in the graph
using the SSSP algorithm.
Note that we are only concerned with the $y$ coordinate in the computation of paths. The $x$ coordinate gets a free
ride!

\begin{proposition}
	Let $\alpha_k$ be an arbitrary path into a node $e_k \in G_0$. Then, we have 
	\begin{align*} 
		\shp{k} \leq \alpha_k \leq \lop{k}.  
	\end{align*}
\end{proposition}
This simply follows from the fact that $\shp{k}$ and $\lop{k}$ are the extreme shortest and longest paths respectively into $e_k$.

~\cref{def:TF} of TRUE/FALSE labels of nodes has an important use. A TRUE node has a vertex point, and thus it has the
\emph{potential} to receive a zero path, whereas a FALSE node has none. Let $e_k$ be a TRUE node. If it turns out that the
shortest path $\alpha_k^{\ast} > 0$ then it can be seen that node $e_k$ can never have a zero path into it. We can then relabel
the node $e_k$ as FALSE. 

\begin{proposition} 
	Let $G = (V,E)$ be the OL graph at some stage of iteration, and let $e_k \in V \smallsetminus \{e_0\}$ be 
	a node in $G$.  A necessary condition that at least one instance of $e_k$ intersects the orbital line is 
	\begin{align*} 
		[0, |e_k|) \cap [\shp{k}, \lop{k}] \neq \emptyset 
	\end{align*}
	\label{prop:nodenec}
\end{proposition}

\begin{proof} 
	Recall that each path into a node $e_k$ corresponds to a sequence of line segments in the geometry. This sequence 
	of segments has $e_0$ as the first segment and $e_k$ as the final segment.  In order for this instance of $e_k$ to 
	intersect the \emph{OL}, the path length must be in $[0, |e_k|)$. Since every path into $e_k$ lies in the interval formed by
	the shortest and longest paths into $e_k$, the condition is clear.
	Suppose all instances of $e_k$ are above the OL. Then the shortest and longest paths are both negative and the above condition 
	leads to an empty interval. Similarly, if all instances of $e_k$ are below the OL, then the shortest and longest paths will 
	be greater than the edge length.  Again, the above condition leads to an empty set.
\end{proof}

Now we are ready to describe \emph{Filtering} of the graph. 

\begin{proposition}(Node Removal) 
	Let $G = (V,E)$ be the OL graph at some stage of iteration, and let $e_k \in V \smallsetminus \{e_0,e_{\infty}\}$ be 
	a node in $G$. Then $e_k$ can be removed if either of the following conditions
	is satisfied.
	\begin{itemize} 
		\item[(i)] The node $e_k$ is not reachable from $e_0$ 
		\item[(ii)]  The forward and reverse shortest and longest paths into $e_k$ satisfy  
			\begin{equation*} 
				[0 , |e_k|) \cap [\shp{k} , \lop{k} ] \cap [\bshp{k} , \blop{k}] =  \emptyset 
			\end{equation*} 
	\end{itemize}
	\label{prop:noderemoval}
\end{proposition}
\begin{proof} 
	Condition (i) is clear: in order for $e_k$ to be in the graph, it must be reachable from $e_0$ via edge interactions. 
	Coming to the first two terms in LHS of condition (ii), we need at least one instance of $e_k$ to intersect the OL, 
	as a necessary condition, by~\cref{prop:nodenec}.  The third term on LHS of (ii) is similar to the middle term,
	but in the reverse direction.  If there is a zero path passing through node $e_k$ reaching $e_{\infty}$, then we need a 
	collision of forward and reverse paths at $e_k$.  A necessary condition for this
	is that these forward and reverse intervals have a non-empty intersection, and the claim follows.
\end{proof}

\begin{figure}
\begin {center}
\begin {tikzpicture}[-latex ,auto ,node distance =4 cm and 5cm ,on grid ,
semithick ,
state/.style ={ circle ,top color =white , bottom color = white!20 ,
draw,black , text=blue , minimum width =1 cm}]
\node[state] (A) {$e_0$};
\node[state] (B) [right =of A] {$e_k$};
\node[state] (C) [right =of B] {$e_{\infty}$};
\path (A) edge [bend left =25] node[above] {forward path} (B);
\path (A) edge [bend left =-25] node[below] {forward path} (B);
\path (C) edge [bend left =25] node[below] {reverse path} (B);
\path (C) edge [bend left =-25] node[above] {reverse path} (B);
\end{tikzpicture}
\end{center}
\caption{Collision of forward and reverse paths at $e_k$} \label{fig:coll}
\end{figure}
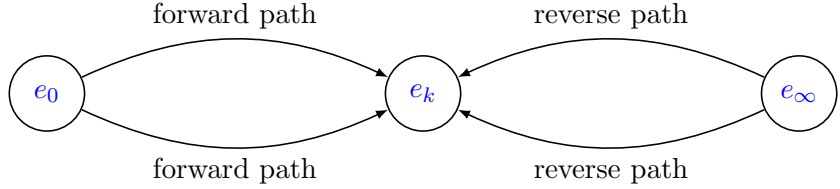

\begin{proposition}(Arc Removal) 
	Let $G = (V,E)$ be the OL graph at some stage of iteration, and let $w_{ij}  \in E$ be an arc.
	Then the  arc $w_{ij}$ can be removed if either of the following conditions is satisfied.  
	\begin{itemize} 
		\item[(i)] the intervals of shortest and longest paths, in the forward direction,  of $e_i$ and $e_j$ do not interact 
			\begin{equation*} 
				[\shp{i} + \overline{w}_{ij} , \lop{i} + \overline{w}_{ij}] \cap [\shp{j} , \lop{j}] =  \emptyset 
			\end{equation*} 
		\item[(ii)] the intervals of shortest and longest paths, in the reverse direction,  of $e_i$ and $e_j$ do not interact  
			\begin{equation*} 
				[\bshp{j} - \overline{w}_{ij} , \blop{j} - \overline{w}_{ij}] \cap [\bshp{i} , \blop{i}] =  \emptyset 
			\end{equation*} 
	\end{itemize}
\end{proposition}

\begin{proof} 
	The basic idea is this: although nodes $e_i$ and $e_j$ interact (with their edge intervals), they may not interact when we use the 
	intervals formed by shortest and longest paths. In such a case, it is straightforward to see that the shortest path at node $e_i$ cannot 
	lead to shortest path at $e_j$. Hence, we can safely remove this arc.
\end{proof}

\begin{algorithm}
	\caption{(Pseudo-code for Filter operation)}
	\label{alg:filter}
\begin{algorithmic}[1]
\Procedure{Filter}{$G = (V,E)$, $e_0$, $z_0$, $G'$}
	\State ${G' \gets G}$ \Comment{Copy the initial graph}
	\For{$i=1$ to $n$} \Comment{run this loop at most $n$ times}
	\State Attach destination node $e_{\infty}$
	\State $G' = REACH(G',e_0,e_{\infty})$ \Comment{reachable graph between $e_0$ and $e_{\infty}$}
	\State $z_{\infty} \gets 0$
	\State Compute $\shp{},\lop{},\bshp{}$ and $\blop{}$ \Comment{Compute extreme paths}
\For{$j=1$ to $|V|$}
	\If{$[0,|e_j|) \cap [\shp{j}, \lop{j}] \cap [\bshp{j}, \blop{j}] = \emptyset$} 
	\State Mark $e_j$ for deletion
	\EndIf
	\If{ $\shp{j} > 0$ \& $e_j$ is a TRUE node} 
	\State Set $e_j$ as FALSE
	\EndIf
\EndFor
\For{$k=1$ to $|E|$}
	\If{$[\shp{i}+\overline{w}_{ij},\lop{i}+\overline{w}_{ij}] \cap [\shp{j}, \lop{j}]  = \emptyset$} 
	\State Mark $w_{ij}$ for deletion
	\EndIf
	\If{$[\bshp{i}-\overline{w}_{ij},\blop{i}-\overline{w}_{ij}] \cap [\bshp{j}, \blop{j}]  = \emptyset$} 
	\State Mark $w_{ij}$ for deletion
	\EndIf
\EndFor
	\If{(no marked nodes and arcs)}
	\State delete dummy node and break
	\EndIf
\State Delete marked nodes and arcs, including dummy node
\EndFor
\State \textbf{return} ${G'}$\Comment{Return the filtered graph $G'$}
\EndProcedure
\end{algorithmic}
\end{algorithm}
The pseudocode for the filtering operation is shown in~\cref{alg:filter}.
In the inner loop of filter operation, there are two calls to SSSP to compute the shortest and longest paths in both directions. To figure out the nodes
and arcs for deletion, we loop over the nodes once, and over the arcs once. Again the overall complexity is linear
in the size of the input graph. The loop is executed at most $n$ times. The reason for this is: when some nodes and arcs are removed, the shortest
and longest path lengths to the next level change. This is due to the greedy nature of the SSSP routine. After the $k$th
execution of the loop, it can be shown that the shortest and longest paths at levels $1$ through $k$ will not change, so that after at most $n$ iterations, we 
will not get any nodes and arcs that can be deleted. 

\begin{remark}
	At the end of each loop of the FILTER step, each surviving node $e_i$ satisfies one of the following properties.
	\begin{itemize}
		\item[(i)] The node $e_i$ has a zero path into it that has been discovered ($\alpha_i^{\ast} = 0$).
		\item[(ii)] The node $e_i$ has a potential for a zero path into it ($\alpha_i^{\ast} < 0$ and $\alpha_i^{\square} > 0$) but it is not clear
			now. This uncertainty will be removed after further filter or refine steps.
		\item[(iii)] The node $e_i$ does not have a zero path into it ($\alpha_i^{\ast} > 0$), but it can potentially support a zero path
			at a downstream node. In this case $e_i$ is labeled as FALSE. If furthermore, $e_i$ is a LEAF node, it will be disconnected
			from the destination node $e_{\infty}$ in the next loop and removed (see step $5$ of~\cref{alg:filter}). Otherwise, it 
			will remain for now.
		\item[(iv)] When a node $e_i$ that is both a LEAF node and a FALSE node is removed, one or more of its immediate parent nodes can become 
			a LEAF node. If such a node is also FALSE, it will be removed in the next loop by above step (iii). Thus, there can be a 
			cascaded removal of nodes (during the $n$ loops of the FILTER operation) that do not support zero paths.
	\end{itemize}
	As the iterations of the algorithm progress, the number of TRUE nodes is non-increasing. Meanwhile the FALSE nodes which do not support zero
	paths get removed.
	\label{rem:filter}
\end{remark}

\subsection{The algorithm at a glance}
Now we are ready to describe the algorithm. With the above detailed description of the \emph{refine} and \emph{filter} operations,
the algorithm (referred to as IHM) is quite simple as shown in~\cref{alg:ihm}. The algorithm consists of $m$ iterations, where in each
iteration we perform the \emph{refine} and \emph{filter} operations.  

\begin{algorithm}
\caption{High level description of the algorithm.}\label{alg:ihm}
\begin{algorithmic}[1]
\Procedure{IHM}{$G_0 = (V_0,E_0)$, $e_0$, $z_0$, $m$, $G_m$}
	\State ${G\gets G_0}$ \Comment{Copy the initial graph}
	\State $G \gets FILTER(G,e_0,z_0)$  \Comment{Apply the filter operation}
\For{$j=1$ to $m$}
	\State $G' \gets REFINE(G,e_0,z_0)$ \Comment{Apply the refine operation}
	\State $G_j \gets FILTER(G',e_0,z_0)$  \Comment{Apply the filter operation}
	\State $G \gets G_j$
	\If{$G_j = \emptyset $} 
	\State return $\emptyset$ 
	\EndIf
\EndFor
\State \textbf{return} ${G_m}$\Comment{Return the final graph $G_m$}
\EndProcedure
\end{algorithmic}
\end{algorithm}

We have the progression of graphs 
\begin{align*}
	G_0, G_1, G_2, \ldots, G_m,
\end{align*}
as we go through the iterations. The refine operation doubles the nodes in each step, whereas the filter operation 
removes some nodes in each step. If the filter step weren't there we would trivially have
\begin{align*}
	|V_m| \le 2^m |V_0|, \\
	\text{and } |E_m| \le 3^m |E_0|,
\end{align*}
making the final graph exponential in $m$.

How many nodes does the filter operation remove? It can be shown quite easily that the filter operation removes
at least a polynomial number (in $n$) of nodes, simply by noting that there are $O(n^3)$ nodes in the first level of the orbital graph, but
only $O(n)$ valid paths into them. To properly quantify the growth of the graph, we have to wait for the next section.

\subsection{Growth factor}
Meanwhile, we define a simple metric to measure the growth rate. For this, we simply use the number of nodes
as a measure. Let 
\begin{align*}
	\eta_i := \frac{|V_i|}{|V_0|}, \text{ for } i \in [m],
\end{align*}
be the growth factor as a function of iterations.

\begin{definition}(peak growth factor)
	Let $\eta_{peak}$ denote the peak growth factor of nodes in IHM. Then
	\begin{align*}
		\eta_{peak} := \max_{i \in [m]} \eta_i = \frac{ \max_{i \in [m]} |V_i|}{|V_0|}.
	\end{align*}
\end{definition}

\begin{proposition}
	The complexity of~\cref{alg:ihm} is $O(\eta_{peak} mn^7)$.
	\label{prop:ihmcomp}
\end{proposition}
\begin{proof}
	First notice that at any iteration of the algorithm, the graph size is $O(\eta_{peak} |G_0|) = O(\eta_{peak}n^6)$.
	The refine operation is simply linear in the size of the graph. The filter operation is also linear in the
	size of the graph, but executed at most $n$ times. Thus, the complexity per iteration is $O(\eta_{peak}n^7)$.
	Since the algorithm consists of $m$ iterations, the overall complexity is $O(\eta_{peak} mn^7)$.
\end{proof}

We will obtain a bound for $\eta_{peak}$ in the next section. It can be readily seen that $\eta_{peak}$ depends on the \emph{maximum 
number of distinct zero paths} ($\max DZP$).  
Regarding the nature of the final graph $G_m$, we have this:
\begin{theorem} 
	The  graph $G_m = (V_m,E_m)$ has the following properties: 
	\begin{itemize} 
		\item[(i)] All nodes have length $\le 1$.  
		\item[(ii)] All arc lengths are zero in the $y$-component.  
		\item[(iii)] All leaf nodes are \emph{TRUE} nodes.  
		\item[(iv)] All paths are zero paths.  
	\end{itemize}
\end{theorem}
\begin{proof}
	With each refine step, the node lengths are halved. Since the maximum possible initial node length is $2^m$, in at most $m$
	iterations, all node lengths become either unity or zero. Since two nodes can be connected by an arc only if they interact,
	using the fact that all node lengths are unity or zero, it follows that arc weights in the $y$ component are zero. This
	takes care of (i) and (ii).

	Recall that a node $e_i \in V_m$ is a TRUE node, if $(x_i^{-},y_i^{-}) \in S_n$. Also recall that a leaf node is a node
	with no outgoing arcs. If some leaf node $e_i$ is a FALSE node, then Step $5$ of the~\cref{alg:filter}, would remove it since it
	will not be connected to $e_{\infty}$. Thus, every leaf node in $G_m$ is a TRUE node, and this takes care of (iii).

	Finally, note that the initial point $y_0 = 0$ in the refined start node $e_0$ (since it is of unit length, and is a half-open
	interval). Since all arc weights are zero in the $y$ component, it follows that all paths are zero paths, and (iv) is true.
\end{proof}
\begin{definition} (Solution Graph)
	If $G_m \neq \emptyset$, we call it as the Solution Graph.
\end{definition}

\subsection{Indices as path lengths}
Recall that in the local reference frame of an edge, the local $x$ coordinate is determined from the local $y$ coordinate(see~\cref{sec:loc_coords}). 
For a given OL, the intersection point in $e_0$ is $z_0 = x_0 + \iota y_0$, and $z_0 = z_0^- + x_0^{loc} + \iota y_0^{loc}$. In general,
$x_0^{loc}$ is a rational number. However, we can ignore this and set $x_0^{loc} = 0$, i.e., treat $x_0$
as $x_0^-$ the global $x$ coordinate of edge $e_0$. We do this for every edge along the path. 
Since the $x$-component of the arc weight is simply the translation along $x$ between the pair of edges, the overall the path length
gives the $x$ coordinate of the point in the geometry. With this provision, we will have only integer values for $x$ and the following result.
\begin{theorem}
	Let $\pi$ be a zero path in $G_m$ starting at $e_0$ and ending at some TRUE node $e_k$. Then the $x$-component of the path length,
	denoted $|\pi|_x$, equals the index of the subset whose subset sum is the target $T$.
\end{theorem}

\section{A Holy Trinity: Number, Size and Additive Structure}
\label{sec:bhairav}

The purpose of this section is to provide an upperbound on the maximum number of distinct zero paths through any equivalence 
class of edges intersecting the OL. In order to do that let's take a contemplative look on the nature of the SSP itself!
Recall that the Subset Sum Problem is specified by two independent parameters: the number of elements $n$, and the maximum size of elements in bits $m$.
However, there is also a third quantity involving the relationship between the elements which we simply call as the \emph{degree of additive structure}
(to be defined later). 

It is the interplay of these three quantities, which we call a \emph{trinity of factors},  
that determine the ease or difficulty of finding the solution to a given instance of the SSP. 
In this section, we examine these three quantities and reveal the precise way in which they control the complexity. 
Finally we provide an upperbound on the maximum number of locally distinct paths through any edge equivalence class intersecting the OL.

The general theme is that the geometric description of the solution space  of SSP has two aspects:
\begin{itemize}
	\item[(i)] a \emph{combinatorial aspect} depending only on $n$,
	\item[(ii)] and a \emph{relational aspect} depending on $m$ and the \emph{degree of additive structure}.
\end{itemize}
Many of the quantities that fall out of symmetries such as the number of paths to level $r$, i.e., $\beta(r)$,
are of \emph{combinatorial} nature, and these are exponential in number. Then there are other quantities like \emph{distinct path lengths}
which have a \emph{relational aspect} to them, and they depend on $m$ and the \emph{degree of additive structure}.

To determine $\max{DZP}$, we first consider \emph{all} the intersection points on the  \ol and  bound
the \emph{maximum distinct valid paths} ($\max{DVP}$) instead; this automatically provides a bound on $\max{DZP}$
since the set of all \emph{zero paths} is a subset of the set of all \emph{valid paths}.
The main reason we look at this larger set  is to understand the structure of points around the OL, and as the reader will soon see,
analyzing this larger set also makes things simpler mathematically.
For this, we define a \emph{configuration graph} that captures both the \emph{combinatorial} and the \emph{relational} aspects of the problem.
This is followed by a rigorous analysis of the number of paths in the configuration graph, from which an upperbound on the $\max{DVP}$
can be extracted. Then a bound for $\max{DZP}$ can be derived (which is a special case) easily.

As a consequence of this analysis, a curious finding that \emph{size doesn't matter beyond a threshold value dependent on} $\log{n}$ is obtained. 
We also provide a precise estimate for this critical size $m_0$.

\subsection{A Configuration Graph}
Let $M := 2^m$. Consider a sequence of $n+1$ vertical segments denoted $(s_0,s_1,s_2,\ldots,s_n)$, where
\begin{align*}
	s_i := \{ (x,y) \in {\mathbb R}^2 : x = i \in [0,n]  \text{ and } y \in [0,M) \},
\end{align*}
arranged along the $x$-axis and are parallel. Thus, each segment has $2^m$ integer points and
these are of interest in what follows.

\begin{figure}[!htbp]
\begin{center}
\begin{tikzpicture}[scale=0.7,
	mydot/.style={
    circle,
    fill=white,
    draw,
    outer sep=0pt,
    inner sep=1.5pt
  }]
\node[outer sep=0pt,circle, fill,inner sep=1.5pt,label={}] (E0a) at (-2,0) {};
\node[mydot] (E0b) at (-2,2.0){};
\node[outer sep=0pt,circle, fill, red, inner sep=1.5pt, label={}] (E0c) at (-2,1.0) {};
        \draw[thick,blue] (E0a) node[below]{$s_0$} -- (E0b);

\node[outer sep=0pt,circle, fill,inner sep=1.5pt,label={}] (E1a) at (0,0) {};
\node[mydot] (E1b) at (0,2.0){};
\node[outer sep=0pt,circle, fill, red, inner sep=1.5pt, label={}] (E1c) at (0,0.5) {};
        \draw[thick,blue] (E1a) node[below]{$s_1$} -- (E1b);

\node[outer sep=0pt,circle, fill,inner sep=1.5pt,label={}] (E1a) at (2,0) {};
\node[mydot] (E1b) at (2,2.0){};
\node[outer sep=0pt,circle, fill, red, inner sep=1.5pt, label={}] (E2c) at (2,1.5) {};
        \draw[thick,blue] (E1a) node[below]{$s_2$} -- (E1b);

\node[outer sep=0pt,circle, fill,inner sep=1.5pt,label={}] (E1a) at (4,0) {};
\node[mydot] (E1b) at (4,2.0){};
\node[outer sep=0pt,circle, fill, red, inner sep=1.5pt, label={}] (E3c) at (4,0.9) {};
        \draw[thick,blue] (E1a) node[below]{$s_3$} -- (E1b);

\node[outer sep=0pt,circle, fill,inner sep=1.5pt,label={}] (E1a) at (6,0) {};
\node[mydot] (E1b) at (6,2.0){};
\node[outer sep=0pt,circle, fill, red, inner sep=1.5pt, label={}] (E4c) at (6,1.0) {};
        \draw[thick,blue] (E1a) node[below]{$s_4$} -- (E1b);

\node[outer sep=0pt,circle, fill,inner sep=1.5pt,label={}] (E1a) at (8,0) {};
\node[mydot] (E1b) at (8,2.0){};
\node[outer sep=0pt,circle, fill, red, inner sep=1.5pt, label={}] (E5c) at (8,1.8) {};
        \draw[thick,blue] (E1a) node[below]{$s_5$} -- (E1b);
\node[outer sep=0pt,circle, fill,inner sep=1.5pt,label={}] (E1a) at (10,0) {};
\node[mydot] (E1b) at (10,2.0){};
\node[outer sep=0pt,circle, fill, red, inner sep=1.5pt, label={}] (E6c) at (10,0) {};
        \draw[thick,blue] (E1a)  -- (E1b);

\node[outer sep=0pt,circle, fill,inner sep=1.5pt,label={}] (E1a) at (12,0) {};
\node[mydot] (E1b) at (12,2.0){};
\node[outer sep=0pt,circle, fill, red, inner sep=1.5pt, label={}] (E7c) at (12,0.45) {};
        \draw[thick,blue] (E1a)  -- (E1b);

\node[outer sep=0pt,circle, fill,inner sep=1.5pt,label={}] (E1a) at (14,0) {};
\node[mydot] (E1b) at (14,2.0){};
\node[outer sep=0pt,circle, fill, red, inner sep=1.5pt, label={}] (E8c) at (14,0.7) {};
        \draw[thick,blue] (E1a) node[below]{$s_r$} node[right] {0} -- (E1b) node[right] {$2^m$};

\draw[thick,red] (E0c) -- (E1c);
\draw[thick,red] (E1c) -- (E2c);
\draw[thick,red] (E2c) -- (E3c);
\draw[thick,red] (E3c) -- (E4c);
\draw[thick,red] (E4c) -- (E5c);
\draw[thick,red] (E5c) -- (E6c);
\draw[thick,red] (E6c) -- (E7c);
\draw[thick,red] (E7c) -- (E8c);
\end{tikzpicture}
\end{center}
\caption{Configuration plot of a valid path $\pi$ of length $r$ along the OL. Although the edge lengths can
vary (but are limited by $M=2^m$), we use this maximum possible length to accomodate all possible local intercepts. }
\label{fig:cplot}
\end{figure}
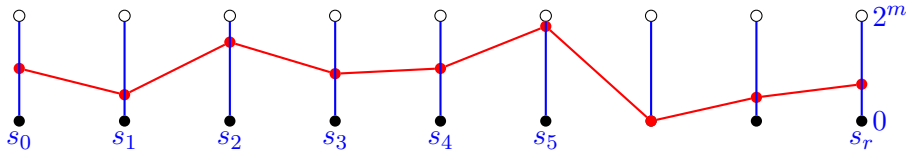

In the graph $G_0$, given a start value $y_0 \in e_0$, let the set of all valid paths of length $r \in [n]$ on the \ol be denoted by
$\Pi_r$. For each path $\pi \in \Pi_r$, since it is valid, we have a sequence of edges $(e_{\pi_1},e_{\pi_2},\ldots,e_{\pi_r})$
that it goes through with local intercepts $(y_{\pi_1}, \ldots, y_{\pi_r})$ respectively, where $y_{\pi_i} \in [0,|e_{\pi_i}|)$ for all $i \in [r]$. 
We mark the point $(0,y_0)$ in the segment $s_0$, $(1,y_{\pi_1})$ in segment $s_1$, and so on with $(r,y_{\pi_r})$ marked in $s_r$. 
We connect points on adjacent segments by a straight line. With this we have a piece-wise linear curve corresponding to the path $\pi$. 
We call this the \emph{configuration plot} of the path $\pi$. An example is shown below in~\cref{fig:cplot}.

\begin{remark}
	For a path $\pi \in \Pi_r$, each segment in the configuration plot corresponds to some arc weight $w_{ij}$ relating
	two interacting edges $e_i, e_j \in G_0$. While $w_{ij}$ has both $x$ and $y$ components, we only care about the
	$y$ component in the configuration plot, and it can be seen that it lies in the interval $[-M+1,M-1]$.
\end{remark}

\begin{definition}(Locally Distinct Configurations)
        Two paths $\pi$ and $\pi'$, both of length $r$, are said to have locally distinct configurations if and only if
        \begin{align*}
                (y_{\pi_1},y_{\pi_2},\ldots,y_{\pi_r}) \neq (y_{\pi_1}',y_{\pi_2}',\ldots,y_{\pi_r}').
        \end{align*}
\end{definition}

\begin{definition}(The set of locally distinct configurations $\Xi_r$)
        Let $\Xi_r$ denote the set of all locally distinct configurations obtained from $\Pi_r$. In other words,
        this is a subset of $\Pi_r$ that produce unique configurations. We denote the size of this set by $|\Xi_r|$.
\end{definition}

Since we are dealing with local values in the configuration plot without regard to which edges they came from, two paths using distinct sequences of nodes in $G_0$
can still result in the same configuration. As a result, we have:
\begin{proposition}
	For any $r \in [n]$, we have $|\Xi_r| \le |\Pi_r|$.
\end{proposition}

If we overlay the configuration plots of all the paths $\Pi_r$ for all $r \in [n]$, in the same configuration graph, 
we will have many distinct piece-wise linear curves emanating from the
point $(0,y_0) \in s_0$, going through the intermediate segments and arriving at some integer point in $s_r$. In general,
this will be a dense thicket of paths corresponding to paths on the OL. Each segment $s_r$ for $r \in [n]$ captures all the local values
of paths entering edges of level $r$ in $G_0$. The configuration plot thus provides a compact view of all the paths on OL.

\begin{definition} (Point set $F_r$ and $\gamma_r$)
	For each $r \in [n]$, let $F_r$ denote the set of distinct points in segment $s_r$ that have at least one incoming path from segment $s_{r-1}$. 
        Each point in $F_r$ has coordinates of the form $(r,y)$ where $y \in [0,2^m)$. Let $\gamma_r := |F_r|$ denote
        the number of distinct points in segment $r$. We have the starting set $F_0$ which consists of a single point $(0,y_0)$, and thus
        $\gamma_0 = 1$. It is also clear that $\gamma_r \le 2^m$, for all $r \in [n]$.
\end{definition}

\begin{definition} (Arc set $C(r,r+1)$)
	For a non-negative integer  $r \in [0,n-1]$ let $C(r,r+1)$ denote the set  of distinct arcs between levels $r$ and $r+1$. 
	Each arc of $C(r,r+1)$ is a line segment connecting
        some point $(r,y) \in F_r$ with some other point $(r+1,y') \in F_{r+1}$. It is clear that
        \begin{align*}
                C(r,r+1) \subseteq F_r \times F_{r+1}.
        \end{align*}
	We denote the size of $C(r,r+1)$ by $|C(r,r+1)|$.
\end{definition}

\begin{definition}(Layered Configuration Graph $\cg$)
        We view the points in $F_r$ for $r \in [n]$ as nodes, and line segments $C(r,r+1)$ as arcs of the configuration graph 
	for $r \in [0, n-1]$. We denote this configuration graph by $\cg$.
        The single point in $F_0$ serves as a root node.
        Thus, $\cg$ is a layered graph with $n+1$ levels and comprised of $n$ bipartite layers. We denote the set of all nodes of $\cg$ by $V(\cg)$ and the set of
        all arcs by $E(\cg)$ using standard notation for graphs. 
\end{definition}
	From the above definitions, it follows that
        \begin{align*}
                \text{\# nodes} & = |V(\cg)| = \gamma_0 + \gamma_1 + \gamma_2 + \ldots + \gamma_n, \\
                \text{and \# arcs} & = |E(\cg)| = |C(0,1)| + |C(1,2)| + \ldots + |C(n-1,n)|.
        \end{align*}

\begin{definition}(point number sequence $\gamma$)
        We define $\gamma := (\gamma_1, \gamma_2,\ldots,\gamma_n)$ as the sequence of number of distinct points in each segment and
        refer to it as the point number sequence.
\end{definition}

\begin{question}
        We are interested in answering the following questions:
        \begin{itemize}
                \item[(i)] What is $\max(\gamma)$?
                \item[(ii)] What is the asymptotic profile of $\gamma$ for large $m$?
        \end{itemize}
\end{question}
It can be seen easily that $\max{DVP} \le \max(\gamma)$, and hence the importance of the first question.

\subsection{Apparent paths in $\cg$}
Once all the valid paths on the OL are plotted in the configuration graph, we can view $\cg$ as a graph on its own.

\begin{definition} (Paths in $\cg$)
	A path of length $r \in [n]$ in $\cg$ is a sequence $(0,y_0), (1,y_1),\\ \ldots, (r,y_r)$ of points one from each of the sets
        $F_0,F_1,\ldots,F_r$, such that every two adjacent points are connected by an arc.
\end{definition}
Note that a connected path in $\cg$ may or may not correspond to a path in $G_0$ as the example plot below shows.

\begin{figure}[!htbp]
\begin{center}
\begin{tikzpicture}[scale=0.7,
	mydot/.style={
    circle,
    fill=white,
    draw,
    outer sep=0pt,
    inner sep=1.5pt
    }]
\node[outer sep=0pt,circle, fill,inner sep=1.5pt,label={}] (E0a) at (-2,0) {};
\node[mydot] (E0b) at (-2,2.0){};
\node[outer sep=0pt,circle, fill, red, inner sep=1.5pt, label={}] (E0c) at (-2,1.0) {};
        \draw[thick,blue] (E0a) node[below]{$s_0$} -- (E0b);
        \node at (-2.4,1.0) {$y_0$};

\node[outer sep=0pt,circle, fill,inner sep=1.5pt,label={}] (E1a) at (0,0) {};
\node[mydot] (E1b) at (0,2.0){};
\node[outer sep=0pt,circle, fill, red, inner sep=1.5pt, label={}] (E1c) at (0,0.5) {};
\node[outer sep=0pt,circle, fill, green, inner sep=1.5pt, label={}] (E1d) at (0,1.5) {};
        \draw[thick,blue] (E1a) node[below]{$s_1$} -- (E1b);

\node[outer sep=0pt,circle, fill,inner sep=1.5pt,label={}] (E1a) at (2,0) {};
\node[mydot] (E1b) at (2,2.0){};
\node[outer sep=0pt,circle, fill, red, inner sep=1.5pt, label={}] (E2c) at (2,1.0) {};
        \draw[thick,blue] (E1a) node[below]{$s_2$} -- (E1b);
\node[outer sep=0pt,circle, fill,inner sep=1.5pt,label={}] (E1a) at (4,0) {};
\node[mydot] (E1b) at (4,2.0){};
\node[outer sep=0pt,circle, fill, red, inner sep=1.5pt, label={}] (E3c) at (4,0.9) {};
\node[outer sep=0pt,circle, fill, green, inner sep=1.5pt, label={}] (E3d) at (4,1.6) {};
        \draw[thick,blue] (E1a) node[below]{$s_3$} -- (E1b);

\node[outer sep=0pt,circle, fill,inner sep=1.5pt,label={}] (E1a) at (6,0) {};
\node[mydot] (E1b) at (6,2.0){};
\node[outer sep=0pt,circle, fill, red, inner sep=1.5pt, label={}] (E4c) at (6,1.0) {};
        \draw[thick,blue] (E1a) node[below]{$s_4$} -- (E1b);

\draw[thick,red] (E0c) -- (E1c);
\draw[thick,red] (E1c) -- (E2c);
\draw[thick,red] (E2c) -- (E3c);
\draw[thick,red] (E3c) -- (E4c);

\draw[thick,green] (E0c) -- (E1d);
\draw[thick,green] (E1d) -- (E2c);
\draw[thick,green] (E2c) -- (E3d);
\draw[thick,green] (E3d) -- (E4c);
        \node at (-0.4,0.5) {$y_1$};
        \node at (-0.4,1.5) {$y_1'$};
        \node at (1.6,1.0) {$y_2$};
        \node at (3.4,0.9) {$y_3$};
        \node at (3.4,1.6) {$y_3'$};
        \node at (5.6,1.0) {$y_4$};
\end{tikzpicture}
\end{center}
\caption{Illustration of real and apparent paths in $\cg$.}
\label{fig:app_paths}
\end{figure}
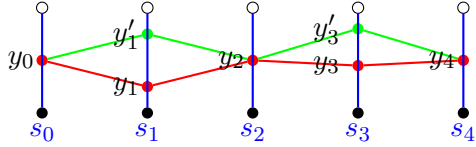

In~\cref{fig:app_paths}, there are two paths $\pi_1 = (y_0,y_1,y_2,y_3,y_4)$ (shown in red) and $\pi_2 = (y_0,y_1',y_2,y_3',y_4)$
(shown in green) coming from the paths in $\Pi_4$. However, two other paths are induced: $\pi_3 = (y_0,y_1,y_2,y_3',y_4)$
and $\pi_4 = (y_0,y_1',y_2,y_3,y_4)$. The paths $\pi_3$ and $\pi_4$ have no existence in $G_0$, and hence are called
\emph{apparent paths}. Thus, the number of paths in $\cg$ can be greater than what we started with!

\subsection{Path magnification in $\cg$}

\begin{definition}(out-degree)
	For a given point $(r,y) \in F_r$, where $r \in [0, n-1]$, the number of distinct outgoing arcs from it to points in $F_{r+1}$ is
	called the out-degree, and denoted by $\mu(r,y)$. 
\end{definition}

\begin{definition} (Max out-degree)
	Let $\mu_r$ denote the maximum out degree over all points in the $r$th level. Thus,
	\begin{align*}
		\mu_r := \max_{y \in F_r} \mu(r,y).
	\end{align*}
\end{definition}

\begin{proposition}
	For a given level $r \in [0,n-1]$, the maximum out-degree $\mu_r$ is at most $\frac{(n-r)^3+3(n-r)^2-13(n-r)+6}{3}$.
\end{proposition}
\begin{proof}
	Suppose that level $r$ corresponds to $Q_{n+1-r}$; then level $r+1$ corresponds to $P_{n-r}$.
        The point $(r,y) \in F_r$ can belong to any of the $|Q_{n+1-r}|$ nodes of level $r$ in $G_0$. 
	The transformation structure of paths going from level $r$ to level $r+1$ is $q_j \xra{\lambda} q_k \xra{\rho} p_k$
	where $j > k$. Each arc weight is between an edge of $q_k$ and an edge of $p_k$ as $k$ ranges in $[n-r]$. Since we are interested
	in distinct arc weights, we consider the worst case where ${\bf a}$ is a dissociated set. Since there are at most $k^2 +k -5$ distinct arc weights
	between $q_k$ and $p_k$ (see~\cref{prop:jmin}), summing over $[n-r]$ we get
	\begin{align*}
		\mu_r & \le \sum_{k=3}^{n-r} k^2 +k -5 \\
		& = \frac{(n-r)^3+3(n-r)^2-13(n-r)+6}{3}
	\end{align*}
	and the claim follows.
	The result remains the same if level $r$ corresponds to $P_{n+1-r}$, and level $r+1$ corresponds to $Q_{n-r}$.
\end{proof}
\begin{definition}
Notice that $\mu_r$ is a decreasing function of $r$, and maximum value is attained at $r=0$ where 
	$\mu_0  = \frac{n^3 + 3n^2 -13n +6}{3} \approx \frac{n^3}{3}$  for $n$ large enough. We will be using $\mu_0 = \frac{n^3}{3}$ from here on.
	\label{rem:mu0}
\end{definition}

\begin{remark}
	The above proposition regarding the  out degree is essentially an estimate in the worst case. When the given
	instance has high degree of additive stucture many arc weights will be the same. While these arcs will be distinct in $G_0$, they will
	not be in $\cg$ since we are only concerned with the $y$ component of arc weight. 
	As a result, the quantity $\mu_r$  which represents the maximum number of distinct out going arcs will be small. For example,
	when ${\bf a} \in CP$, $\mu_r =1$ for all $r \in [0,n-1]$. 
\end{remark}

While $\mu_r$ is more aptly a representative of the \emph{degree of additive disorder}, we can view it as a representative of the \emph{degree of
additive structure} in an inverse way. Thus, the smaller is the value of $\mu_{r}$ the larger is the degree of additive structure and vice versa.

\begin{claim} 
	The configuration graph $\cg$ includes both the combinatorial and relational aspects by capturing
	\begin{itemize}
		\item[(i)] the combinatorial aspect through the number of paths,
		\item[(ii)] the size $m$ through segment lengths,
		\item[(iii)] and the degree of additive disorder/structure through $\mu_r$.
	\end{itemize}
\end{claim}

\subsection{Maximum number of paths in $\cg$}
\begin{proposition} 
	Given a start point $(0,y_0) \in F_0$, the maximum number of distinct paths into level $r \in [n]$, denoted by $\zeta_r$ satisfies $\zeta_r < \mu_0^r$.
\end{proposition}
\begin{proof}
	We feign ignorance regarding the edge in $G_0$ from which the local value $y_0$ comes from, and allow it from all edges of $p_{n}$. Hence the maximum
	number of distinct out going arcs from $(0,y_0)$ is at most $\mu_0$. Thereafter the maximum out degrees are at most $\mu_1, \mu_2, \ldots$ and so on.
	Thus, the maximum number of distinct paths to level $r$ is 
	\begin{align*}
		\zeta_r & \le \mu_0\mu_1\ldots\mu_{r-1} \quad \text{ (since some path lengths may coincide, we use $\le$)} \\
		& < \mu_0^r \quad \text{ (since $\mu_r < \mu_0$ for all $r \ge 1$)}.
	\end{align*}
\end{proof}

\subsection{Some simple relations on $\gamma_r$ and $|C(r,r+1)|$}

\begin{proposition}
	Let $r$ and $r+1$ be two adjacent levels where $r \in [n-1]$, with $\gamma_r$ and $\gamma_{r+1}$ points in each respectively. Then it is true that
        \begin{align}
		\gamma_{r+1}  \le  \gamma_r \mu_r. 
                \label{ineq:G1}
        \end{align}
\end{proposition}
\begin{proof}
	The inequality follows from the observation that the number of distinct points in level $r+1$ cannot be larger than the number
	of distinct outgoing arcs emanating from all the distinct points of the previous level. 
\end{proof}

\begin{proposition}
	Let $r$ and $r+1$ be two adjacent levels where $r \in [n-1]$, with $\gamma_r$ and $\gamma_{r+1}$ points in each respectively. Then it is true that
        \begin{align}
                \max(\gamma_r, \gamma_{r+1}) \le |C(r,r+1)| \le \gamma_r \gamma_{r+1}.
                \label{ineq:C}
        \end{align}
\end{proposition}
\begin{proof}
        Since $F_r$ as well as $F_{r+1}$ are obtained from the paths in $G_0$, it follows that these sets are connected by line
        segments. The minimum number of line segments needed to have $\gamma_r$ points in level $r$, and $\gamma_{r+1}$ points
        in level $r+1$ is easily seen to be $\max(\gamma_r,\gamma_{r+1})$. If we have less than this number of segments, then at least
        one point will be isolated.
        The upper limit shows the other extreme, where every point in $F_r$ is connected to every point in $F_{r+1}$. The actual value
        $|C(r,r+1)|$ must necessarily be in between these limits.
\end{proof}

\begin{proposition}
	Let $\gamma_r$ be the number of distinct points in level $r \in [0,n-1]$. Then the maximum number of arcs between levels
        $r$ and $r+1$ is
        \begin{align}
                |C(r,r+1)| \le \gamma_r \mu_r \le \gamma_r\mu_0.
                \label{ineq:gain}
        \end{align}
\end{proposition}

\begin{proof}
	Since the maximum out degree of a point $(r,y) \in F_r$ is at most $\mu_r$, and there are $\gamma_r$ distinct points in level $r$,
	the claimed inequality follows. 
\end{proof}

\subsection{The importance of size}
Recall that $\Xi_n$ denotes the set of distinct configurations produced by the paths $\Pi_n$ in the
configuration graph $\cg$.
\begin{proposition}
        The maximum number of locally distinct configurations of paths in $\cg$ is
        \begin{align*}
                |\Xi_n| \le \min(2^{mn}, \mu_0^n).
        \end{align*}
	\label{prop:mrole}
\end{proposition}
\begin{proof}
	Since each segment $s_r$ has $2^m$ available integral points, the maximum number of possibilities for distinct configurations
	is $2^m \times 2^m \times \ldots \times 2^m$ ($n$ times) which is equal to $2^{nm}$. On the other hand, we know that the maximum
	number of distinct paths in $\cg$ is bounded by $\mu_0^n$. Since each distinct path results in a distinct configuration,
	putting them together, the claim follows.
\end{proof}

Let $n$ be fixed and consider instances of SSP with varying $m$. For each instance, we look at the corresponding configuration graph.
If $m$ is small, the segment lengths will be small and many paths will have the same configuration. If we 
use larger numbers i.e., increase $m$, the available number of configurations increase and some paths will start to have distinct configurations.
As we continue to increase $m$, the number of distinct configurations in the geometry cannot increase arbitrarily since they cannot exceed the number of
paths (see~\cref{ex:varym} for an example). 

\begin{definition}
	For all $i \in [n]$, we define $\Gamma_i := \gamma_1 \gamma_2 \ldots \gamma_i$.
\end{definition}

\begin{corollary} It is true that
        \begin{align*}
                |\Xi_n| \le \min(\Gamma_n, \mu_0^n).
        \end{align*}
	\label{prop:distconfigs}
\end{corollary}
\begin{proof}
	Whatever the sequence $\gamma$ may be, it is clear that $\gamma_i \le 2^m$ for all $i \in [n]$.
	Furthermore, it is easily seen that the maximum number of distinct configurations 
	$|\Xi_n| \le \gamma_1\gamma_2\ldots\gamma_n$, and the claim follows.
\end{proof}

While the number of distinct configurations cannot be increased by increasing $m$ beyond $\log\mu_0$, the distribution of $\gamma$ can be affected.
Suppose that $m > \log\mu_0$. Then it can be seen that there are many distributions $\gamma$ that support $\mu_0^n$ distinct
configurations. For example,
\begin{align*}
        \gamma = (\mu_0,\mu_0,\ldots,\mu_0) & \text{ with } \max(\gamma) = \mu_0, \text{ and }\\
        \gamma = (\mu_0,\mu_0^2,\ldots,\mu_0^n) & \text{ with } \max(\gamma) = \mu_0^n,
\end{align*}
both support $\mu_0^n$ distinct configurations. In the first case each point in  $F_r$ connects with all points in $F_{r+1}$, whereas in the
second case each point in $F_r$ connects with $\mu_0$ distinct points in $F_{r+1}$. The first case corresponds to \emph{complete bi-partite layers}
while the second corresponds to a $\mu_0$-\emph{regular} tree. 

Also while the number of distinct configurations cannot exceed $\mu_0^n$, the above propositions don't tell us anything
about $\max(\gamma)$. To obtain a bound for $\max(\gamma)$, our approach is to first obtain an upper bound for $\Gamma_n$, and then extract
the desired upper bound. First of all, notice that $\Gamma_n$ \emph{need not be} bounded by $\mu_0^n$. In fact it can well exceed it
since the sequence $\gamma$ can be wasteful, i.e., not all points in one level need to connect with all points in the next level. So the
product $\Gamma_n$ can be much larger than the number of paths.

\subsection{A Product Inequality}
In this section we derive a useful inequality, simply called a \emph{Product Inequality} (PI), that would enable us to  obtain an upper bound on $\max(\gamma)$.

\begin{definition}(Graph $\cj$)
	Let the graph $\cj := (V(\cj), E(\cj))$ be a layered graph with $n+1$ levels, and $n$ complete bipartite layers, defined by
	\begin{align*}
		V(\cj) & := V(\cg) = \cup_{i=0}^n F_i, \\
		E(\cj) & := \cup_{i=0}^{n-1} F_i \times F_{i+1}.
	\end{align*}
	Alternatively, $\cj$ is a copy of the $\cg$ graph, but where we add more arcs so that all points in level $i$ are connected to 
	all points in level $i+1$ for all $i \in [0,n-1]$. Thus, the number of arcs between levels $i$ and $i+1$ is 
	simply $|F_i| \times |F_{i+1}| = \gamma_i \gamma_{i+1}$.
	\label{def:jgraph}
\end{definition}

\begin{definition}(Arc Growth Factor (AGF) $\alpha_i$)
	Let $X$ be a layered graph with $l+1$ levels and bipartite layers. Then for any $i \in [l]$ we define the arc growth 
	factor $\alpha_i(X)$ as the ratio of the number of arcs between levels $i$ and $i+1$ to the number of arcs between levels $i-1$ and $i$, i.e.,
	\begin{align*}
		\alpha_i(X) = \frac{\text{\# arcs between levels $i$ and $i+1$}}{\text{\# arcs between levels $i-1$ and $i$}}.
	\end{align*}
	\label{def:AGF}
\end{definition}

\begin{proposition}
	For each $i \in [n-1]$, the AGF in $\cj$ satisfies 
	\begin{align}
		\alpha_i(\cj) \le \mu_{i-1}\mu_i < \mu_0^2.
	\end{align}
	\label{prop:agf_j0}
\end{proposition}
\begin{proof}
	By~\cref{def:jgraph} and by~\cref{def:AGF}, we have
	\begin{align*}
		\alpha_i(\cj) & = \frac{\gamma_i \gamma_{i+1}}{\gamma_{i-1} \gamma_{i}}  = \frac{\gamma_{i+1}}{\gamma_{i-1}} \\
		& \le (\mu_{i-1}) (\mu_i) \quad \text{ (using $\gamma_i \le \gamma_{i-1}\mu_{i-1}$ and $\gamma_{i+1} \le \gamma_i \mu_i$)} \\
		& < \mu_0^2 \quad \text{ (since $\mu_j < \mu_0$ for all $j \in [n-1]$)}
	\end{align*}
	and the claim follows.
\end{proof}

\begin{remark}
	While the above~\cref{prop:agf_j0} shows that $\alpha_i(\cj) < \mu_0^2$, it is possible for $\alpha_i(\cj)$ to be less than
	one, for example,  when $\gamma_{i+1} < \gamma_{i-1}$. This kind of a situation happens when paths fanout upto the middle levels, but later converge to
	a small number of points in the final level. 
	In such cases $\alpha_i(\cj) < 1$ for the latter levels.
	\label{rem:alpha_less_than_one}
\end{remark}

The following proposition contrasts the graph $\cg$ with $\cj$ with regard to AGF.
\begin{proposition}
	For any $i \in [n-1]$, the AGF in $\cg$ satisfies 
	\begin{align}
		\alpha_i(\cg) \le \mu_i < \mu_0.
	\end{align}
	\label{prop:agf_c0}
\end{proposition}
\begin{proof}
	By the definition of AGF, we have
	\begin{align*}
		\alpha_i(\cg) & = \frac{|C(i,i+1)|}{|C(i-1,i)|} \\
		& \le \frac{\gamma_i \mu_i}{|C(i-1,i)|} \quad \text{ (since $|C(i,i+1)| \le \gamma_i \mu_i$) } \\
		& \le \frac{\gamma_i \mu_i}{\gamma_i} \quad \text{ (since $|C(i-1,i)| \ge \max(\gamma_{i-1},\gamma_i)$ by~\cref{ineq:C}) } \\
		& \le \mu_i < \mu_0.
	\end{align*}
\end{proof}

\begin{remark}
	From the above two propositions we see that $\alpha_i(\cg) < \mu_0$ and $\alpha_i(\cj) < \mu_0^2$. Thus, the maximum magnification factor
	per layer in $\cj$ is a square of that of $\cg$ to account for the presence of more arcs. In general, for any random instance the magnification
	factor in $\cj$ is expected to be in between $\mu_0$ and $\mu_0^2$.
	\label{rem:actual_gain_j0}
\end{remark}

\begin{proposition}
	Given a positive integer $r \in [n]$, let  $\hat{\Upsilon}_r$ denote the number of distinct paths to level $r$ in the $\cj$ graph originating from
	the point $(0,y_0) \in F_0$ in the zeroth level. Then
	\begin{align*}
		\hat{\Upsilon}_r = \max_{j \in [r]} \quad \gamma_1 \prod_{i=1}^{j-1} \alpha_i(\cj).
	\end{align*}
	\label{prop:formulaJ}
\end{proposition}
\begin{proof}
	For $r=1$ the number of paths to level one is simply the number of arcs between levels zero and one, which is $\gamma_0 \gamma_1 = \gamma_1$
	since $\gamma_0 = 1$.
	These paths result in $\gamma_1$ points from which new arcs fan out into the second level. The number of arcs between levels one
	and two, by definition is $\gamma_1 \gamma_2$. This can be likened to as the arcs to first level branching out to level two. Thus, the 
	path magnification from level one to level two is $\frac{\gamma_1 \gamma_2}{\gamma_0 \gamma_1} = \alpha_1(\cj)$ by~\cref{def:AGF}.
	Thus, the number of paths to level two is
	$\gamma_1 \alpha_1(\cj)$ which equals $\gamma_1\gamma_2$. 

	As we go to higher levels the arcs branch out into more arcs and we continue multiplying with AGFs $\alpha_i(\cj)$. 
	However, by~\cref{rem:alpha_less_than_one}, it is possible for some of the $\alpha_i(\cj)$s to be less than one, particularly if all paths converge to
	a single point (or a small number of points) in level $r$. In this case, the number of distinct points in level $r$ is small but the 
	number of distinct paths is still the maximum obtained over all the previous levels. The $\max_{j \in [r]}$ term correctly preserves the 
	maximum number if the maximum happens somewhere in the middle levels. 
\end{proof}

\begin{corollary}
	For $r \in [n]$, let  $\hat{\Upsilon}_r$ denote the number of distinct paths to level $r$ in the $\cj$ graph originating from
	the point $(0,y_0) \in F_0$ in the zeroth level. Then $\hat{\Upsilon}_r < \mu_0^{2r-1}$.
	\label{cor:j0_bound}
\end{corollary}
\begin{proof}
	We have 
	\begin{align*}
		\hat{\Upsilon}_r & = \max_{j \in [r]} \quad \gamma_1 \prod_{i=1}^{j-1} \alpha_i(\cj) \quad \text{(from~\cref{prop:formulaJ})} \\
		& < \gamma_1 (\mu_0^2)^{r-1} \quad \text{ (from~\cref{prop:agf_j0})} \\
		& < \mu_0^{2r-1} \quad \text{ (since $\gamma_1 \le \mu_0$)}
	\end{align*}
	and the claim follows.
\end{proof}

Thus, the maximum number of paths in $\cj$ is at most $\mu_0^{2r-1}$ which for algebraic simplicity and 
convenience (as will be seen later) is loosely bounded as $\mu_0^{2r}$.

\begin{theorem}(Product Inequality (PI))
	Let $({\bf a}, T)$ be an instance of the SSP with dimensions $(n,m)$, and let $\gamma$ be the associated point number
	sequence.  Then for any $r \in [n]$, it is true that $\Gamma_r < \mu_0^{2r}$.
	\label{thm:Gbound}
\end{theorem}
\begin{proof}
	The number of distinct paths in $\cj$ is simply the number of distinct combinations taking one point from each of the sets $F_i$ for $i \in [r]$, 
	which is clearly $\Gamma_r = \gamma_0 \gamma_1 \ldots \gamma_r$. 
	On the other hand we know from~\cref{cor:j0_bound} that the number of distinct paths is less than $\mu_0^{2r}$. Putting them together the claim follows.
\end{proof}

\subsection{An upper bound on $\max(\gamma)$}
In this section we obtain an upper bound on $\max(\gamma)$ for any given instance of the SSP. We also obtain 
an asymptotic formula for the profile of $\gamma$ in the absence of any additive structure and for large $m$. These
results depend crucially on the \emph{Product Inequality}.

Taking the $r$-th root on both sides of $\Gamma_r < \mu_0^{2r}$, we obtain that the geometric mean $\gamma_{GM} := \Gamma_r^{1/r} < \mu_0^2$. 
Thus, the PI simply says that the geometric mean $\gamma_{GM}$ is bounded by a polynomial. However, it doesn't immediately reveal anything about
$\max(\gamma)$ since one can envision an arbitrary number of sequences $\gamma$ with varying maximal values that all have a given bounded geometric mean. 
On the other hand, the sequence $\gamma$ cannot be arbitrary as it has to correspond to paths on the OL and satisfy the growth condition
given in~\cref{ineq:G1}.
So what is the worst case distribution that results in maximum possible $\max(\gamma)$? A momentary thought shows that the worst case must correspond
to a Geometrical Progression.

\begin{theorem}
	For any instance of SSP of size $n$, we have $\max(\gamma) < \mu_0^4$.
	\label{thm:gp}
\end{theorem}
\begin{proof}
	To maximize $\max(\gamma)$  we consider the
	sequence $(\gamma_1,\gamma_2,\ldots,\gamma_r)$ to be a $GP(r,\theta)$ for some $1 < \theta \le \mu_0$.
	In particular, we let $\gamma_i = \theta^i$ for $i \in [k]$. Using the product inequality (~\cref{thm:Gbound})
	we have $\Gamma_r < \mu_0^{2r}$. Thus,
	\begin{align*}
		\theta^1 \theta^2 \ldots \theta^r & < \mu_0^{2r}, \\
		\theta^{\frac{r(r+1)}{2}} & < \mu_0^{2r}, \\
		\theta^{\frac{(r+1)}{2}} & < \mu_0^2, \text{(after taking the $r$ th root on both sides)} \\
		\theta^{(r+1)} & < \mu_0^4, \text{(after squaring on both sides)} \\
		\theta^r & < \frac{\mu_0^4}{\theta} < \mu_0^4, \text{ since $\theta > 1$}.
	\end{align*}
	Thus, the maximum element $\max(\gamma) = \gamma_r < \mu_0^4$.
\end{proof}
It should be noted that this is an extreme case that cannot be realized in practice since $\mu_r$ is decreasing function of $r$, but does provide us
with a (loose) upper bound for $\max(\gamma)$.
Next we show that the above result is still true even if we consider a symmetric geometric progression.
\begin{definition}(Symmetric Geometric Progression $SymGP(r,\theta)$)
        Let $r$ be an even positive integer, and let
        \begin{align*}
                SymGP(r,\theta) := (\theta^1, \theta^2, \ldots, \theta^{r/2}, \theta^{r/2},\theta^{r/2-1},\ldots,\theta^1)
        \end{align*}
        be a $r$-term Symmetric Geometric Progression centered at $r/2$, where $\theta > 1$.
\end{definition}

\begin{proposition}
        If $\gamma$ is a $SymGP(r, \theta)$ where $\theta > 1$, and $\gamma_{GM} < \mu_0^2$,
        then $\max(\gamma) \le \mu_0^4$.
	\label{thm:sgp}
\end{proposition}
\begin{proof}
        For simplicity we assume that $r$ is even. We have
        \begin{align*}
                \Gamma_r  & = (\theta^{\frac{(r/2)(r/2+1)}{2}})^2, \\
                 & = \theta^{(r/2)(r/2+1)} = \theta^{\frac{r(r+2)}{4}}
        \end{align*}
        \begin{align*}
                \gamma_{GM} = \Gamma_r^{1/r} = \theta^{\frac{r+2}{4}} < \mu_0^2.
        \end{align*}
        Squaring both sides, we get
        \begin{align*}
                 \theta^{r/2+1} < \mu_0^4, \\
                 \text{which implies } \theta^{r/2} < \frac{\mu_0^4}{\theta} < \mu_0^4
        \end{align*}
        since $\theta > 1$. Thus, $\max(\gamma) < \mu_0^4$.
\end{proof}
Note that even if $r$ in the above proposition is odd the result will be unaffected.

\begin{remark}
	We have seen above that for a Geometric Progression as well as a Symmetric Geometric Progression, we have $\max(\gamma) < \mu_0^4$. 
	Even if we use any of the following forms,
	\begin{itemize}
		\item[(i)] A Geometric Progression followed by a flat region,
		\item[(ii)] A Geometric Progression followed by a flat region that is followed by a decreasing Geometric Progression,
	\end{itemize}
	we will still get the same bound of $\mu_0^4$. 
	Since the proofs simply mimic the proof of~\cref{thm:gp}, they are not given here. 
\end{remark}

\begin{proposition}
	For any given instance of SSP, if we use~\cref{alg:ihm}, then the peak growth factor satisfies $\eta_{peak} < \mu_0^4$.
	\label{prop:firstbound}
\end{proposition}
\begin{proof}
	(By contradiction) Suppose that $\eta_k > \mu_0^4$ at some iteration $k \in [m]$ where $k > 4\log{\mu_0}$. 
	Then there must be at least 
	one edge $e_i \in G_0$ with more than $\mu_0^4$ disjoint parts that survived the FILTER steps. Since the number of distinct valid paths is less 
	than $\mu_0^4$ by~\cref{thm:gp}, we have more parts than distinct paths. By the
	Pigeon Hole Principle, there must be at least one part of $e_i$, with no valid paths through it. Such an edge will either be
	above or below the OL but does not intersect it. In other words, either the shortest path is larger than the edge length, or the longest path is negative.
	In either case the FILTER step (see~\cref{alg:filter}, step 9) removes the node by~\cref{prop:nodenec} providing a contradiction, and the claim follows.
\end{proof}
The above bound will be significantly improved later when we analyze the Zero Paths. For now, we already have a polynomial bound of 
$\mu_0^4 = \frac{n^{12}}{81}$ (from~\cref{rem:mu0}), and thus a polynomial time complexity for~\cref{alg:ihm}.

\subsection{Profile of $\gamma$}
Given an instance of the \emph{SSP}, what is the shape or profile of $\gamma:[n] \ra \mathbb{N}$ as a function of level? 
The short answer is that it depends on $n$, $m$ and the \emph{additive structure} of the given instance. 
For instance, if $m \le \log{\mu_0}$, then all the segments in the graph $\cg$ have a length at most $\mu_0$,
and the profile of $\gamma$ in the worst case is capped by a uniform distribution with a maximum of $\mu_0$. 
For larger $m$, if there is a high degree of additive structure, as in when ${\bf a} \in CP$ or ${\bf a} \in AP$, then many arc weights will be the same
and $\mu_r$ will be very small. Thus, the number of distinct configurations in $\cg$ will be small, and hence $\max(\gamma)$ will also be small.

For the case when $m > \log{\mu_0}$, and when there is no additive structure, the profile of $\gamma$ is a Gaussian with peak at $r = n/2$.
This should not be surprising, since after all the number of locally distinct paths to level $r$ is a subset of the number
of transformation paths to level $r$, which as we have already seen follows a Binomial distribution (see ~\cref{thm:binomial}).
However, we also noted earlier that
only a subset of the transformation paths result in valid wormhole paths for a given target value $T$, 
and even a smaller number of distinct OLs.  In other words, we have the sequence of inequalities
\begin{align*}
	\gamma_r \le \hat{\beta}_T(r) \le \beta_T(r) \le \beta(r) = \binom{n}{r}.
\end{align*}

While $\gamma_r$ is loosely bounded by the combinatorial quantity $\binom{n}{r}$ on one side, it is not determined by it. 
The \emph{main controlling factor} is the number of unique arc weights $\mu_r$ that reflects the degree of additive structure. 
As we have noted before, the simplest proof of this is furnished by the ${\bf a} \in CP$ case, where $\mu_r =1$ and $\max(\gamma)=1$. 
We will see other cases in~\cref{sec:adds}.

Thus, for any given instance of SSP with sufficiently large $m$ and no additive structure we have 
\begin{align*}
	\gamma_r = Ke^{\frac{-(r-n/2)^2}{2\sigma^2}}, \forall r \in [n],
\end{align*}
where $K=K(n)$ is a positive scaling factor, and $\sigma^2$ is the variance. We will determine bounds on $K$ and $\sigma^2$ below which depend on $\mu_0$.

For intermediate values of $m$ and differing degrees of additive structure, the profile of $\gamma$ can take several different shapes,
still bounded by Gaussian distribution from above.

\subsection{Asymptotic profile of $\gamma$}
Consider an instance $({\bf a},T)$ with large $m$ and large $\mu_r$. When $m$ is large, say $m > n$, the segment lengths in $\cg$ are large which allows for the possibility
to have more distinct points in each segment. When $\mu_r$ is large we have larger gain in the configuration graph allowing for more distinct paths. 
This is the case of primary interest, where we are interested in the profile of $\gamma$ as $n \ra \infty$.

\begin{definition}(Gaussian Distribution $Gauss(n,n/2,\sigma^2)$)
        For a positive integer $n$, let
        \begin{align}
		f(i) = Ke^{-\frac{(i-n/2)^2}{2\sigma^2}}, \text{ for } i \in [n],
		\label{eq:gauss}
        \end{align}
        be a Gaussian distribution with scaling factor $K > 0$, mean $n/2$ and variance $\sigma^2$.
\end{definition}

\begin{proposition}
	Let $\gamma \sim Gauss(n,n/2,\sigma^2)$ be a Gaussian distribution with scaling factor $K$.
	Then $K < \mu_0^3$ and $\sigma^2 = \frac{n^2}{8\log{K}}$.
\end{proposition}
\begin{proof}
	Taking $\log$ on both sides of~\cref{eq:gauss}, we have
\begin{align*}
        \ln(\gamma_i) = \ln{K} -\frac{(i-n/2)^2}{2\sigma^2}.
\end{align*}
	Summing over $i \in [n]$ on both sides,
\begin{align*}
        \sum_{i=1}^n \ln(\gamma_i) & = n\ln{K} -\sum_{i=1}^n\frac{(i-n/2)^2}{2\sigma^2}, \\
        & = n\ln{K} - \frac{1}{2\sigma^2} \frac{2n+n^3}{12}, \\
        & = n\ln{K} - (\frac{n(n^2+2)}{24\sigma^2}). 
\end{align*}
Applying the product inequality, we can write
\begin{align*}
        n\ln{K} -  (\frac{n(n^2+2)}{24\sigma^2}) < 2n\ln{\mu_0}, \\
        \ln{K} - (\frac{(n^2+2)}{24\sigma^2}) < 2\ln{\mu_0}, \\
        \ln{K}  < 2\ln{\mu_0} +  (\frac{(n^2+2)}{24\sigma^2}),
\end{align*}
leading to
\begin{align}
        K < \mu_0^2 e^{\frac{(n^2+2)}{24\sigma^2}}.
        \label{kineq}
\end{align}

To determine the variance we use the initial value of $\gamma_0 =1$, i.e.,
\begin{align*}
        1 = \gamma_0 & = K e^{-\frac{(-n/2)^2}{2\sigma^2}}, \\
        \implies \frac{1}{K} & = e^{-\frac{n^2}{8\sigma^2}}. 
\end{align*}
Taking $\log$ on both sides and simplifying, we get
\begin{align*}
        2\sigma^2 = \frac{n^2}{4\log{K}}.
\end{align*}
Using this in~\cref{kineq}, we get
\begin{align*}
        K  < \mu_0^2 e^{\frac{(n^2+2)}{24\sigma^2}} = \mu_0^2 e^{\frac{n^2+2}{12 (\frac{n^2}{4\log{K}})}}.
\end{align*}
After simplification, we get
\begin{align*}
	K < \mu_0^2 e^{(\frac{n^2+2}{n^2}) \frac{\log{K}}{3}} \approx \mu_0^2 e^{\frac{\log{K}}{3}}, \quad \text{(for large $n$)}.
\end{align*}
Taking $\log$ on both sides, we get
\begin{align*}
        \log{K} & < 2\log{\mu_0} + \frac{\log{K}}{3}, \\
	\text{and } K & < \mu_0^{3}.
\end{align*}
\end{proof}

Notice that this bound of $\max(\gamma) < \mu_0^3$ is better than what we obtained for a geometric progression above.

\begin{example}
	The asymptotic distribution of $\gamma$ for a random $n=36, m=72$ instance and a random $n=40, m=80$ instance with no additive structure is shown below
	in~\cref{fig:asym_gamma}.
\begin{figure}[!htbp]
\begin{center}
\epsfxsize=5in \epsfysize=3in {\epsfbox{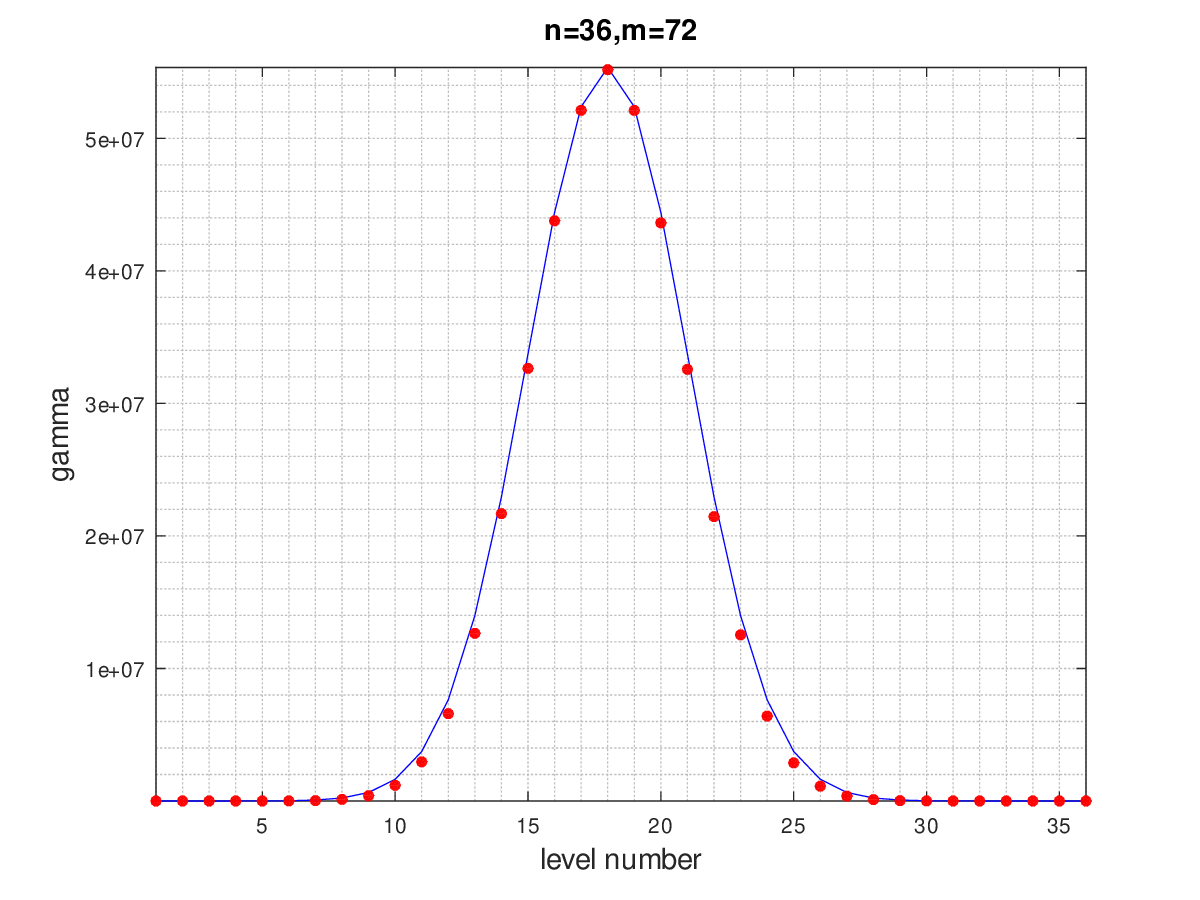}}
\epsfxsize=5in \epsfysize=3in {\epsfbox{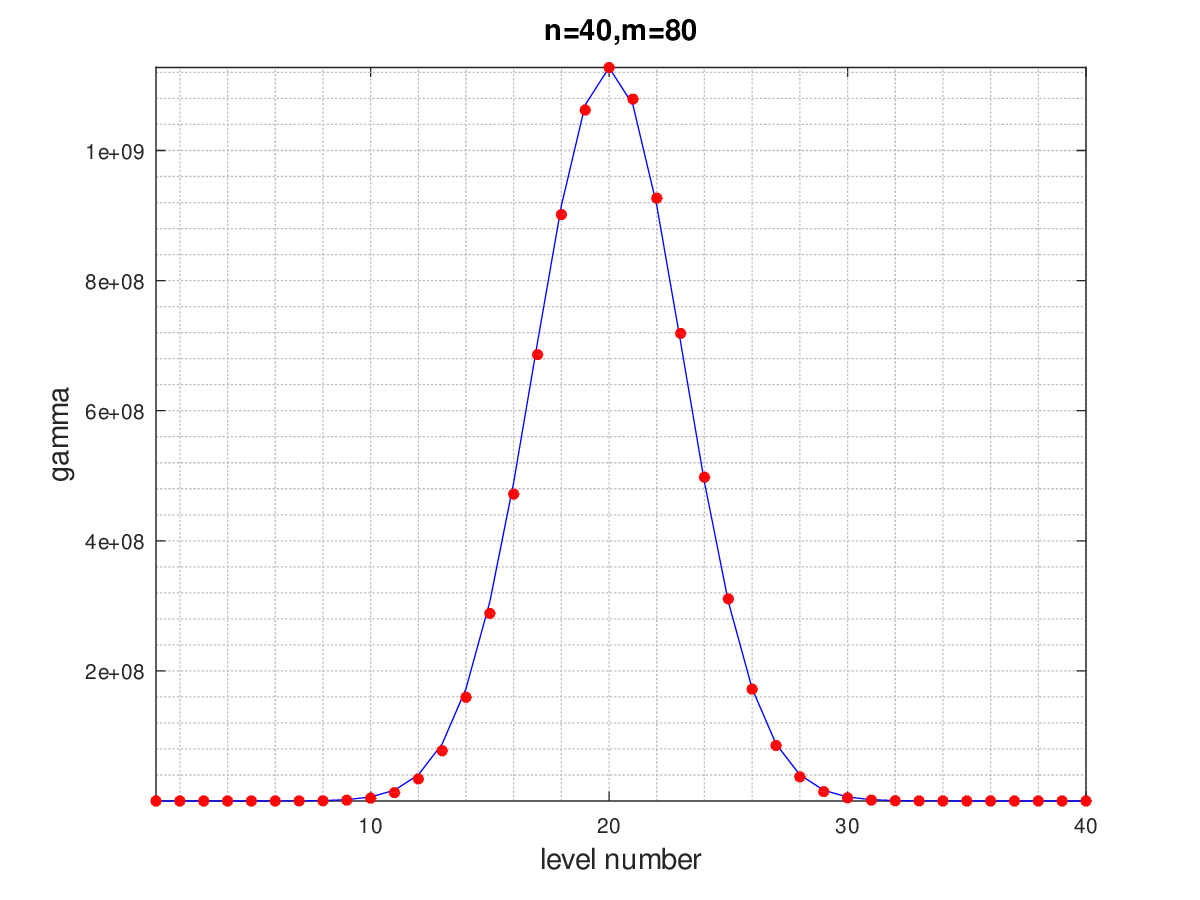}}
        \caption{The top plot is for a $n=36, m=72$ example, where the profile of $\gamma$ is shown by red dots.  
	The blue curve corresponds to a Gaussian fit where we used $\mu_0^{\delta}$ instead of $\mu_0^2$ as the gain in $\cj$ graph 
	(see~\cref{rem:alpha_less_than_one} and ~\cref{rem:actual_gain_j0}), where $\delta = 1.2225$ and $\mu_0 = \frac{n^3+3n^2-13n+6}{3} = 16,694$. 
        The bottom plot is for a $n=40, m=80$ example,  
	where we used $\delta = 1.385$ with $\mu_0 = \frac{n^3+3n^2-13n+6}{3} = 22,762$. As $n$ increases, we expect $\delta \ra 2$.
        }
\label{fig:asym_gamma}
\end{center}
\end{figure}
\end{example}

\begin{theorem}
	For an instance of SSP without any additive structure and with large $m$, the associated sequence $\gamma$ 
	satisfies the asymptotic profile given by 
	\begin{align} 
		\gamma_r & = \mu_0^{3} e^{-\frac{6(\log{\mu_0^2})(r-n/2)^2}{n^2}}, \quad \forall r \in [n],
	\end{align}
	and $\gamma_0 =1$.
\end{theorem}

\begin{corollary}
	For any instance of SSP with dimensions $(n,m)$, no edge equivalence class can intersect the orbital line at 
	more than $\mu_0^3$ distinct points.
\end{corollary}

The above corollary suggests that there is an \emph{occupancy limit} in the segments of $\cg$ graph! 
\subsection{Geometric Interpretation}
We provide a rough geometrical interpretation on what the above result means, in this section.
However, it is only a qualitative picture aimed at complementing the rigorous quantitative answer given above. 
\begin{definition}($ken((T-M,T])$)
	Consider the geometric structure as in~\cref{fig:mainexample} and assume that we have a window along the $y$-axis with limits $(T-M, T]$.
	We further assume that all the lower vertices of edges that intersect the OL are lighted up, and the rest of the points in the geometry are dark.
	When we look through this window horizontally, the number of distinct lighted points we see is called $ken((T-M,T])$, which is nothing but the
	projection of these points on to the $y$ axis.
\end{definition}

If the given instance ${\bf a} \in CP$, then it is clear that $ken((T-M,T]) = 1$ since all edges that intersect the OL do so at their lower vertex
if there is a solution, otherwise all of them will be offset by the same quantity (see~\cref{fig:thin_fat}).
Thus, there is only one lighted point we can see through the window.
If ${\bf a} \in AP$, then $ken((T-M,T]) \le n$ which is proved in~\cref{sec:adds}.
For a random instance the above result suggests that $ken((T-M,T]) \le n\mu_0^3$ since there are $n$ levels in the configuration graph. 
To understand this intuitively, recall from~\cref{subsec:seg_mult} that there are many segments which have an exponential path multiplicity, i.e, 
an exponential number of NDPs 
pass through them. Also note that in the case when all elements of ${\bf a}$ are distinct, the graph of NDPs slowly rises from left to right, and many subspaces 
go out of range with respect to a fixed OL. This becomes clear when we consider the more drastic example of ${\bf a} = {\bf b}$. In this case only one of either the 
left or right half subspaces intersect the given OL, but not both. To summarize, when there is lot of additive structure many points have the same projection; When
there is no additive structure the number of distinct intersections is fewer because of the rise in the graph, and segments having an exponential path multiplicity. 

\subsection{The case of Zero Paths}
So far we have been looking at the case of all valid paths on the OL. However, in the SSP we are primarily interested in paths
on the OL that lead to a solution, which is clearly a subset of all valid paths on the OL. While we could simply use the bounds
obtained for \emph{all valid paths}, there is benefit in considering the case of \emph{all zero paths} since it results
in a significant improvement in the bound for  $\max(\gamma)$.

In this section, we investigate the configuration graph where every path is a \emph{zero path}. 
Recall that a \emph{zero path} is a valid path that ends at the vertex point of an edge in $G_0$,  or equivalently the point $(j,0)$
in the $\cg$ for some $j \in [n]$. When only zero paths are plotted in the configuration graph, we call it as the ZP-configuration graph.

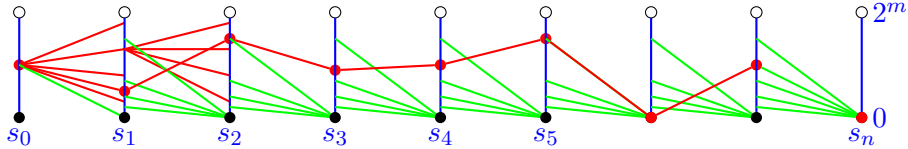
\begin{figure}[!htbp]
\begin{center}
\begin{tikzpicture}[scale=0.7,
	mydot/.style={
    circle,
    fill=white,
    draw,
    outer sep=0pt,
    inner sep=1.5pt
    }]
\node[outer sep=0pt,circle, fill,inner sep=1.5pt,label={}] (E0a) at (-2,0) {};
\node[mydot] (E0b) at (-2,2.0){};
\node[outer sep=0pt,circle, fill, red, inner sep=1.5pt, label={}] (E0c) at (-2,1.0) {};
        \draw[thick,blue] (E0a) node[below]{$s_0$} -- (E0b);
        \draw[thick,red] (-2,1.0) -- (0,1.8);
        \draw[thick,red] (-2,1.0) -- (0,1.3);
        \draw[thick,red] (-2,1.0) -- (0,0.8);
        \draw[thick,red] (-2,1.0) -- (0,0.3);

        \draw[thick,green] (-2,1.0) -- (0,0);
\node[outer sep=0pt,circle, fill,inner sep=1.5pt,label={}] (E1a) at (0,0) {};
\node[mydot] (E1b) at (0,2.0){};
\node[outer sep=0pt,circle, fill, red, inner sep=1.5pt, label={}] (E1c) at (0,0.5) {};
        \draw[thick,blue] (E1a) node[below]{$s_1$} -- (E1b);
        \draw[thick,red] (0,1.3) -- (2,1.8);
        \draw[thick,red] (0,1.3) -- (2,1.3);
        \draw[thick,red] (0,1.3) -- (2,0.8);
        \draw[thick,red] (0,1.3) -- (2,0.3);

        \draw[thick,green] (0,1.5) -- (2,0);
        \draw[thick,green] (0,0.7) -- (2,0);
        \draw[thick,green] (0,0.4) -- (2,0);
        \draw[thick,green] (0,0.2) -- (2,0);
\node[outer sep=0pt,circle, fill,inner sep=1.5pt,label={}] (E1a) at (2,0) {};
\node[mydot] (E1b) at (2,2.0){};
\node[outer sep=0pt,circle, fill, red, inner sep=1.5pt, label={}] (E2c) at (2,1.5) {};
        \draw[thick,blue] (E1a) node[below]{$s_2$} -- (E1b);

        \draw[thick,green] (2,1.5) -- (4,0);
        \draw[thick,green] (2,0.7) -- (4,0);
        \draw[thick,green] (2,0.4) -- (4,0);
        \draw[thick,green] (2,0.2) -- (4,0);
\node[outer sep=0pt,circle, fill,inner sep=1.5pt,label={}] (E1a) at (4,0) {};
\node[mydot] (E1b) at (4,2.0){};
\node[outer sep=0pt,circle, fill, red, inner sep=1.5pt, label={}] (E3c) at (4,0.9) {};
        \draw[thick,blue] (E1a) node[below]{$s_3$} -- (E1b);

        \draw[thick,green] (4,1.5) -- (6,0);
        \draw[thick,green] (4,0.7) -- (6,0);
        \draw[thick,green] (4,0.4) -- (6,0);
        \draw[thick,green] (4,0.2) -- (6,0);
\node[outer sep=0pt,circle, fill,inner sep=1.5pt,label={}] (E1a) at (6,0) {};
\node[mydot] (E1b) at (6,2.0){};
\node[outer sep=0pt,circle, fill, red, inner sep=1.5pt, label={}] (E4c) at (6,1.0) {};
        \draw[thick,blue] (E1a) node[below]{$s_4$} -- (E1b);

        \draw[thick,green] (6,1.5) -- (8,0);
        \draw[thick,green] (6,0.7) -- (8,0);
        \draw[thick,green] (6,0.4) -- (8,0);
        \draw[thick,green] (6,0.2) -- (8,0);
\node[outer sep=0pt,circle, fill,inner sep=1.5pt,label={}] (E1a) at (8,0) {};
\node[mydot] (E1b) at (8,2.0){};
\node[outer sep=0pt,circle, fill, red, inner sep=1.5pt, label={}] (E5c) at (8,1.5) {};
        \draw[thick,blue] (E1a) node[below]{$s_5$} -- (E1b);

        \draw[thick,green] (8,1.5) -- (10,0);
        \draw[thick,green] (8,0.7) -- (10,0);
        \draw[thick,green] (8,0.4) -- (10,0);
        \draw[thick,green] (8,0.2) -- (10,0);
\node[outer sep=0pt,circle, fill,inner sep=1.5pt,label={}] (E1a) at (10,0) {};
\node[mydot] (E1b) at (10,2.0){};
\node[outer sep=0pt,circle, fill, red, inner sep=1.5pt, label={}] (E6c) at (10,0) {};
        \draw[thick,blue] (E1a)  -- (E1b);

        \draw[thick,green] (10,1.5) -- (12,0);
        \draw[thick,green] (10,0.7) -- (12,0);
        \draw[thick,green] (10,0.4) -- (12,0);
        \draw[thick,green] (10,0.2) -- (12,0);
\node[outer sep=0pt,circle, fill,inner sep=1.5pt,label={}] (E1a) at (12,0) {};
\node[mydot] (E1b) at (12,2.0){};
\node[outer sep=0pt,circle, fill, red, inner sep=1.5pt, label={}] (E7c) at (12,1.0) {};
        \draw[thick,blue] (E1a)  -- (E1b);

        \draw[thick,green] (12,1.5) -- (14,0);
        \draw[thick,green] (12,0.7) -- (14,0);
        \draw[thick,green] (12,0.4) -- (14,0);
        \draw[thick,green] (12,0.2) -- (14,0);
\node[outer sep=0pt,circle, fill,inner sep=1.5pt,label={}] (E1a) at (14,0) {};
\node[mydot] (E1b) at (14,2.0){};
\node[outer sep=0pt,circle, fill, red, inner sep=1.5pt, label={}] (E8c) at (14,0) {};
        \draw[thick,blue] (E1a) node[below]{$s_n$} node[right] {0} -- (E1b) node[right] {$2^m$};

\draw[thick,red] (E0c) -- (E1c);
\draw[thick,red] (E1c) -- (E2c);
\draw[thick,red] (E2c) -- (E3c);
\draw[thick,red] (E3c) -- (E4c);
\draw[thick,red] (E4c) -- (E5c);
\draw[thick,red] (E5c) -- (E6c);
\draw[thick,red] (E6c) -- (E7c);
\draw[thick,green] (E7c) -- (E8c);
\end{tikzpicture}
\end{center}
\caption{A ZP-Configuration graph consisting of zero paths. Only some sample paths are shown. The incoming paths into lower
end of a segment from the previous segment are shown in green. The other paths are shown in red.}
\label{fig:zp_cplot}
\end{figure}

In~\cref{fig:zp_cplot} we show a rough sketch of the structure of ZeroPaths on the OL. Every path must end at $(r,0)$ for some $r \in [n]$.

In what follows and for the rest of this subsection we assume that we have \emph{only zero paths} in the configuration graph. As in the case of
all valid paths we use $F_r$, with $r \in[n]$ to denote the set of distinct points in each segment.

\begin{proposition}
	For a positive integer $r \in [n]$, let $(r,y) \in F_r$ be a point in segment $s_r$. Then either $y=0$ or there is at least one zero path from $(r,y)$ to
        some $(k,0)$ where $k \in \{r+1,r+2,\ldots,n\}$.
\end{proposition}
Since any zero path must end at a vertex point, the above proposition is clear.
Instead of considering zero paths of all lengths, we first consider zero paths of a fixed length $k$ for some $k \in [n]$ for convenience.
Later we will generalize to the case of all lengths.
\begin{definition}(Zero Paths of length $k$)
	For a given positive integer $k \in [n]$, let $ZP_k$ denote the set of all zero paths of length equal to $k$.  
	The associated point number sequence is denoted by 
	\begin{align*}
		(\zgamma(k,0),\zgamma(k,1),\ldots,\zgamma(k,k))
	\end{align*}
	where $\zgamma(k,0) = \gamma_0 = 1$. For the sequence we use double index, where the first index corresponds to path length
	and second the level number.  Since all these paths end at $(k,0) \in F_k$ and we have only
	zero paths, it follows that $\zgamma(k,k) = 1$. The set of zero paths $ZP_k$ have a common source point $(0,y_0) \in s_0$ and a
	common destination point $(k,0) \in s_k$. Let $|ZP_k|$ denote the number of zero paths of length $k$.
	\label{def:ZPdef}
\end{definition}

\begin{proposition}
	The sequence $(\zgamma(k,1), \zgamma(k,2), \ldots, \zgamma(k,k))$ satisfies
	\begin{itemize} 
		\item[(i)] $\zgamma(k,r+1)  \le \zgamma(k,r) \mu_r \le \zgamma(k,r) \mu_0$, for all $r \in [k-1]$
		\item[(ii)] $\zgamma(k,0)  = 1$ and  $\zgamma(k,k)  = 1$.
	\end{itemize}
\end{proposition}
\begin{proof} 
	Since the set $ZP_k$ is a subset of all valid paths, it follows that the sequence $(\zgamma_0, \ldots, \zgamma_k)$ obeys 
	the same growth bounds as in~\cref{ineq:G1}. This takes care of (i).
	Item (ii) directly follows from~\cref{def:ZPdef}. 
\end{proof}

\begin{proposition}
	The number of distinct zero paths of length $k$ is at most $\mu_0^{k/2}$.
	\label{prop:maxZPk}
\end{proposition}
\begin{proof}
	Consider the segment $s_r$ where  $r \in [k-1]$ is an intermediate level in the ZP-configuration graph.
	Then the number of paths into $s_r$ from the zeroth level is at most $\mu_0\mu_1\ldots\mu_{r-1} \le \mu_0^r$.
	The number of (reverse) paths into $s_r$ from the point $(k,0) \in s_k$ is at most ${\mu}_{k}\mu_{k-1}\ldots{\mu}_{r+1} \le (\mu_0)^{k-r}$.

	Since we are considering only zero paths, it follows that the each path from the left must match with a path from the right; otherwise
	we would have non-zero paths. Thus,
	\begin{align*}
		\text{\# distinct zero paths through $s_r$} \le \min(\mu_0^r, \mu_0^{k-r}).
	\end{align*}
	To maximize the number of distinct zero paths through $s_r$ we equate both to obtain that $r = k/2$.
	Thus, the maximum number of distinct zero paths of length $k$ is at most $\mu_0^{k/2}$.
\end{proof}

\begin{lemma}(Product Inequality for Zero Paths)
	Consider the configuration graph of the set of all zero paths of length $k \in [n]$ with source point $(0,y_0) \in F_0$ and destination
	point $(k,0) \in F_k$. Let the associated point number sequence be $\zgamma_k:= (\zgamma(k,0),\ldots,\zgamma(k,k))$ and let
	$\mathring{\Gamma}_k := \zgamma(k,0)\ldots\zgamma(k,k)$.
	Then $\mathring{\Gamma}_k < \mu_0^k$.
	\label{lemma:PI_zp}
\end{lemma}
\begin{proof}
	We will mimic the proof as in~\cref{cor:j0_bound}, by first treating the product $\mathring{\Gamma}_k = \zgamma(k,0)\ldots\zgamma(k,k)$ 
	as the number of distinct paths in a configuration graph 
	(denoted by $\cj$) where each point in a level is connected to all points in the neighboring levels.
	Furthermore, we have the arc growth factors (AGF defined in~\cref{def:AGF}) $\alpha_1, \alpha_2, \ldots, \alpha_{k-1}$. 
	Since the sequence $\zgamma$ follows the same growth bounds as the parent sequence $\gamma$ (the sequence corresponding to all
	valid paths), it follows from~\cref{prop:agf_j0} that $\alpha_i < \mu_0^2$ for all $i \in [k/2 - 1]$. For levels greater than
	or equal to $k/2$ the AGFs will be less than one since the paths converge to a single point in level $k$. We therefore set
	\begin{align*}
		\alpha_i & = \begin{cases}
			\mu_0^2 \text{ for all } i \in [k/2 - 1], \\
		1 \text{ for all } i \in \{k/2, k/2 + 1, \ldots, k-1\}.
		\end{cases}
	\end{align*}
	The number of arcs between the zeroth and first levels is at most $\mu_0$. With this the number of paths in $\cj$
	can be estimated using AGFs as
	\begin{align*} 
		\text{\# paths in $\cj$} \le \mu_0 (\mu_0^2)^{k/2 -1} = \mu_0^{k-1} < \mu_0^k. 
	\end{align*}
	Since $\mathring{\Gamma}_k$ represents all possible combinations (paths) it follows that
	\begin{align*}
		\mathring{\Gamma}_k < \mu_0^k.
	\end{align*}
\end{proof}

\begin{lemma}
	Consider the set of all zero paths of length $k$, with associated point number sequence $\zgamma_k:= (\zgamma(k,0),\ldots,\zgamma(k,k))$.
	Then $\max(\zgamma_k) < \mu_0^2$.
	\label{lem:zpk}
\end{lemma}
\begin{proof}
	To maximize $\zgamma(k,k/2)$ we consider the
	sequence $(\zgamma(k,1),\ldots,\zgamma(k,k))$ to be a $SymGP(k,\theta)$ for some $1 < \theta \le \mu_0$.
	In particular we let $\zgamma(k,i) = \theta^{\min(i,k-i)}$ for $i \in [k]$. From~\cref{lemma:PI_zp}
	we have $\mathring{\Gamma}_{k} < \mu_0^{k}$. Thus,
	\begin{align*}
		(\theta^1 \theta^2 \ldots \theta^{k/2})^2 & < \mu_0^{k}, \\
		\theta^{\frac{k(k+2)}{4}} & < \mu_0^{k}, \\
		\theta^{\frac{k+2}{2}} & < \mu_0^2, \text{(after taking the $k/2$ th root on both sides)} \\
		\theta^{k/2+1} & < \mu_0^2,  \\
		\theta^{k/2} & < \frac{\mu_0^2}{\theta} < \mu_0^2, \text{ since $\theta > 1$}.
	\end{align*}
	Thus, the maximum element $\zgamma(k,k/2) < \mu_0^2$.
\end{proof}

\begin{theorem}
	Consider the set of all zero paths of all lengths $k \in [n]$ for a given instance of the SSP, denoted by $ZP_{\ast}$. Let the 
	associated point number sequence by $\bg := (\bg_0,\bg_1,\bg_2,\ldots,\bg_n)$, where $\bg_0  = 1$. Then
	\begin{align*}
		\max(\bg) < n\mu_0^2.
	\end{align*}
\end{theorem}
\begin{proof}
	It is clear that
	\begin{align*}
		ZP_{\ast} = \cup_{k=1}^n ZP_k.
	\end{align*}
	Since for each fixed $k \in [n]$, the maximum number of distinct points in $s_{k/2}$ is less than $\mu_0^2$, it follows that
	the number of distinct points in any segment will be at most $n$ times $\mu_0^2$ and the claim follows.
\end{proof}
The following theorem follows from the above result.
\begin{theorem}
	The maximum number of distinct zero paths through any node $e_i \in G_0$, is at most $\min(2^m, n\mu_0^2)$.
	\label{thm:zp_bound}
\end{theorem}
Note that we cannot claim a Gaussian profile for $\gamma$ in the case of zero paths since there may not even be a solution to a given instance. As a result,
we used the symmetric geometric progression to get a worst case estimate in~\cref{lem:zpk}. 
Since we are bounding the number of distinct zero paths into any node using the above result (which includes contributions from all nodes in a given level
represented in a single segment of the configuration graph), it should be clear that the above bound is not tight.

\section{Main Results \& Complexity Calculations}
\label{sec:guru}
In this section we prove the main claims of the paper, which follow easily given all the 
results up to this point. Recall that the complexity of the algorithm depends on the bound for $\eta_{peak}$(~\cref{prop:ihmcomp}). 
We have already proved a bound in~\cref{prop:firstbound} using the bound for maximum number of valid paths through a node. 
We provide a much better bound next using the results of our analysis on distinct zero paths.

\begin{theorem}
	$\eta_{peak} \le \min(2^m, \frac{n^7}{9})$.
\end{theorem}
\begin{proof}(Sketch)
	We have already shown that $\eta_{peak} < \mu_0^4$ (in~\cref{prop:firstbound}) using the bound for the maximum number of distinct valid paths through
	any node. This bound is too loose for $\eta_{peak}$, since the FILTER step of~\cref{alg:ihm} not only removes nodes that have only invalid paths into them,
	but also removes nodes that have valid but only non-zero paths.
	Only those nodes that can support zero paths are preserved by the FILTER step.  To show the claim we make the following arguments. 

	There are three types of paths in the orbital graph: (1) Invalid paths, (2) Valid but non-zero paths and
	(3) Zero paths. For an arbitrary node $e_i \in G_0$, all the three types of paths can co-exist and go through the node. We know from~\cref{thm:zp_bound}
	that the maximum number of distinct zero paths cannot exceed $n\mu_0^2$. Thus, after $k := \myceil{\log{(n\mu_0^2)}}$ refinements, we can expect, in general, that
	that all the distinct zero paths are separated into the $2^k$ different parts of $e_i$. Each part may have paths of types (1) and (2) along with the zero
	path. Let $e_j$ be one such part of $e_i$, that has all the three types of paths. In the next refinement, the node $e_j$ is split into $e_j'$ and $e_j''$.
	The single distinct zero path can only be in one of $e_j'$ or $e_j''$ but not both. The other part will now have only paths of types (1) or (2). If it
	is of type (1), the node is removed since it does not intersect the OL. If it is of type (2), the node is either already labeled FALSE or will now be labeled
	FALSE. The FILTER routine disconnects this node from the destination node $e_{\infty}$, and will be removed when all the FALSE LEAF nodes are removed
	during the $n$ loops of the FILTER procedure (see~\cref{rem:filter}). It is possible for a few FALSE nodes that do not support zero paths to survive the
	current FILTER step due to the way they are connected to next level, but the uncertainty will be removed in one or two refinements. These small fluctuations
	do not affect, since one can see that the nodes in the first few levels of the orbital graph
	will have only a very small number of surviving parts (and they cannot increase with more refinements), as the number of paths into them is very small. 
	In other words, whatever the amount by which the number of nodes increases due to 
	the REFINE step beyond iteration $k$, about the same number get removed in general by the FILTER procedure, since the distinct zero paths have been separated. 
	Thus, the growth factor of the graph cannot increase beyond $n\mu_0^2 = \frac{n^7}{9}$ (from~\cref{rem:mu0}). 
	Taking into account the edge length limit of $2^m$ also, the claim follows.
\end{proof}

\begin{corollary}
	For any positive number $n$, there is a positive number $m_0 = \lceil 7\log{n} \rceil$ such that for all $m > m_0$, size of elements in ${\bf a}$
	do not have any affect on the complexity of the SSP.
\end{corollary}

\begin{corollary}
	The maximum number of nodes in the refined graph $G_m$ is at most $\frac{n^{11}}{216}$.
	The total number of arcs in $G_m$ is at most $\frac{n^{13}}{6480}$.
\end{corollary}

\begin{proof}
	The maximum number of nodes in the refined graph $G_m$ is,
	\begin{align*}
		|V(G_m)| \le \eta_{peak} |V_0| = \min(2^m, \frac{n^7}{9}) \frac{n^4}{24} \le \frac{n^{11}}{216}.
	\end{align*}
	Every arc in the initial graph $G_0$ can become at most $\eta_{peak}$ arcs in $G_m$.
	Thus, 
	\begin{align*}
		|E(G_m)| \le \eta_{peak} |E_0| = \min(2^m,\frac{n^7}{9}) \frac{n^6}{720} \le \frac{n^{13}}{6480}.
	\end{align*}
\end{proof}

With this we finally have the main result:
\begin{theorem}
	${\mathcal P} = \mathcal{NP}$.
\end{theorem}

From a result of Stockmeyer~\cite{ls77}, the following corollary follows, where $\mathcal{PH}$ is the polynomial hierarchy 
(the set of classes above $\mathcal{NP}$ and \emph{co}-$\mathcal{NP}$ with increasing number of alternating quantifiers $\exists$ and $\forall$).
\begin{corollary}
	$\mathcal{P} = \mathcal{PH}$.
\end{corollary}

\subsection{Finding the number of solutions}
With our algorithm, finding the number of solutions to a given instance is easy, and practically free. Once we obtain the \emph{Solution Graph} $G_m$,
we use a possibly well-known trick from Graph Theory, to find the number of solutions $N_{sols}$. If $G_m = \emptyset$, then
$N_{sols} = 0$. Otherwise, we follow these steps below.
\begin{itemize}
	\item[(i)] Attach a counter $\xi_i$ to each node $e_i \in G_m$.
	\item[(ii)] Set $\xi_0 = 1$ for the root node.
	\item[(iii)] For each node $e_i \in G_m$, let
		\begin{align*}
			\xi_i = \sum_{j \in parents(e_i)} \xi_j.
		\end{align*}
	\item[(iv)] Traverse the graph using a Breadth-First-Search (BFS), applying (iii) at each node.
	\item[(v)] Recall that the node $e_{\infty}$ is connected to all the \emph{TRUE} nodes in the graph, and that
		it acts as a collector. Thus, the number of solutions $N_{sols}$ to the given instance is simply
		\begin{align*}
			N_{sols} = \xi_{\infty}.
		\end{align*}
\end{itemize}

Since $G_m$ is polynomial sized, the complexity of running the above steps is simply $O(|V_m|+|E_m|)$, which is a polynomial. 
This proves the third claim of this paper.
\begin{theorem}
	$\mathcal{FP} = \# {\mathcal P}$.
\end{theorem}

Since $\mathcal{P}$ with a $\#\mathcal{P}$ oracle is simply $\mathcal{P}$ by the above theorem, we have that
\begin{corollary}
	$\mathcal{P} = \mathcal{P}^{\#\mathcal{P}}$.
\end{corollary}

Bernstein and Vazirani~\cite{bv93}, defined the complexity class $\mathcal{BQP}$ of problems that can be solved
in polynomial time using quantum computers. They further showed the containment that
$\mathcal{BQP} \subseteq \mathcal{P^{\#P}}$.  Therefore the following (somewhat surprising) result readily follows.
\begin{corollary}
	$\mathcal{BQP} \subseteq \mathcal{P}$.
\end{corollary}

\subsection{Classifying the solutions}
The layered structure of the  graph $G_0$ (or $G_m$), with a unique $5$-tuple identification for the nodes,
provides a natural classification of the solutions. This classification may be used possibly in several ways, 
for example: It is quite conceivable that in some problem (that could be reduced to \ssp), additional 
requirements among equivalent solutions, could be important. Suppose that there are two solutions, with the first
zero path terminating at $e_i \in G_0$, and the second terminating at $e_j \in G_0$. While both solutions solve the
original problem, one solution may be \emph{less expensive or easier} to implement than the other. Mapping the solution 
classes back to the original problem will provide the needed information. A second application is where one is interested
in a particular equivalence class (e.g in level $10$, on a $q_7$ curve, etc.), and whether there are any solutions into it.
In such a case, we can consider the smaller graph reachable from $e_0$ in $10$ hops with a fixed destination node, and compute paths.

\subsection{Numerical results}
We provided several numerical examples to illustrate the workings of the algorithm, and also point out some
applications of the algorithm to some combinatorial problems. These are put 
at the end of the paper (see~\cref{sec:appendix}) so as to maintain the flow of the paper. We urge the reader
to take a look at it,  as it helps to understand the algorithm better.

\section{Instances \& Additive Structure}
\label{sec:adds}

We have seen that given any instance, we can find the number of solutions using a polynomial amount of resources. 
This raises the following questions:
\begin{question}
Does every instance $({\bf a}, T)$ have the same complexity? If not, then what controls the complexity?
\end{question}
We have already shown that the \emph{degree of additive disorder} (equivalently \emph{degree of additive structure} in an inverse way)
of ${\bf a}$, as measured by the number of unique arc weights $\mu_r$ where $r \in [n]$, is the main determiner 
of the \emph{complexity}. We have shown that when $\mu_r$ is small, such as in the case of ${\bf a} \in CP$, the complexity is low.
In this section we consider more commonly used measures of \emph{additive structure} given below. Then we consider some instances with varying degrees of 
additive structure and provide the complexity estimates, while also attempting to make a connection between the \emph{additive structure} and \emph{geometric structure}.

\subsection{Two quantities of interest}
There does not seem to be a standard definition for the \emph{degree of additive structure}, but several possibilities are listed in the book
by Tao and Vu~\cite{tv06}.
To understand the degree of \emph{additive structure}, we consider two quantities mentioned in~\cite{tv06}, that are appropriate for the \ssp. 

\begin{definition}(Unique sums)
Given all the $2^n$ subset sums for a given ${\bf a}$, we denote the set of all unique sums by $\mcu$,
and the number of unique sums by $U = |\mcu|$. Clearly $1 \le U \le {\min}(2^n, A_n)$.
\end{definition}

\begin{definition}(Maximum Concentration)
Given all the $2^n$ subset sums for a given ${\bf a}$, the maximum concentration denoted by $N_{LO}$ is defined as
\begin{equation*}
N_{LO} = {\max}_{s \in \mcu} |\{{\bf x} \in \{0,1\}^n : {\bf a} \cdot {\bf x} = s \}|
\end{equation*}
i.e., the maximum number of subsets that have the same sum. This type of quantity was first studied by Littlewood and Offord in what is
	now called \emph{Littlewood-Offord Theory}~\cite{tv06}, and hence the subscript $_{LO}$.
\end{definition}

It would be useful to generalize the notion of \emph{Maximum Concentration} to simply \emph{concentration} for each unique sum
$s \in \mcu$. For this, we order the unique sums in increasing order, and denote and define  $L_i$ as the set of all subsets that
have the same $i$th unique sum. Formally,
\begin{equation*}
L_i := \{{\bf x} \in \{0,1\}^n : {\bf a} \cdot {\bf x} = s_i, s_i \in \mcu\}
\end{equation*}
where $s_i$ is the $i$th unique sum. Thus, $N_{LO} = {\max}_i|L_i|$. Clearly, we have
\begin{equation*}
{\sum}_{i=1}^U |L_i| = 2^n.
\end{equation*}

Note that $\mathcal{L} := \{|L_1|, |L_2|, \ldots, |L_{U-1}|,|L_U|\}$ is nothing but the distribution of sums in $[0 , A_n]$ into
$U$ bins, and this captures more information about the point set compared to simply $N_{LO}$. 

The quantities $U$ and $N_{LO}$ are inversely proportional. If $U$ is large, $N_{LO}$ will be small and vice versa. 
When there is additive structure, $|L_i|$s are expected to be large and $U$ is expected to be small. 
Given an instance ${\bf a}$, the subset sums will be distributed according to $L$. For a given $T$, the only possible
number of solutions is from the set $\{0, |L_1|, |L_2|, \ldots, |L_{U-1}|, |L_U|\}$ and nothing else.
For a given instance $({\bf a}, T)$, suppose that there are $|L_i|$ solutions for some $i \in [U]$. What is the relation between
all these subsets apart from being solutions? One ready answer is provided by the \emph{Solution Graph}, where all of these are
connected to a common root node.

\subsection{Dependence of Solution Graph size  on $U$}
Next, we show how the size of the final graph, in particular $|V_m|$ can be related to the number of unique sums $U$.
The geometric structure we discussed earlier enables this relation. Although it is a rather loose bound, it sheds
light on how the \emph{additive structure} (via $U$) affects the complexity.
\begin{lemma}
\label{lem:vmbound}
	$|V_m| \le \min (\frac{2U}{n(n+1)},2^m)|V_0|$.
\end{lemma}

\begin{proof} 
	Let ${\zeta}_k$ be the number of unit intervals of a 
	node $e_k \in V_0$ that appear in the final graph. Then it can be seen that 
	\begin{equation*} 
		|V_m| = {\sum}_{k=1}^{|V_0|}{\zeta}_k. 
	\end{equation*} 
	Note that for each node $e_k$, we have $|e_k| \le 2^m$. 
	Since each edge is on a non-decreasing path from $(0,0)$ to $(B_n,A_n)$ with $n(n+1)/2$ points, 
	every unique intersection point with the OL contributes $n(n+1)/2$ points (i.e., the points on the curve) to 
	the count $U$ of unique sums.  
	Since there are $U$ unique sums,  it follows that ${\zeta}_k n(n+1)/2 \le U$. Alternatively, 
	\begin{equation*} 
		{\zeta}_k \le \frac{2U}{n(n+1)}. 
	\end{equation*} 
	Using the fact that ${\zeta}_k \le |e_k|$, we get 
	\begin{equation*} 
		{\zeta}_k \le \min (\frac{2U}{n(n+1)}, 2^m). 
	\end{equation*} 
	Thus,
	\begin{equation*} 
		|V_m| = {\sum}_{k=1}^{|V_0|}{\zeta}_k \le {\sum}_{k=1}^{|V_0|}\min (\frac{2U}{n(n+1)},2^m) \\ 
		= \min (\frac{2U}{n(n+1)},2^m) |V_0|, 
	\end{equation*} 
	and the claim is proved.
\end{proof}

\begin{corollary}
	For a given instance $({\bf a},T)$, if the number of unique sums $U$ is bounded by a polynomial, then
	the size of the solution graph, i.e, $|G_m|$ is also bounded by a polynomial.
\end{corollary}

\begin{corollary}
	Let ${\bf a} \in CP$. Then given $({\bf a}, T)$, we have $|V_m| \le \frac{2|V_0|}{n}$.
\end{corollary}
Since  $U = n+1$ when ${\bf a} \in CP$, the above corollary follows from~\cref{lem:vmbound}.

\begin{proposition}
Let ${\bf a} = AP(n,c,d)$. Then the number of unique sums $U$ in the solution space is at most $\frac{n^3+5n+6}{6}$.
\label{prop:ap}
\end{proposition}
\begin{proof}
We have $a = \{c+(n-1)d, c+(n-2)d, \ldots, c+2d, c+d, c\}$.
A sum with $k$ elements has a sum in the range $[kc + \frac{(k-1)k}{2}d \quad kc + (\frac{(n-1)n}{2}-\frac{(n-k-1)(n-k)}{2})d]$. Thus the
number of distinct sums a $k$ element subset can have is $\frac{(n-k-1)(n-k)}{2} - \frac{(k-1)k}{2} + 1 = kn - k^2 + 1$. To get the total
number of unique sums we carry out the summation as $k$ varies from $1$ to $n$ to get
\begin{equation*}
{\sum}_{k=1}^{n}(kn-k^2+1) = \frac{n^2(n+1)}{2} - \frac{n(n+1)(2n+1)}{6} + n = \frac{n^3+5n}{6}.
\end{equation*}
Now including the empty set sum of $0$, we have $\frac{n^3+5n}{6} + 1 = \frac{n^3+5n+6}{6}$ distinct sums as claimed.
\end{proof}

From~\cref{lem:vmbound}, it follows that
\begin{align*}
	|V_m| & \le \frac{2U}{n(n+1)}|V_0| \\
	& \le \frac{n^3+5n+6}{3n^2 + 3n}|V_0| \approx \frac{n}{3}|V_0|.
\end{align*}
Again the peak growth factor is significantly smaller than the worst case complexity.

The above two cases can be generalized to cases such as $2$-$CP$ (disjoint union of two $CP$s where $U = O(n^2)$) 
and $2$-$AP$ (disjoint union of two $AP$s where $U = O(n^6)$). As $U$ is made to increase by reducing the 
additive structure (by the union of a larger number of CPs or APs), will the complexity continue to increase?
No, since we showed the maximum growth factor $\eta_{peak} \le  n^7/9$. So, for all instances with $U > O(n^7)$,
we can expect no further increase. 

\subsection{A Special Case: ${\bf a} = {\bf b}$}

Next, let's consider the special case of a Geometric Progression where ${\bf a} = {\bf b}$.
In this case, there is no additive structure at all, since the elements are related by geometric growth. We have
$U = 2^n$ and  $N_{LO}  = 1$.
The loose bound of~\cref{lem:vmbound} is not useful in this case as it suggests that this could be a worst case.
However, it turns out that this special case is trivially easy with regard to our algorithm.
In this case, first note that all the NDPs fall onto a single straight line connecting the points $(0,0)$ and $(B_n,B_n)$.
See the bottom plot of~\cref{fig:thin_fat}.
Consider some edge $e_i \in q_k$ through which the OL passes through. This edge may interact with several edges of $P_{k-1}$ in
the next level. However, only one of these nodes will have a valid path. All others will have invalid paths, i.e., paths outside
the bounds of the corresponding edges. This is so, since ${\bf a = b}$ is \emph{super-increasing} ($a_i > {\sum}_{j=1}^{i-1}a_j$, for all $i > 1$), 
leading to the non-overlap of subspaces $S_j$ at $(0,0)$ and $S_j$ at $(b_{j+1},a_{j+1})$ for any $j \in [n-1]$. 
As a result, the FILTER step prior to the iterations in Algorithm $5$, will remove all paths except the solution path. 
The final graph is just a single path consisting of a sequence of at most $n$ nodes, i.e., $|V_m| \le n$.

\subsection{A snapshot of instances}
\begin{figure}[!htbp]
\begin{center}
      \epsfxsize=7in \epsfysize=4.7in {\epsfbox{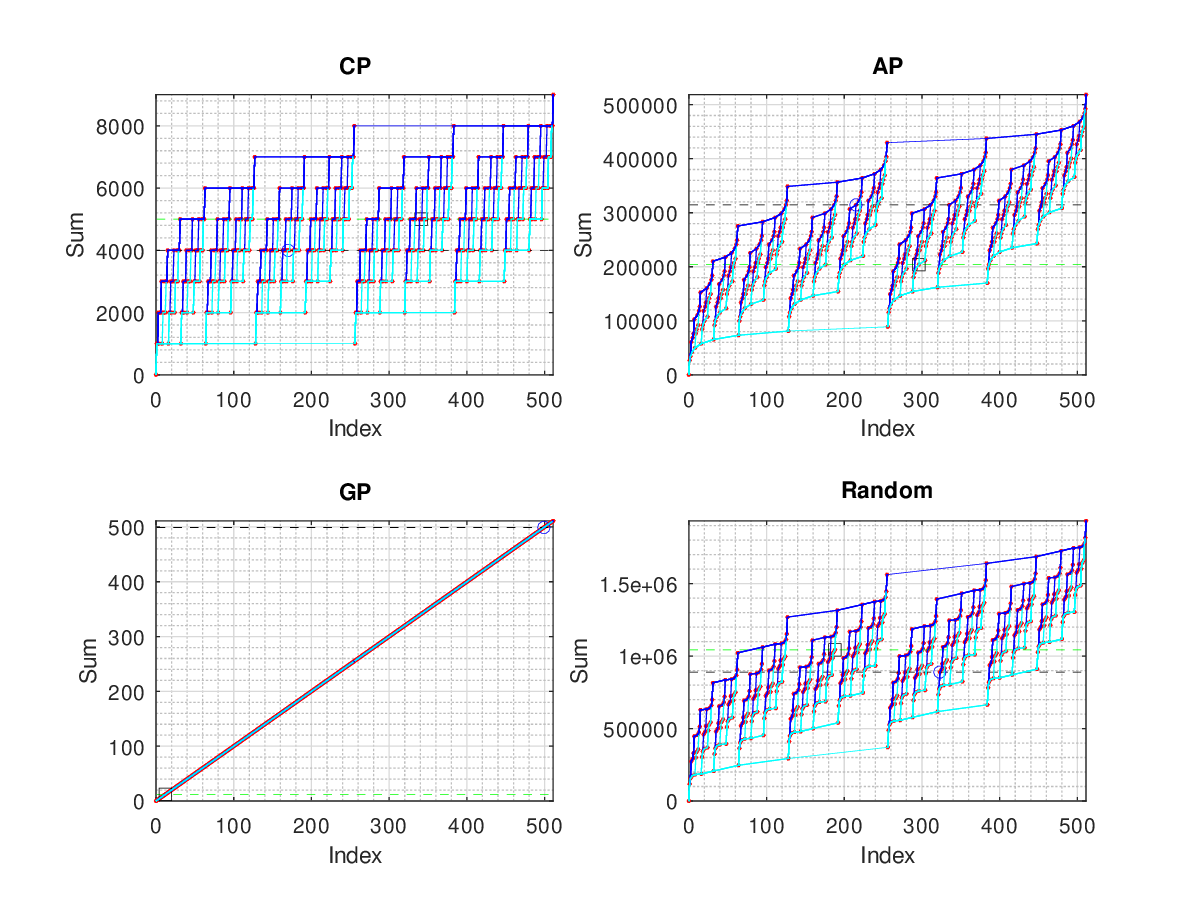}}
        \caption{The geometric structures for four instances with varying degrees of additive structure.
	For visual clarity we used small sized instances ($n = 9$).  Notice the shearing of the geometric structure in the \emph{random} case compared to a $CP$. 
        }
       \label{fig:thin_fat}
\end{center}
\end{figure}

A rough understanding of the \emph{degree of additive structure} can be obtained visually by looking at the geometry of
the instance. Of course this is useful only for small instances.
When the degree of additive structure is high the geometric structure is wide and short, and when the additive structure is low 
it is thin and long. Below we show the geometric structures of four instances (a $CP$, an $AP$, a $GP$ and a random case)
to provide a visual understanding at a glance in~\cref{fig:thin_fat}.
Comparing the random case to that of $CP$, we notice a \emph{shearing} of the geometric structure. 
This \emph{shearing} provides a rough qualitative reason on why the distinct local intersections are polynomially bounded; if we view
the entire point set as a collection of smaller subspaces, for a given OL, many subspaces go out of range due to shearing. Thus the number
of distinct intersections is smaller.

\newpage
\section{Epilogue}
\label{sec:epi}

In this section we make some philosophical observations on the main topic of the paper, and
offer some personal opinions. It should be emphasized that these are not precise statements,
but at best some rambling personal thoughts.
These are divided into two parts: in the first part we discuss some reasons that
\emph{likely influenced} the majority opinion on the $\mathcal{P}$ vs. $\mathcal{NP}$ question,
and in the second part the historical antecedents of the problem are discussed.

\subsection{A combinatorial facade and a reconciliation}
In many mathematical problems and particularly in combinatorics, the power set of some relevant base set
is the universe.  Whenever we think of all possible combinations (of elements of a set), we are implicitly
thinking of them as \emph{equal} in some sense.  Each is just a combination out of the many possibilities, and
contributing an unit to the total count. 
We are often required to estimate the number of combinations
(satisfying some property) associated with a problem, but less concerned with how these are related to each other in the first place.
This is particularly true when faced with a tremendously large number of combinations.
In other words, we are intuitively coaxed into believing that either \emph{no structure is possible}, or it is \emph{futile to look for a structure}, 
but be content with the count.  Thus, the \emph{combinatorial explosion} eclipses any attempts to understand the structure. 

For any \npc\ee problem the \emph{combinatorial aspect} presents itself readily 
by its hypercube with an exponential number of vertices.
From the phrasing of the \ssp, one sees it as an \emph{one-dimensional problem} with the $2^n$ scalar numbers corresponding to the subset sums. 
For any given instance, at first glance we can rule out only a small number of vertices from the search space; 
after that almost all of the hypercube still remains.  Furthermore, the hypercube doesn't 
reveal why one subset is more \emph{likely} than another to be the solution. 
All of them appear equally likely and
one is led to believe that the structure is \emph{indeed} that of a combinatorial type. 
Perhaps these are some reasons why so many researchers intuitively felt that an exhaustive search cannot be avoided.

In our quest to understand the structure, we converted the \ssp\ee into a \emph{two-dimensional problem} by attaching a unique index to each element of the problem. 
With this the hypercube is transformed into a point set in the Cartesian plane, and suddenly we see some structure that was not evident in the \emph{one-dimensional}
view\footnote{Regarding the power of an extra dimension, one can recall from Calculus how several integrals of a real variable become amenable to exact evaluation \emph{only}
when transferred to the complex plane.}. 
Further efforts to fully understand the point set enabled us to characterize the structure of the \ssp.
We were able to find enough structure so that we can always solve the decision problem.
Once we understood the structure we realized that it is the much smaller \emph{relational aspect} that is important, 
and it was being shielded all along by a seemingly impregnable combinatorial facade.

\subsection{Historical significance}
The \emph{Theory of Computing} is a vast subject spanning centuries, with many crises and triumphs, and featuring contributions from some of the greatest minds.
There is no way to trace a short path through history without missing out on many important developments that eventually paved the way to 
the ${\mathcal P}$ vs. ${\mathcal NP}$ question.
In this short section we only sample some key moments through history to provide an overly simplistic and rough historical perspective in a highly informal way.

The great mathematician David Hilbert, in 1928~\cite{ha28},  asked the question: ``Is there a finite procedure to determine the truth or falsity of
any mathematical statement''? What Hilbert had in mind was a ``mechanical process'' of applying inference rules
to a set of axioms and checking the truth of a given statement.
In modern terminology, his question is about the existence of an \emph{algorithm}
to \emph{decide} the truth of any mathematical statement in a given axiomatic system. This famous question is known as the \emph{Entsheidungsproblem} (decision
problem). Hilbert believed that every mathematical question must have a definite answer, either by a solution or by a proof of impossibility.
His guiding view was: \emph{``Wir m{\"u}ssen wissen, wir werden wissen''} (We must know, we will know),
that there can be no ignorance in mathematics, and that we can know everything.

It turns out that Hilbert was not the first to have such a grand ambition. More than 250 years prior, the great mathematician 
and philosopher Gottfried Wilhelm Leibniz, dreamt
of mechanization of reason, which he called a ``wonderful idea''(see Davis~\cite{md2018}). 
According to Davis~\cite{md2018}, Leibniz wanted to collect all human knowledge, extract key concepts and attach
a distinct character(symbol) to each. Then, using rules of deductive logic one could mechanically manipulate the symbols 
to arrive at truths.
This extraordinarily ambitious project, could not be realized particularly given the times of Leibniz. But evidently Leibniz continued thinking 
and writing about it till the end.

However, the great logician Kurt G\"odel in 1931 showed by his famous \emph{incompleteness theorems} that in any consistent axiomatic system, there exist
truths that are not provable (see for e.g the translation~\cite{kg92}). Given an axiomatic system, let $J_1$ be the set of all statements that are provable,
and let $J_2$ be the set of statements that are not provable. Then G\"odel's theorems imply $J_2 \neq \emptyset$,
but do not preclude the existence of a classification procedure, which given a statement $s$ can tell if $s \in J_1$ or $s \in J_2$.
Turing in 1936~\cite{at36}, succeeded in formalizing the ``mechanical process'' of Hilbert, by defining \emph{computability}
using \emph{Turing machines}. 
Our current digital world has been made possible by this singularly powerful result of Turing!
Turing  showed the equivalence of \emph{computability} and \emph{halting} of \emph{Turing Machines(TMs)}.
Turing's result says that the set $J_1$ corresponds to the set of \emph{TM}s that \emph{halt} in a finite
amount of time. In a way, the same sets are relabeled into 
$J_1 \cong \emph{HALT}$, and $J_2 \cong \emph{DoNotHALT}$.  He also showed that given a statement, whether the
corresponding \emph{TM} halts or not, is \emph{undecidable}.  In other words, given a statement, there is no
procedure that can tell us if it is provable or not. These results effectively put an end to Hilbert's dream of placing all of mathematics in
a formal axiomatic system. \\

Out of the problems that are decidable, we are interested in those problems where a claimed solution can be verified in polynomial
time, which is the class $\mathcal{NP}$. 
Cook~\cite{c71} discovered the first $\mathcal{NP}$-complete problem in \emph{Boolean Satisfiability Problem (SAT)} 
which has the property that every problem in $\mathcal{NP}$ can be reduced to it in polynomial time, and which is atleast as hard as any
problem in $\mathcal{NP}$. 
The class $\mathcal{NP}$, as it turns out is extremely rich in problems of interest to humanity
covering diverse areas such as circuit design, mathematics, computer science, industry, artificial intelligence, cryptography, biology, 
economics, medicine, physics and more~\cite{aw19,mm15}.
Cook's amazing discovery reduced thinking about a multitude of problems in $\mathcal{NP}$ to a single $\mathcal{NP}$-complete  problem!
Thus, if we can efficiently solve \emph{SAT} or equivalently any one of the thousands of other \npc\ee problems, we can
solve all problems in $\mathcal{NP}$ also efficiently.
What is the worst case complexity of \emph{SAT}?  Thus, the $\mathcal{P}$ vs. $\mathcal{NP}$ problem was born. 

All the \npc\ee problems admit trivial exponential algorithms involving a brute force search over the exponential search space.
As many efforts over several decades to find efficient algorithms to \npc\ee problems weren't successful researchers started to believe that exhaustive search cannot be
avoided. Amazingly enough, Kurt G\"{o}del anticipated the $\mathcal{P}$ vs. $\mathcal{NP}$ problem $15$ years earlier in $1956$ even before these classes were
formally defined. He viewed it as a bounded version of Hilbert's \emph{Entsheidungsproblem}\cite{sa16}.
In his letter to John von Neumann, G\"{o}del wrote: ``Now, it seems to me, however, to be totally within the realm of possibility 
that $\varphi(n)$ grows slowly''\cite{ms92}. Here $\varphi(n)$ refers to the number of steps taken by a Turing machine to decide the truth of a propositional
formula limiting to proofs of length $n$. We can understand this to be the time complexity of a \npc\ee problem in equivalent terms.

With our result that $\mathcal{P} = \mathcal{NP}$, certainly G\"{o}del's expectation has come true. 
However, G\"{o}del also hoped for a small degree polynomial $n^c$ where $c \le 2$. We are not there yet, but with adequate improvements to the
complexity, there is hope to realize the ``wonderful idea'' of Leibniz and fulfill Hilbert's quest.
We await to see what the future beholds!  
Going back to Hilbert's optimism, our result serves as a vindication to a degree (after the negative results of G\"{o}del and Turing), 
showing that he was correct after all as \emph{we can know all the knowable truths}, at least those of the easily verifiable kind, in principle!

\section*{Dedication}
To the memory of my loving father and high school mathematics teacher B. Venkateswarlu, 
who lit the initial flame, my first philosopher and personal hero, and to whom I will remain indebted forever;
To the many great mathematicians of past and present who by their work, and the many selfless teachers who by their exposition,
inspired in me a love for mathematics; To the Eternal One who inspires, enables and fulfills everything, I dedicate this 
work with a deep sense of gratitude. \emph{\'S{r}\=i Gurudeva-cara\d{n}\=a{r}pa\d{n}amastu. Om Tat Sat}.

\newpage
\bibliography{ssp_geometry}

\newpage
\section{Appendix}
\label{sec:appendix}

In this section we provide some numerical results to illustrate the behavior of the algorithm and highlight
key metrics for which upper bounds have been derived in the main paper. As pointed out at the beginning of the paper
the algorithm has a fairly highly complexity. 
As a result of this, we could not simulate instances with large $n$ due to high memory demands. 
Furthermore, we have not made adequate efforts to optimize the code.
Nevertheless, the claims of the paper can be understood even with smaller instances.

\subsection{Example $1$: Multiple Solutions}
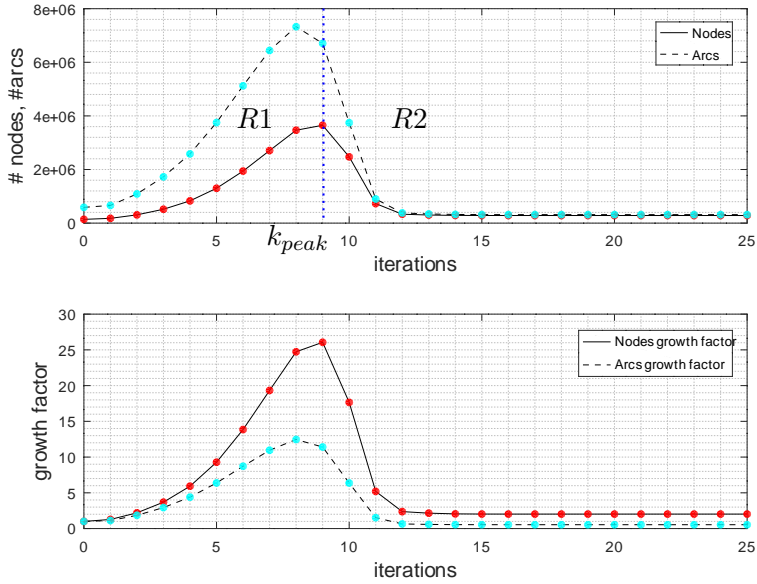
\begin{figure}[!htbp]
       \begin{tikzpicture}
             \node (image) [anchor=south west, inner sep=0pt] {\includegraphics[scale=.56]{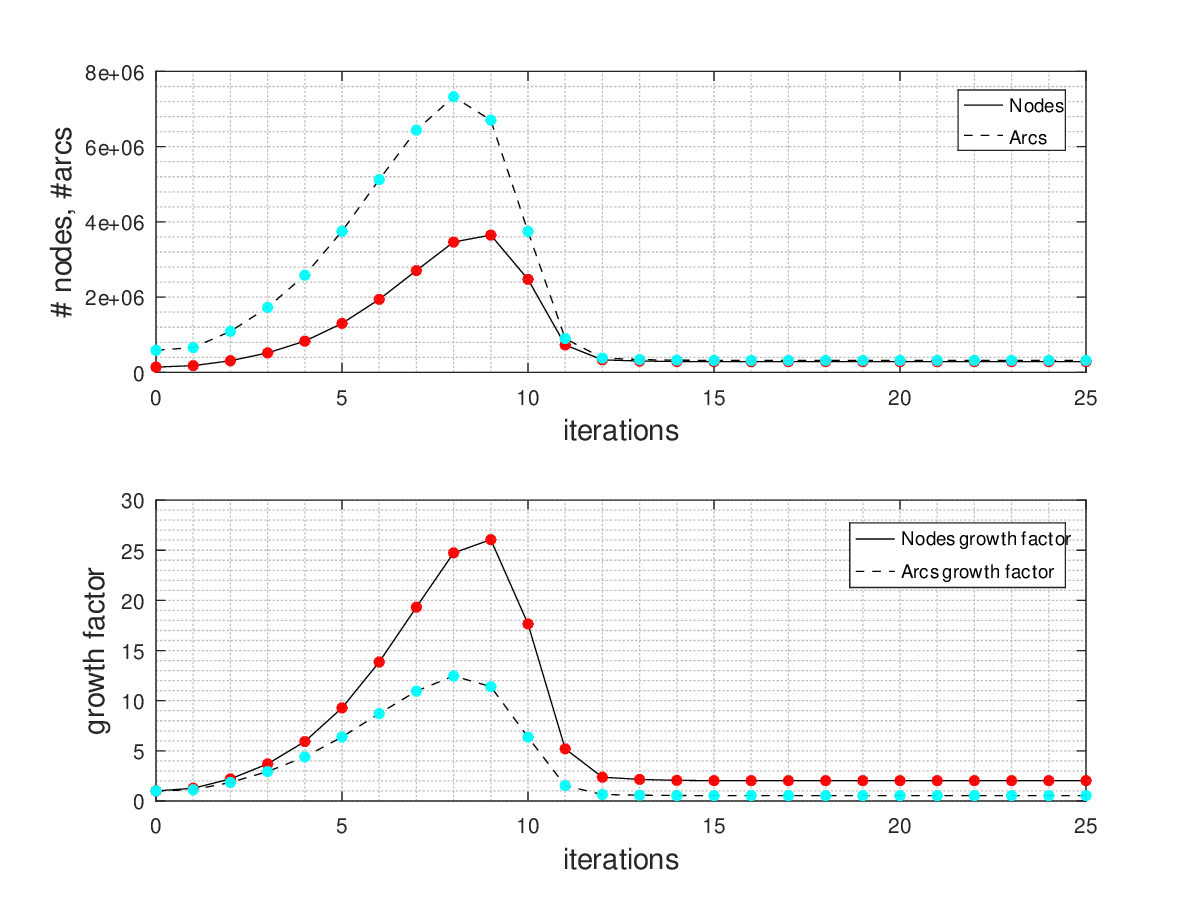}};
             \begin{scope}[x={(image.south east)}, y={(image.north west)}]
                \node  at (.33,.75) {$R1$};
                \node  at (.51,.75) {$R2$};
		\node  at (.38,.56) {$k_{peak}$};
		\draw[thick,dotted,blue] (0.41,0.92) -- (0.41,0.59);
            \end{scope}
          \end{tikzpicture}
	\caption{The number of nodes and arcs as a function of iteration is shown in the top plot. 
	The bottom plot shows the growth factors $\frac{|V_k|}{|V_0|}$ and $\frac{|E_k|}{|E_0|}$
	as a function of iteration $k$. }
	\label{fig:lf}
\end{figure}

We first present an example with $n=40$ and $m=23$, where we have multiple solutions. The sequence ${\bf a}$ is 
\begin{align*}
\begin{array}{cccccccc}
	(102347  & 237178  & 278185  & 428923  & 492462  & 508288  & 678198  & 706739 \\
   914229  & 921016  & 1069771 & 1117479 & 1135293 & 1161956 & 1320170 & 1349570 \\
   1563847 & 1645714 & 1710619 & 1726111 & 1778124 & 1997780 & 2129472 & 2132052 \\
   2351467 & 2537993 & 2613230 & 2878081 & 2943934 & 2992315 & 3096988 & 3580746 \\
	3633163 & 3758382 & 4064504 & 4274011 & 4638683 & 4914859 & 5518984 & 6399285),
\end{array}
\end{align*}
the target number $T = 43,665,189$, and $A_n = 87,302,148$. 
For this instance of SSP, the initial orbital graph has $|V_0| = 140,090$ nodes and $|E_0|= 587,913$ arcs. Recall that
we have the progression of graphs $G_0, G_1, \ldots, G_m$ for the $m$ iterations of the algorithm. 
The number of nodes and arcs as a function of iterations 
is shown in the top plot of~\cref{fig:lf}. The final graph $G_m$ has $283,577$ nodes and $316,691$ arcs, and thus
about $2$ times larger compared to $G_0$. 
The bottom plot shows the overall growth factor of nodes and arcs as 
a function of iteration. The \emph{first} iteration at which we have the maximum number of nodes in the graph is denoted $k_{peak}$.
In this example $k_{peak} = 9$ which is marked by a vertical dotted line. 
At iteration $k_{peak}$ we have $|V_9| = 3,650,487$ and $|E_9| = 6,704,902$, and the maximum growth factor for nodes is about $26$.
The region to the left of the dotted line
is denoted $R1$ where the graph size is increasing compared to the previous iteration. In region $R1$, more nodes are created (by REFINE) than
removed (by FILTER) causing an increase in the graph size.
However, the efficiency of filtering increases with each iteration, and
at iteration $k_{peak}$ the filtering step removes as many nodes as generated by refining. The region to the right of
the dotted line is denoted $R2$ where the graph is non-increasing until only the zero paths remain. In this region, the filtering
step removes at least as many nodes as those created by refining.

Some remarks follow:
\begin{itemize}
	\item[(i)] The plots of~\cref{fig:lf} are typical for any instance of SSP, and not for this example only. However,
		the location of $k_{peak} \in [m]$ and maximum growth factor depend on the instance and degree of additive
		structure.
	\item[(ii)] In this example, the final graph $G_m$ has a size larger than the initial graph. This is typical when
		there are many solutions. It turns out that this example has $47,187$ solutions (zero paths). If a given instance
		has just a few solutions, the region $R2$ is lowered significantly and gets closer to the $x$-axis.
		In the case where there are no solutions (the final graph $G_m = \emptyset$) it coincides with the $x$-axis. 
		When there are large number of solutions,
		as in the case when $m$ is small, the region $R2$ rises and becomes level with the peak graph size.
	\item[(iii)] Note that the peak growth factor $\eta_{peak} \le 2^{k_{peak}}$. In this example $\eta_{peak} =26.058$
		which is much smaller than $2^{k_{peak}} = 512$. Also this much smaller than the theoretical maximum growth factor
		given by $\myceil{7\log{40}} = 38$.
	\item[(iv)] The plots show that an \emph{energy barrier} must be crossed in order to discover the zero paths. This is an intrinsic
		and typical behaviour of our algorithm.
\end{itemize}

\subsection{Example $2$: A Single Solution}
Next, we consider an instance with $n=40$, $m=50$ that has only one solution. The target value for this instance
is $T = 11,734,810,597,199,265$, and the sequence ${\bf a} $ is given below:
\begin{align*}
\begin{array}{cccc}
	{\bf a} = (5340838300510 &  72879613659014 &  106133278417124  & 188686744784933 \\
	196231635078036 &  204742082291972 &  213403962557486 & 244279857200551 \\
	255857319212728 &  273576871530631  & 289708754284101 &  334920772207496 \\
	372914717341816 & 429101169184127 & 441841927212719 & 450764719411565  \\
	461837234112443 & 496024035526859 & 514816372113660 & 537046917450189 \\ 
	562997054764154 & 640773760680379 & 643885354170988 & 705244168633211 \\ 
	707689919002051 & 787375951122724 & 806333455912404 & 815636078884372 \\ 
	854263299158483 & 878538656073276 & 883649606688863 & 915253093666743 \\ 
	917298576163677 & 964114532211852 & 1008165187539165 & 1014126287123623  \\
	1015148570166785 & 1033545047017119 & 1103557815472065 & 1123705011761031).
\end{array}
\end{align*}
Recall that the $x$-coordinate of a \emph{zero path} gives the \emph{index}
of the subset that sums to the target. In this example, the solution $index = 251872521694$, whose binary representation
shows the elements:
\begin{align*}
	[index]_2 & = (0 0 1 1 1 0 1 0 1 0 1 0 0 1 0 0 1 1 0 0 0 1 0 1 1 0 1 0 1 1 0 1 1 1 0 1 1 1 1 0).
\end{align*}

\begin{figure}[!htbp]
\begin{center}
      \epsfxsize=5.0in \epsfysize=3.0in {\epsfbox{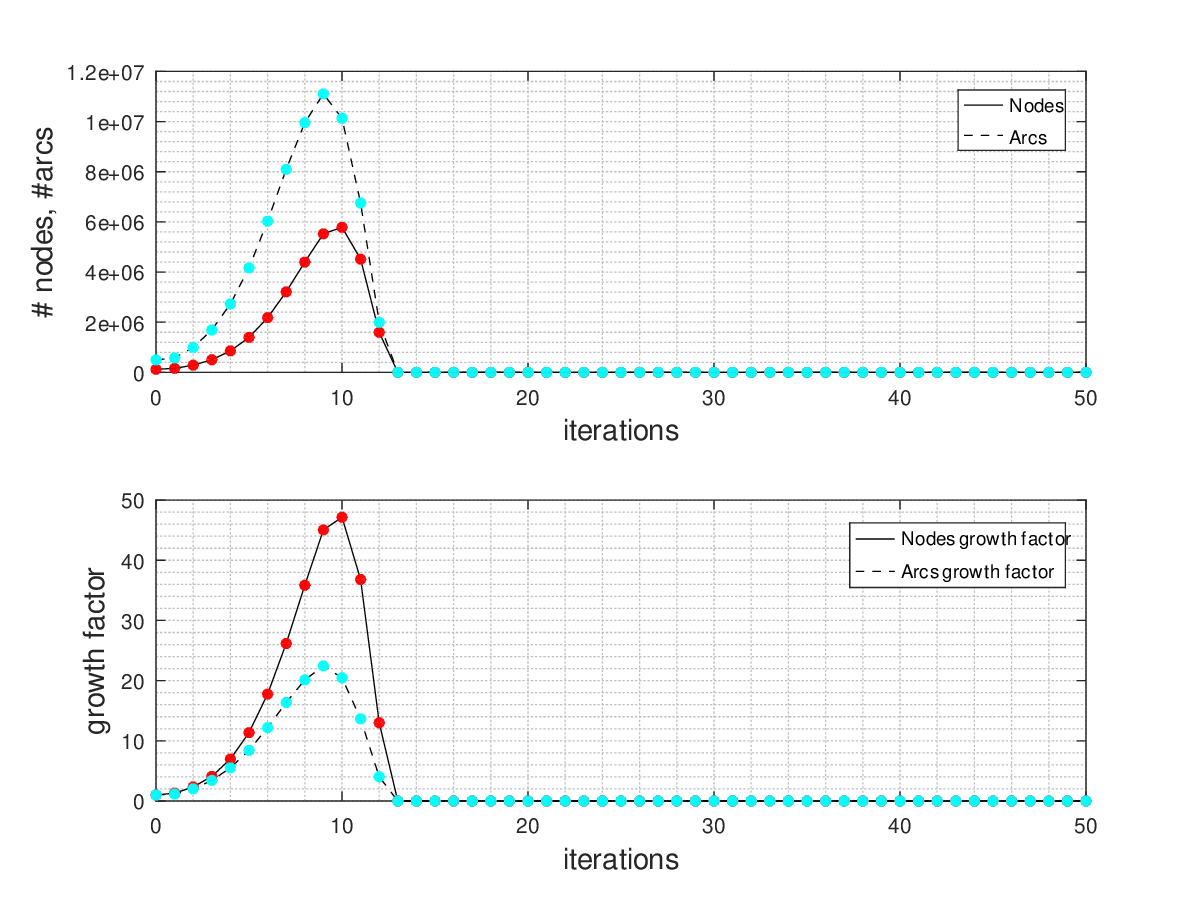}}
        \caption{The graph size and growth factor as a function of iteration for an instance with only one solution.}
       \label{fig:T40}
\end{center}
\end{figure}

The various computed quantities are as follows:
\begin{align*}
	|V_0| = 122695,  \quad   |E_0| = 494892, \\
	|V_{10}| = 5785430, \quad |E_{10}| =  10132432 \\
	|V_m| = |V_{13}| = 47,  \quad      |E_m| = |E_{13}| =    46 \\
	k_{peak} = 10, \quad \text{and } \eta_{peak} = 47.153 < 2^{10} = 1024.
\end{align*}
Again the peak growth is significantly below the theoretical estimates, and the plots are shown in~\cref{fig:T40}.

\subsection{Example $3$: Growth of $k_{peak}$ with $n$}
\begin{figure}[!htbp]
\begin{center}
      \epsfxsize=5in \epsfysize=3in {\epsfbox{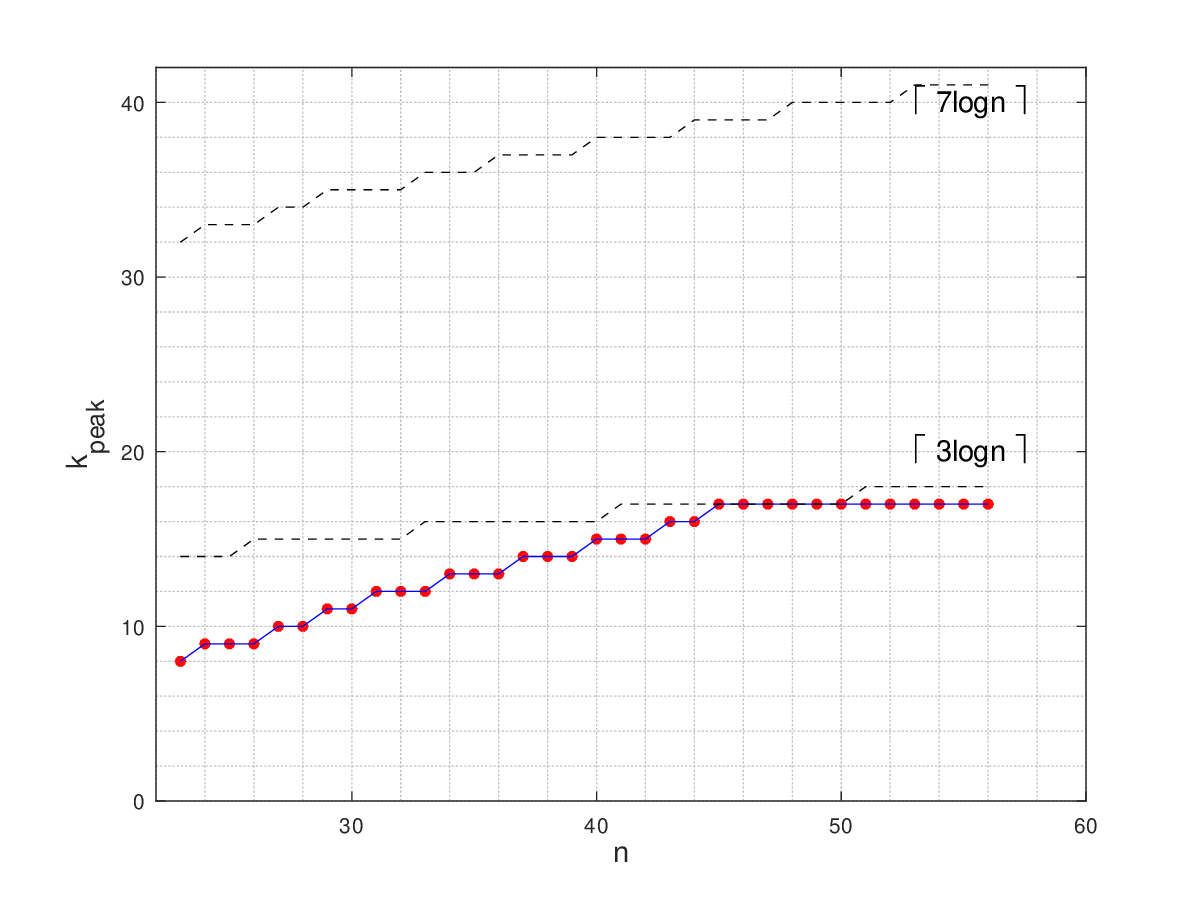}}
	\caption{The variation of $k_{peak}$ with $n$ is shown by the blue curve. The theoretical bound of $\lceil7\log{n}\rceil$ is 
	shown at the top for reference.
	The curve $\lceil3\log{n}\rceil$ is also shown for reference which is closer to data at these small values of $n$. 
        }
       \label{fig:kpeak}
\end{center}
\end{figure}

As noted from the previous example, $k_{peak}$ is a reasonable indicator of the maximum growth factor 
in the algorithm. In~\cref{fig:kpeak} we show the variation of $k_{peak}$ as a function of $n$. For each $n$ we varied
$m \in [10, 2n]$, and for each $(n,m)$ pair we generated a random instance where each $a_i, i \in [n]$ was chosen randomly
from the interval $[1, 2^m]$. The target value was chosen so that $T \approx \frac{A_n}{2}$ since the geometrical structure
would have the widest span, in general, at half the total sum. This also corresponds to maximum possible orbital graph size.
We took the $\max \{k_{peak}\}$ for each $n$ and plotted it. We notice that it is a slow
growing function much like a $c\log{n}$. For reference, we plotted $\lceil 7\log{n}\rceil$ which is the theoretical
bound we obtained for maximum number of distinct zero paths through any node. The plot of $k_{peak}$ for these smaller
values of $n$ is much closer to $\lceil 3 \log{n} \rceil$ than the theoretical bound.

\subsection{Example $4$: Varying $m$}
\label{ex:varym}
\begin{figure}[!htbp]
\begin{center}
      \epsfxsize=6in \epsfysize=4in {\epsfbox{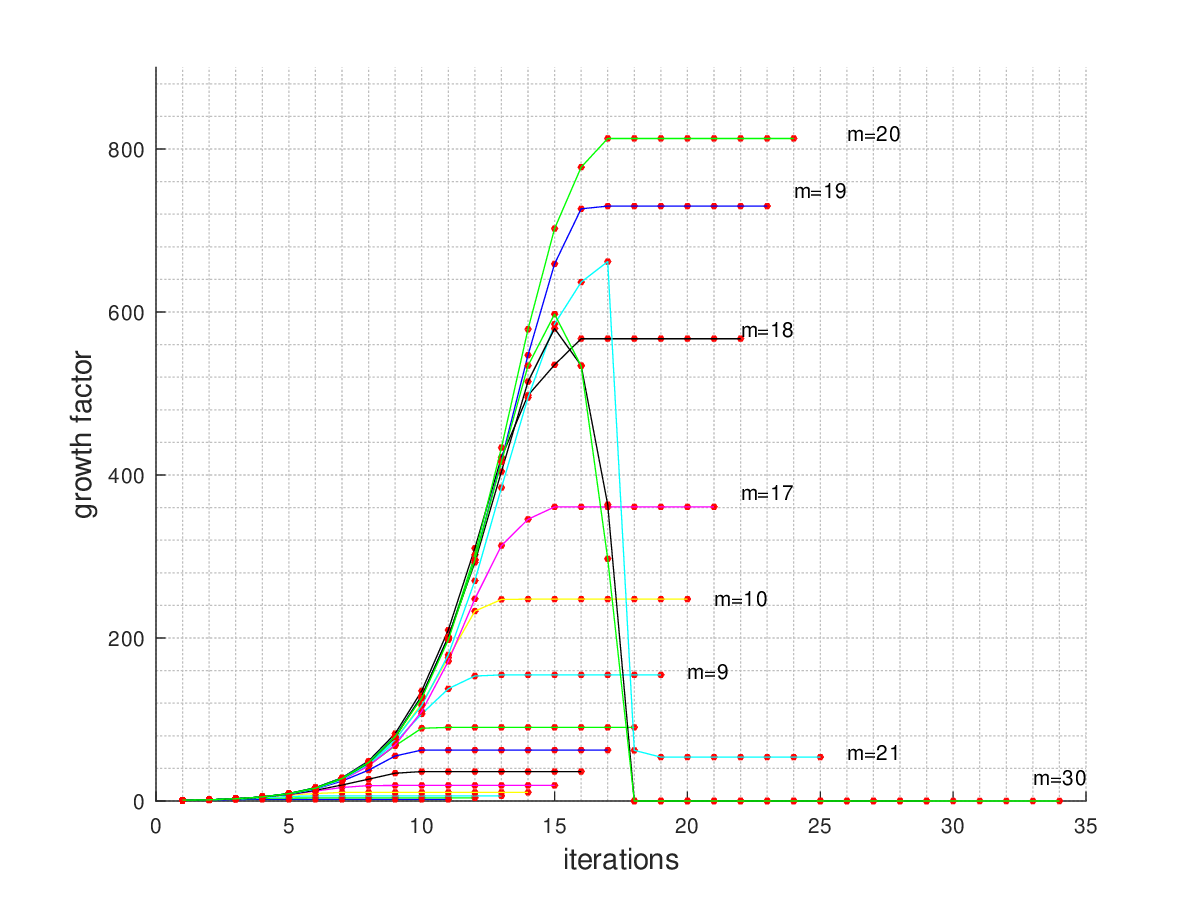}}
	\caption{Growth curves with varying $m$ for $n=50$.}
       \label{fig:ms}
\end{center}
\end{figure}

Next we consider multiple instances of size $n=50$, but varying $m$. In each case the instance was generated by randomly
picking numbers in $[1,2^m]$. We plot the growth curves for the various cases in the same plot to show the effect of $m$
in~\cref{fig:ms}. Some remarks follow:
\begin{itemize}
	\item[(i)] Recall that the peak growth factor corresponds to the maximum number of distinct zero paths through
		an edge. When $m$ is very small, the peak growth factor is limited. As $m$ increases, the growth factor 
		increases as the edges can have more number of distinct zero paths.
	\item[(ii)] As $m$ continues to increase beyond a threshold, the number of solutions decrease. Some of the paths
		will be non-zero paths which will be filtered away. Thus the curve drops and flattens out to support the
		number of zero paths for the given instance. In this plot $k_{peak} \approx 18$ which is less than the 
		theoretical bound $\myceil{7\log{50}} = 40$.
\end{itemize}

\subsection{Example $5$: High degree of additive structure}
\begin{figure}[!htbp]
\begin{center}
      \epsfxsize=5.6in \epsfysize=3in {\epsfbox{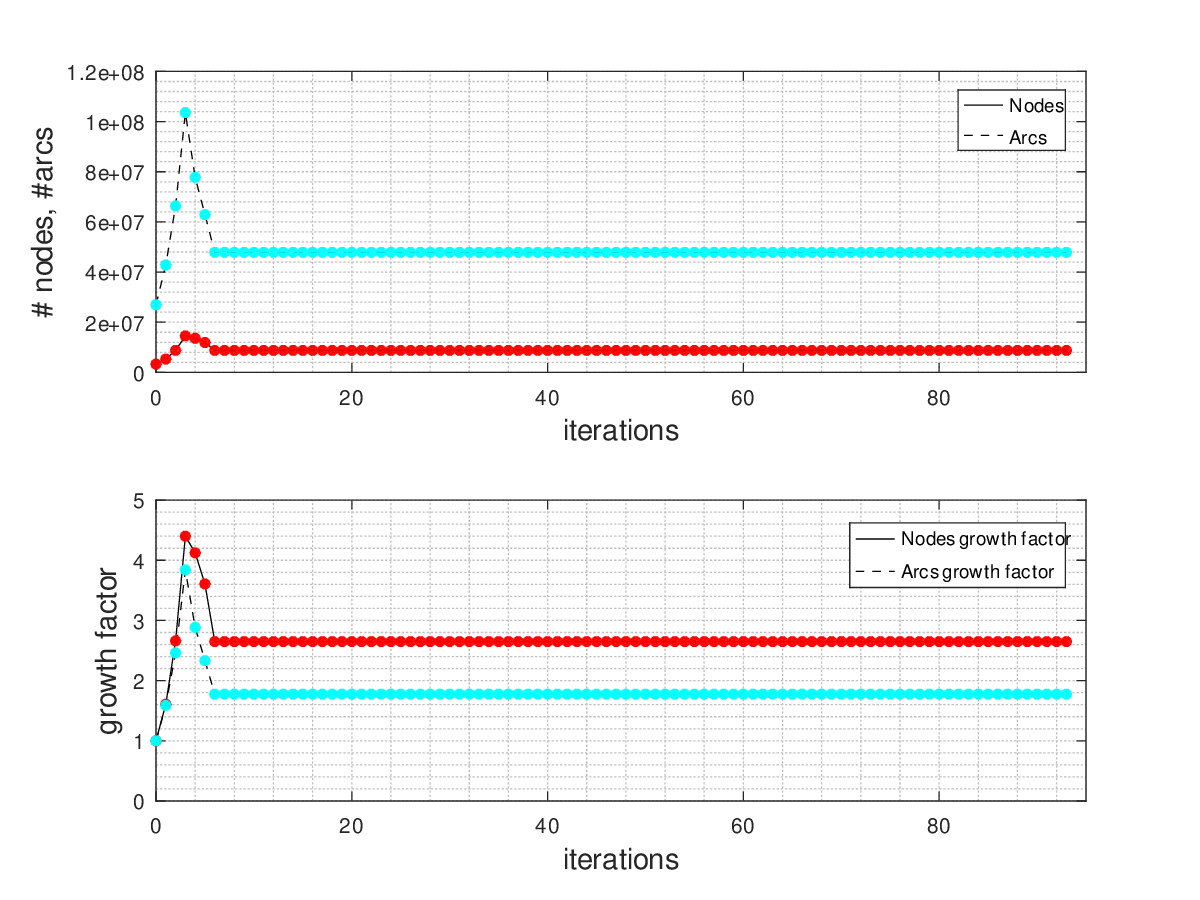}}
	\caption{Progression of graph size and growth factor for an AP, where $k_{peak}=3$. }
      \label{fig:ap90}
\end{center}
\end{figure}

In this example we consider the case ${\bf a} = AP(90,c,d)$ with $n=90$, $m=89$, and where
\begin{align*}
	c & = 19,329,079,171,151,820,605,874,590, \\
	d & = 5,835,760,507,709,645,161,289,732 \\
	\text{and } T & = 12,559,207,227,207,892,175,122,963,296.
\end{align*}
The graph size and the growth factor are shown in~\cref{fig:ap90}.
Inspite of the larger size of the problem, the peak growth factor is small due to the high degree of additive structure.
The number of solutions is $120,723,382,126,197,997,657,472$.

\subsection{Some Combinatorial Applications}
The algorithm's ability to determine the number of solutions is very valuable. We can use it to determine many combinatorial
quantities. Two examples are shown below.
These examples also serve as additional proof of the correctness of the algorithm.

\subsubsection{Binomial Coefficients}
While there are well known efficient methods to computing Binomial coefficients, we can also use the SSP algorithm.
To compute $\binom{n}{k}$, simply choose ${\bf a} = {\mathbf 1}_n$ and $T = k$. As an example, for $n = 80$ and $k = 40$, using the
SSP algorithm, we obtained the number of zero paths to be $107,507,208,733,336,176,461,620$, which is the correct result.

\subsubsection{Restricted Partitions I}
Consider the problem of finding the number of distinct partitions of a given positive integer $N$ where each part is 
at most $K \in {\mathbb N}$. We denote the number of partitions by $R(N,K)$. This is a well known \emph{restricted partitions}
problem for which there is no known closed form expression. Such quantities can be calculated by recursion following a 
\emph{divide and conquer} approach.
We can determine $R(N,K)$ using the SSP, by identifying ${\bf a} = [K]$
and $T = N$. Considering the problem of determining $R(915,60)$, we obtained the number of zero paths to be
\begin{align*}
	R(915,60) = 3,360,682,669,655,028.
\end{align*}
One can verify the correctness of the result by other methods. 

\subsubsection{Restricted Partitions II}
This is like the previous case but we are interested in decomposing a given positive integer $N$ into sum of distinct cubes,
where each part is less than $K^3$ for some $K \in {\mathbb N}$. We denote this quantity by $R_{cubes}(N,K)$.
Considering the problem of determining $R_{cubes}(12345,50)$ we obtain that there are only $7$ distinct decompositions into cubes.
The solution indices are $76790$,  $79382$,  $80038$,  $90506$,  $141210$,  $142491$ and $527286$, whose binary form will give
the elements that sum to the target. The solutions are shown explicitly (in the same order) below:  
\begin{align*}
	12345 & =  2^3  + 3^3 +  5^3 +  6^3 +  7^3 +  8^3 +  9^3 + 10^3 + 12^3 + 14^3 + 17^3 \\
	12345 & =  2^3  + 3^3 +  5^3 + 10^3 + 11^3 + 13^3 + 14^3 + 17^3 \\
	12345 & =  2^3  + 3^3 +  6^3 +  8^3 + 12^3 + 13^3 + 14^3 + 17^3 \\
	12345 & =  2^3  + 4^3 +  8^3 +  9^3 + 14^3 + 15^3 + 17^3 \\
	12345 & =  2^3  + 4^3 +  5^3 +  8^3 +  9^3 + 10^3 + 11^3 + 14^3 + 18^3 \\
	12345 & =  1^3  + 2^3 +  4^3 +  5^3 +  8^3 + 11^3 + 12^3 + 14^3 + 18^3 \\
	12345 & =  2^3  + 3^3 +  5^3 +  6^3 +  8^3 +  9^3 + 10^3 + 12^3 + 20^3.
\end{align*}
Looking just at the first two solutions, we see that
\begin{align*}
	6^3 + 7^3 + 8^3 +9^3 + 12^3 = 11^3 + 13^3.
\end{align*}
It should be clear that innumerable such relations can be extracted, in general,  by having access to all solutions, and that 
many other combinatorial applications are possible. 

\subsection{A remark on size}
In both the above examples, the complexity of finding $R(N,K)$ using other approaches such as Dynamic Programming or Recursion,
will depend on the size of $N$ for a given $K$, and hence can increase without bound. However, in our algorithm the
complexity is limited to a polynomial in $K$, since as we showed, \emph{size} does not matter beyond a threshold!

\end{document}